\documentclass[submission, Phys]{SciPost}

\usepackage{amsmath}
\usepackage{amssymb}
\usepackage{mathtools}
\usepackage{booktabs}
\usepackage{tikz}
\usetikzlibrary{arrows.meta}

\begin{document}

\begin{center}{\Large \textbf{
Macroscopic fluctuation theory of stochastic systems undergoing monitoring
}}\end{center}

\begin{center}
João Costa\textsuperscript{1,2$\star$},
Pedro Ribeiro\textsuperscript{1},
Andrea De Luca\textsuperscript{3}

\end{center}

\begin{center}
{\bf 1} CeFEMA, Instituto Superior Técnico, Universidade de Lisboa, Av. Rovisco Pais, 1049-001 Lisboa, Portugal
\\
{\bf 2} 
Laboratoire de Physique Th\'eorique et Mod\'elisation, CY Cergy Paris Universit\'e, CNRS, F-95302 Cergy-Pontoise, France
\\
{\bf 3} 
Laboratoire de Physique de l'\'Ecole Normale Sup\'erieure, ENS, Universit\'e PSL, CNRS, Sorbonne Universit\'e, Universit\'e Paris-Cité, 75005 Paris, France

${}^\star$ {\small \sf joao.sanfins.costa@tecnico.ulisboa.pt}
\end{center}

\begin{center}
\today
\end{center}

\section*{Abstract}
{\bf
We develop a macroscopic fluctuation theory for diffusive classical lattice gases subject to continuous coarse-grained monitoring. Treating measurements in the typical-outcome (quenched) regime, we show that observation adds a quadratic confining term to the MFT action that pins the density to its hydrodynamic profile. We derive sufficient conditions under which, in the presence of monitoring, the current large-deviation function remains convex, the additivity principle holds, and dynamical phase transitions are excluded. For particle-hole-symmetric one-dimensional models we find that monitoring suppresses dynamical phase transitions and can change their order from second to first, as we demonstrate in the Katz--Lebowitz--Spohn model. In the strong-measurement limit, we find that the current distribution becomes Gaussian and determine the scaling of its cumulants: they follow simple power laws when the mobility remains nonzero throughout the system, but exhibit anomalous logarithmic scaling when the mobility vanishes at the boundaries.}

\vspace{10pt}
\noindent\rule{\textwidth}{1pt}
\tableofcontents\thispagestyle{fancy}
\noindent\rule{\textwidth}{1pt}
\vspace{10pt}

\section{Introduction} 

Measurements play a central role in shaping the dynamics and information structure of quantum systems. In recent years, the discovery of measurement-induced phase transitions (MIPTs)~\cite{PhysRevB.100.134203, PhysRevX.12.041002, PhysRevB.101.104302, PhysRevB.109.174307} in quantum circuits has revealed that monitoring can trigger qualitative changes in the nature of quantum correlations and in the information acquired by an observer. These transitions—initially characterized by shifts in entanglement scaling—have also been reframed as learnability transitions, in which the measurement outcomes undergo a qualitative change in the information they reveal about the system’s initial state or conserved quantities.

In the quantum setting these information-theoretic effects are conditioned on Born-distributed measurement outcomes; their classical counterpart, considered throughout this work, is the conditioning of the observer's probability distribution on a measurement record.

This shift in perspective raises a broader question: To what extent are learnability transitions uniquely quantum, and which of their essential features persist in classical systems? Quantum mechanics is not the only setting in which measurements shape an observer’s description of a system. In classical systems, although the dynamics of observables are fundamentally deterministic and unaffected by whether they are being monitored, an observer typically cannot access or track the full microscopic state or exact trajectory of the system. As a result, they must adopt a statistical description, representing the state of the system as described by the observer—that is, their knowledge—as a probability distribution over configurations. Measurements—understood here as partial observations—serve to refine this distribution by incorporating new information. Thus, even in the absence of quantum indeterminacy, monitoring a subset of observables can significantly affect the observer’s evolving state description, suggesting a natural classical counterpart to the measurement-induced phenomena seen in quantum settings.
Yet, while the role of measurements in quantum systems has been extensively investigated, their impact in classical settings has received comparatively little attention until recently~\cite{nahum2025_ClassicalMeasurements, gopalakrishnan2025_ClassicalMeasurements, gerbino2025measurement, PhysRevE.111.024135}. In classical systems, measurements are often framed as problems of inference—where one seeks to reconstruct hidden variables or trajectories from partial or noisy observations~\cite{PhysRevE.111.024135, 10.5555/897831, DBLP:journals/pami/YuilleC00}, a perspective particularly prominent in machine learning. A representative example of such inference problems is found in~\cite{PhysRevE.111.024135}, where the authors considered reconstructing the trajectory of a random walker from a sequence of noisy images, a task that maps onto a generalized directed polymer problem in statistical physics.

More broadly, such inference problems often admit elegant formulations in terms of equilibrium statistical mechanics~\cite{NishimoriBook2001, Ricci_Tersenghi_2019, gerbino2025measurement}. Bayesian reconstruction from noisy data, for example, is known to correspond to a random-bond Ising model along the Nishimori line~\cite{H.Nishimori_1980}, where the correspondence between temperature and data noise yields exact identities and thermodynamic consistency. These connections reveal a deep structural analogy between inference and equilibrium phenomena.

Measurement-induced and learnability transitions have also been studied in classical settings, including chaotic many-body systems~\cite{2022PhRvB.106b4305W}, diffusive dynamics~\cite{ha2024measurement}, inference and hidden-Markov problems~\cite{2025arXiv250408888K}, learning transitions in classical lattice models~\cite{2025arXiv250412385P}, and Bayesian criticality in the Ising model~\cite{wiese2026bayesian}, with the underlying inference$\leftrightarrow$statistical-mechanics correspondence reviewed in~\cite{zdeborova2016statistical,iba1999nishimori}.

Building on these insights, recent work has begun to explore directly how measurements affect classical systems in equilibrium. In particular,~\cite{nahum2025_ClassicalMeasurements} examined a broad class of monitored classical lattice models, revealing rich phenomenology such as charge-sharpening transitions and intricate phase diagrams. These transitions, previously predicted in monitored quantum systems~\cite{PhysRevX.12.041002}, reflect a dramatic restructuring of the charge profile conditioned on measurement outcomes. The study also proposed an effective field theory describing these transitions in non-equilibrium settings—a framework that was independently developed and analyzed in~\cite{gopalakrishnan2025_ClassicalMeasurements}. In~\cite{gopalakrishnan2025_ClassicalMeasurements}, the authors also considered monitored ensembles of prototypical lattice models such as the Symmetric and Asymmetric Simple Exclusion Processes (SSEP and ASEP). These models describe particles stochastically hopping on a lattice with exclusion constraints, with ASEP introducing a directional bias to the dynamics. While SSEP and ASEP belong to distinct universality classes in the absence of monitoring, it was shown that under measurement, their large-scale dynamics flow to the same fixed point, illustrating how measurements can unify otherwise disparate non-equilibrium behaviors.

A crucial distinction between our setup and the works above is the \emph{coarse-grained} nature of the monitoring: we measure the particle density averaged over a mesoscopic region containing a large number of sites, rather than site-resolved occupations. This choice is what makes monitoring compatible with the macroscopic fluctuation theory description. We emphasize that, due to this choice of coarse-grained measures, our protocol differs from those that give rise to charge-sharpening transitions associated with an on-site particle-number measure with discrete values~\cite{gopalakrishnan2025_ClassicalMeasurements, nahum2025_ClassicalMeasurements}. 

In this work, we shift the focus to transport properties in monitored driven classical lattice systems—a regime not addressed in previous studies, which primarily considered closed systems and bulk observables. To this end, we study systems with open boundary conditions, where particles are exchanged with reservoirs at the edges, and analyze how continuous monitoring affects the statistics of particle currents. To study this, we consider the coarse-grained (hydrodynamic) limit, where large-scale behavior is governed by Macroscopic Fluctuation Theory (MFT)~\cite{RevModPhys.87.593, PhysRevLett.94.030601}—a powerful framework for describing non-equilibrium steady states and their fluctuations in diffusive systems. MFT enables the computation of large deviation functions, such as the cumulant generating function of the current, and provides variational principles that capture the most likely evolution conditioned on rare events. Within this framework, we introduce measurement protocols compatible with the large-scale description, allowing us to study how monitoring interacts with boundary-driven transport at the hydrodynamic scale.

Formally, averaging observables over measurement outcomes can be seen as a disorder average in an appropriate stat-mech problem. A standard way to deal with this problem is to introduce $N$ copies, or replicas, of the system. Eventually, restoring the appropriate statistical weights requires the formal replica limit $N\to1$. This differs from the $N\to0$ limit relevant to quenched disorder or forced (post-selected) measurements~\cite{nahum2025_ClassicalMeasurements,gopalakrishnan2025_ClassicalMeasurements}. In this work, the use of MFT allows us to avoid the use of replicas to study quenched disorder as it induces a concentration of measure around a typical set of measurement outcomes.

The remainder of this section is structured as follows. We first introduce the underlying microscopic stochastic lattice gas and the observer's a-priori description (subsection~\ref{subsec:micro}), of which MFT is the coarse-grained hydrodynamic limit. We then review the key principles of MFT relevant to our setting (subsection~\ref{subsec:revMFT}). We finally define the notion of coarse-grained measurements and explain how they can be systematically incorporated into the MFT description of classical transport (subsection~\ref{sec:MeasForm}).

\subsection{Microscopic stochastic lattice gases and the observer's description}\label{subsec:micro}
We begin from the underlying microscopic stochastic dynamics; Macroscopic Fluctuation Theory, reviewed in the following subsection, is its coarse-grained hydrodynamic description, and monitoring will then be introduced directly at this microscopic level.
The models we focus on in this work are stochastic lattice gases~\cite{Landim_Bertini_2007,Eyink1990,spohn2012large,kipnis2013scaling,kipnis1989hydrodynamics}, the class of Markovian models from which Macroscopic Fluctuation Theory (MFT) originates. These capture essential features of particle transport and interactions on a discrete spatial lattice. To be concrete, we consider a bounded physical connected domain in $d$ spatial dimensions $\Lambda \subset \mathbb{R}^d$ and define a discretized version $\Lambda_L = \Lambda \cap \left(\mathbb{Z}/L\right)^d$, such that the thermodynamic limit corresponds to $L \to \infty$. The microscopic state of the system at time $t$ is denoted by a configuration $C_t$, which evolves according to a continuous time Markov process. Updates occur via Poisson processes, with rates that depend locally on $C_t$, reflecting interaction rules such as particle hopping, reactions, or exchanges. These models are particularly attractive for our purposes not only due to their tractability and universality, but also because they naturally arise in other contexts. For instance, they can be interpreted as the fully dephased limit of quantum channels, corresponding to quantum systems undergoing strong decoherence~\cite{Marcuzzi_2014,Bauer_2015,10.21468/SciPostPhys.3.5.033}. In this sense, stochastic lattice gases provide a natural bridge between classical and quantum regimes, making them an ideal setting for studying the effects of monitoring.

We further allow the system to interact with external particle reservoirs through its boundary $\partial \Lambda$, modeled by creation and annihilation processes at boundary sites. In this work, for simplicity, we restrict attention to a single species of particles subject to a local conservation law for particle number.

We can now make explicit the protocol that underlies our discussion. The system follows a specific trajectory in time $\mathcal{C}^{(0)} = \left\{C^{(0)}_t \right\}_t$, where $C^{(0)}_t$ denotes a microscopic configuration at time $t$, specifying the occupation number at each site of the discretized domain $\Lambda_L$. In principle, the underlying microscopic dynamics can be deterministic but controlled by inaccessible microscopic quantities. Thus, we assume a probabilistic description, where the configuration $C^{(0)}_t$ jumps stochastically with transition rates $H_{C,C'}$. We assume the observer does not have access to the exact trajectory $\mathcal{C}^{(0)}$ but has full knowledge of the Markov transition rates (this hypothesis can also be relaxed in the context of inference, see for instance~\cite{PhysRevE.111.024135}). Therefore, in the absence of monitoring, their best estimate of the system state is encoded in the probability measure $d\mathbb{P}_{\Lambda_L,[-t,0]}(\mathcal{C})$ over trajectories $\mathcal{C}$ in the configuration space
of the Markov process generated by $H_{C,C'}$. This represents the a-priori distribution over trajectories by the observer. The marginal at a given time satisfies the Fokker-Planck equation
\begin{equation}
\label{eq:PXevol}
P_t(C') = \int_{C_t = C'} d\mathbb{P}_{\Lambda_L,[-t,0]}(\mathcal{C}) \quad \Rightarrow \quad \frac{dP_t(C)}{dt} = - \sum_{C'} H_{C,C'} P_t(C'),
\end{equation}

\subsection{Macroscopic fluctuation theory as the coarse-grained description}\label{subsec:revMFT}
We now recall how MFT arises as the coarse-grained, hydrodynamic large-deviation description of the microscopic process introduced above. MFT provides a framework for describing the leading-order fluctuations around the deterministic hydrodynamic equations governing general diffusive systems~\cite{Landim_Bertini_2007}. A key assumption underpinning the macroscopic description of such systems is local equilibrium: within any mesoscopic region of space—large compared to the lattice scale, but small compared to the system size—the system rapidly equilibrates to a local stationary state. That is, in a region containing $l^d$ sites such that $l \gg 1$ and $l/L \ll 1$, local thermodynamic variables can be well defined. This assumption is well justified and rigorously established in various models~\cite{spohn2012large,kipnis2013scaling}.
In the diffusive regime relevant to our discussion, the macroscopic limit is obtained by rescaling space and time as $\mathbb{Z}^d \ni i \to x = i/L \in \Lambda_L$ and $t \to \tau = t/L^2$. Under this diffusive scaling, the coarse-grained particle density $\rho_L(x,\tau)$ becomes a smooth function of space and time. Conservation of particle number implies the existence of an associated current $\mathbf{j}_L(x,\tau)$ satisfying the continuity equation 
\begin{equation}
\label{eq:conslaw}
\partial_\tau \rho_L(x,\tau) + \nabla \cdot \mathbf{j}_L(x,\tau) = 0 \;.
\end{equation}
At finite $L$, the operator $\nabla \cdot$ should be understood as a suitably discretized or regularized version of the divergence, which converges to the continuum operator in the limit $L \to \infty$. Additionally, throughout this work, vector fields such as the current are denoted in boldface (e.g., $\mathbf{j}$) to distinguish them from scalar fields.

Fluctuations around the typical macroscopic profiles of density and current over the time window $[\tau_0,\tau_1]$, in the large-$L$ limit, are described by a probability distribution $\mathbb{P}_{\Lambda_L,[\tau_0,\tau_1]}$ over the space of empirical trajectories $\rho_L(x,\tau)$ and $\mathbf{j}_L(x,\tau)$. At this point we make the connection with MFT, which is built on the assumption that $\mathbb{P}_{\Lambda_L,[\tau_0,\tau_1]}$ satisfies a Large Deviation Principle (LDP) as $L \to \infty$, with exponentially suppressed weights for atypical trajectories governed by a rate functional playing the role of a nonequilibrium action~\cite{Touchette_2009,Touchette_2012}. Indeed, for lattice gases with a single "type" of particle~\cite{Landim_Bertini_2007}, it can be expressed in the form
\begin{equation}\label{eq:BigProb}
    \mathbb{P}_{\Lambda_L,[\tau_0,\tau_1]} \left(\mathbf{j}_{L}\approx \mathbf{j}\left|\,\rho_{L}(\tau_0)=\rho_{\tau_0}\right.\right) \asymp\exp\big\{-L^{d}\,\mathcal{I}_{[\tau_0,\tau_1]}\left(\rho[\rho_{\tau_0},\mathbf{j}],\mathbf{j}\right)\big\},
\end{equation}
where the symbol $\asymp$ denotes asymptotic equivalence in the large-$L$ limit and $\rho[\rho_{\tau_0},\mathbf{j}](x,\tau) = \rho_{\tau_0}(x) - \int_{\tau_0}^{\tau} ds\; \nabla \cdot \mathbf{j}(x,s)$. 
Note that, because of the conservation law \eqref{eq:conslaw}, the manifold of particle-conserving trajectories $(\rho, \mathbf{j})$ between times $\tau_0$ and $\tau_1$ can be parametrized by the initial density profile $\rho_{\tau_0}$ and the current $\mathbf{j}$  since the density at later times is determined via $\rho = \rho[\rho_{\tau_0}, \mathbf{j}]$.
The rate functional $\mathcal{I}$ is given by
\begin{equation}\label{eq:BigRate}
    \mathcal{I}_{[\tau_0,\tau_1]}(\rho,\mathbf{j}) = \int_{\tau_0}^{\tau_1} d\tau\;  \mathcal{U}(\rho(\tau),\mathbf{j}(\tau)) ,\text{ for  }   \;  \mathcal{U}(\rho,j) = \frac{1}{2}\int_\Lambda dx\; [\mathbf{j}-\mathcal{J}(\rho)] \cdot \sigma(\rho)^{-1}[\mathbf{j}-\mathcal{J}(\rho)],
\end{equation}
where, for general isotropic diffusive systems subjected to a weak external field $\mathbf{E}$,
\begin{equation}\label{eq:constrel}
    \mathcal{J}(\rho) = - D(\rho) \nabla_x \rho +  \sigma(\rho) \mathbf{E}.
\end{equation}
Remarkably, within this framework, the whole microscopic properties of the system are encoded in the density functions $D(\rho)$ and $\sigma(\rho)$, known respectively as the diffusion constant and the mobility. Eq.~(\ref{eq:constrel}) is termed the constitutive relation and it specifies the deterministic current profile in the thermodynamic limit $L \to \infty$ as a function of the density $\rho(x,\tau)$. 

The object $ \mathbb{P}_{\Lambda_L,[\tau_0,\tau_1]} \left(\mathbf{j}_{L}\approx \mathbf{j}\left|\,\rho_{L}(\tau_0)=\rho_{\tau_0}\right.\right) $ should thus be understood as a probability distribution over this manifold of coarse-grained trajectories, conditioned on the initial density $\rho_{\tau_0}$. This macroscopic object plays the role of an effective transition probability—directly analogous to the Markov transition functions that govern the underlying microscopic dynamics. In this coarse-grained description, the density encodes the relevant features of the microscopic configuration, while the current captures the effective transition process between configurations.
On the boundary $\partial \Lambda$, the suitable boundary conditions need to be imposed conditioned on how the system interacts with its surroundings. 

\subsection{Large deviation and cumulant generating functionals}\label{sec:LD&CGF}
Eqs.~(\ref{eq:BigProb}-\ref{eq:BigRate}) provide a powerful formula which enables the extension of some standard techniques from equilibrium physics to an out of equilibrium setting. A typical application of Eqs.~(\ref{eq:BigProb}-\ref{eq:BigRate}) is in the characterization of the system's steady state, which generally consists in the study of the density and current fluctuations. This is achieved by noting that, in the limit $\tau_0 \to -\infty$, the system will have had time to reach the steady state by $\tau_1=0$, allowing us to write the steady state density's probability distribution as
\begin{equation}\label{eq:Pprobdistr}
    \mathbb{P}_{\Lambda_L,\infty}(\widetilde{\rho}) = \lim_{\tau \to \infty}\int_{\rho[\rho_{-\tau},\mathbf{j}](0)=\widetilde{\rho}} D[\mathbf{j}] \;\mathbb{P}_{\Lambda_L,[-\tau,0]}\left(\mathbf{j}_L \approx\mathbf{j}\left|\;\rho_L(-\tau)=\rho_{-\tau}\right.\right),
\end{equation}
for $D[\mathbf{j}]$ the uniform integration measure on the space of current profiles. Of course from Eq.~(\ref{eq:BigProb}), $\mathbb{P}_{\Lambda_L,\infty}(\widetilde{\rho})$ also satisfies a large deviation principle with 
\begin{equation}\label{eq:ratefunctionofthedensity}
    \varphi(\widetilde{\rho}) = \min_{\mathbf{j}:\rho[\rho_{-\infty},\mathbf{j}](x,0)=\widetilde{\rho}(x)} \mathcal{I}_{[-\infty,0]}(\rho[\rho_{-\infty},\mathbf{j}],\mathbf{j})\Longrightarrow \mathbb{P}_{\Lambda_L,\infty}(\widetilde{\rho}) \asymp e^{-L^d \varphi(\widetilde{\rho})}.
\end{equation}
Note that even though, in the previous two equations, we had to specify some initial condition $\rho_{-\tau}$, after the limit $\tau \to \infty$ the steady state should be independent of $\rho_{-\tau}$.
On the other hand, steady-state current fluctuations are typically analyzed via the probability distribution of the time-averaged (in time) current $\mathbf{J}_{\tau}[\mathbf{j}](x) =  \frac{1}{\tau} \int_{-\tau}^0 d\tau' \; \mathbf{j}(x,\tau')$,
\begin{equation}\label{eq:Qprobdistr}
    \mathbb{Q}_{\Lambda_L,\tau}(\mathbf{J}) = \int_{\mathbf{J}=\mathbf{J}_{\tau}[\mathbf{j}]} D[\mathbf{j}] \;\mathbb{P}_{\Lambda_L,[-\tau,0]}\left(\mathbf{j}_L \approx\mathbf{j}\left|\;\rho_L(-\tau)=\rho_{-\tau}\right.\right),
\end{equation}
which also admits a large deviation description,
\begin{equation} \label{eq:DefPhiJ}
     \phi(\mathbf{J}) = \lim_{\tau \to \infty}\frac{1}{\tau}\min_{\mathbf{j}:\mathbf{J}=\mathbf{J}_{\tau}[\mathbf{j}]} \mathcal{I}_{[-\tau,0]}(\rho[\rho_{-\tau},\mathbf{j}],\mathbf{j}) \Longrightarrow  \mathbb{Q}_{\Lambda_L,\tau}\left(\mathbf{J}\right) \asymp e^{-\tau L^d \phi(\mathbf{J})}.
\end{equation}
In this context, $\asymp$ means asymptotic equivalence in the large $L$ and $\tau$ limit.
Note that, because of conservation of the number of particles, the rate functional $\phi(\mathbf{J})$ is defined over the space of divergence free vector fields on $\Lambda$, i.e. $ \lim_{\tau \to \infty} \nabla \cdot \mathbf{J}_{\tau}[\mathbf{j}] = 0$, as it would otherwise imply either negative density or particle condensation.

The celebrated Gärtner-Ellis theorem (see for instance \cite{Touchette_2009}) allows us to relate these rate functionals with the respective cumulant generating functionals (CGF) through a Legendre-Fenchel transform. Explicitly for the current, one has that $\phi(\mathbf{J})$ can be computed from the CGF,
\begin{equation}\label{eq:BigCurrGenv}
     \lambda(\mathbf{v}) = \lim_{\tau \to \infty} \lim_{L \to \infty} \frac{1}{L^d \tau} \log \mathbb{E}_{\Lambda_L,\tau}^{(\mathbb{Q})}\left[e^{\tau L^d \int_{\Lambda} dx\; \mathbf{v}(x) \cdot \mathbf{J}_{\tau}(x)}\right],
\end{equation}
through the Legendre transform
\begin{equation}
\label{eq:legphilambda}
\phi(\mathbf{J}) = \max_{\mathbf{v}} \left(\int_{\Lambda} dx\; \mathbf{v}(x) \cdot \mathbf{J}(x) - \lambda(\mathbf{v})\right)
\end{equation}
provided $\lambda(\mathbf{v})$ is differentiable (with $\mathbf{v}(x)$ a divergence free vector field, $\nabla \cdot \mathbf{v}(x) = 0$). Here, $\mathbb{E}_{\Lambda_L,\tau}^{(\mathbb{Q})}[\ldots]$ corresponds to the expectation value with respect to the measure $\mathbb{Q}_{\Lambda_L,\tau}$, defined in Eq.~(\ref{eq:Qprobdistr}). Similarly, one could also define a rate/cumulant generating function for the density in the steady state, taking instead as large parameter the system size $L$ and setting $\tau \to \infty$ from the outset. More explicitly, we write the density's CGF as 
\begin{equation}\label{eq:DensGenF}
     \mathcal{G}(\alpha) = \lim_{L \to \infty} L^{-d}\log \mathbb{E}_{\Lambda_L,\infty}^{(\mathbb{P})}\left[e^{L^d \int_{\Lambda} dx\; \alpha(x) \tilde{\rho}(x)}\right],
\end{equation}
which, in similarity to the current case, can be used to compute the density's rate function through $\varphi(\tilde{\rho}) = \max_{\alpha} \left(\int_{\Lambda} dx\; \alpha(x)\tilde{\rho}(x) -\mathcal{G}(\alpha)\right)$. Again, here we define $\mathbb{E}_{\Lambda_L,\infty}^{(\mathbb{P})}[\ldots]$ to be the expectation value with respect to the measure $\mathbb{P}_{\Lambda_L,\infty}$ introduced in Eq.~(\ref{eq:Pprobdistr}).

Both $\lambda(\mathbf{v})$ and $\mathcal{G}(\alpha)$ have been extensively studied in the literature for distinct lattice gas models driven out of equilibrium both by interaction with reservoirs at different chemical potentials and influence of an external field. We now briefly summarize key results for MFT models in one spatial dimension ($d=1$), with particular emphasis on the symmetric simple exclusion process (SSEP), which plays a central role in this work due to its exceptional analytical tractability. SSEP describes a system in which each site can be occupied by at most one particle, and particles hop to neighboring sites at uniform rates, as long as the target site is empty (the exclusion constraint). Indeed, SSEP is a diffusive model that admits a MFT description in the large-$L$ limit, with transport coefficients given by
\begin{equation}\label{eq:SSEPtranspcoeff}
    D(\rho)=1,\;\sigma(\rho)=2\rho \left(1-\rho\right).
\end{equation}
In $d=1$, the system's spatial domain can always be rescaled to the unit interval $\Lambda=[0,1]$. 
Additionally, in 1D,  the fact that $\mathbf{J}(x)$ and $\mathbf{v}(x)$ are divergence free imposes that their derivatives vanish identically, and thus these vector fields become space independent: $\mathbf{J}(x)=J$ and $\mathbf{v}(x)=v$, for $J,v\in \mathbb{R}$. Considering that the system interacts with reservoirs through both of its edges $x=0$ and $x=1$, conservation of particles implies that, for very large times, the total net amount of particles transferred to the right reservoirs, $Q_{\text{R}}(\tau)$, satisfies $\mathbf{J}_{\tau}[j]\asymp \frac{Q_{\text{R}}(\tau)}{\tau L}$ (the same applies to the net amount of particles transferred from the left reservoirs $Q_{\text{L}}(\tau)$). This enables us to recast Eq.~(\ref{eq:BigCurrGenv}) into a form more commonly encountered in the literature,
\begin{equation}\label{eq:BigCurrGens}
    \lambda(v) =  \lim_{\tau \to \infty} \lim_{L \to \infty} \frac{L^{-1}}{\tau} \log \mathbb{E}_{\Lambda_L,\tau}^{(\mathbb{Q})}\left[e^{v Q_{\text{R}}(\tau)}\right].
\end{equation}
In fact, in the particular case of SSEP, one can derive a closed analytical expression for $\lambda(v)$ (see~\cite{Derrida2004,Derrida_2007}),
\begin{equation}\label{eq:analyticalssep}
    \lambda_{\text{SSEP}}(v) = \left(\text{arccosh}\left(\sqrt{\left(1 + \left(e^v - 1\right) n_0\right) \left(1 + \left(e^{-v} - 1\right) n_1\right) }\right)\right)^2,
\end{equation}
where $\rho(0) = n_0$ and $\rho(1) = n_1$ are the boundary conditions imposed by the reservoirs. 
Remarkably, the exact same formula holds, up to an overall geometry-dependent prefactor, for a general region $\Lambda$ in higher dimensions where the reservoirs couple through two distinct boundary points, while the rest of the boundary is impermeable, preventing any current from flowing out, see~\cite{Akkermans_2013}. Eq.~(\ref{eq:analyticalssep}), and quantities directly extracted from it, will serve as a point of comparison throughout the paper, especially in section~\ref{sec:LMS}. Of particular relevance are the cumulants of the current, defined through $\kappa_n=\left.\partial_v^n\lambda(v)\right|_{v=0}$.

More generally, although a closed-form expression for the CGF is not always available, one can nevertheless derive a system of two coupled equations (of course dependent on the model through $D(\rho)$ and $\sigma(\rho)$) that implicitly determine $\lambda(v)$ for all MFT models—under the boundary conditions described above and in arbitrary dimensions—provided the so-called additivity principle holds, see~\cite{DerridaAdditivityPrinciple,Akkermans_2013}.
In essence, the additivity principle corresponds to the assumption that, for any fixed integrated current $\mathbf{J}(x)$, satisfying $\nabla \cdot \mathbf{J}(x)=0$, the current profile $\mathbf{j}(x,\tau')$, $\tau'\in[-\tau,0]$, that minimizes $\mathcal{I}_{[-\tau,0]}$ can be written as
\begin{equation}\label{eq:APjJ}
    \mathbf{j}(x,\tau')=\delta \mathbf{j} (x, \tau') + \mathbf{J}(x),
\end{equation}
where $\delta \mathbf{j}$ accounts for the transient current, which vanishes long after the initial time $-\tau$, i.e. $\delta \mathbf{j}(x, \tau') \to 0$ for $\tau'+\tau\to\infty$. 
Although it vanishes at large times, the transient current can still be responsible for a non-trivial density profile. More explicitly, by conservation of particles, the density profile has an asymptotic limit as
\begin{equation}
\label{eq:APrho}
\rho[\rho_{-\tau}, \mathbf{j}](x,\tau') = \rho_{-\tau}(x) - \int_{-\tau}^{\tau'} ds\, \nabla\cdot \delta \mathbf{j}(x,s) \longrightarrow \rho(x) \quad \text{for } \tau'+\tau\to\infty,
\end{equation}
where we used that $\nabla \cdot \mathbf{J}(x) = 0$. Then,
\begin{equation}
\label{eq:largetauItoU}
    \lim_{\tau \to \infty} \frac{1}{\tau} \mathcal{I}_{[-\tau, 0]} (\rho[\rho_{-\tau}, \mathbf{j}], \mathbf{j})  = \mathcal{U}(\rho(x), \mathbf{J}(x)).
\end{equation}
Using Eq.~\eqref{eq:DefPhiJ}, this implies that $\phi(\mathbf{J}) =  \min_{\rho}  \mathcal{U}\left(\rho,\mathbf{J}\right)$, where the minimization is performed over the stationary profile $\rho(x)$ from Eq.~\eqref{eq:APrho}. 
Note that time-independent profiles as those in Eq.~(\ref{eq:APjJ}) and Eq.~(\ref{eq:APrho}) can always be considered, even if not minimal, thus leading to the generic upper bound, i.e. $\phi(\mathbf{J}) \leq  \min_{\rho} \mathcal{U}\left(\rho,\mathbf{J}\right)$.

The additivity principle is a key assumption that enables significant simplifications in the analysis of MFT models. Later in this paper, we will revisit this principle and discuss its domain of applicability, extending previous results to encompass the monitored dynamics considered here. In the next subsection, we introduce the concept of coarse-grained classical weak measurements and explore their impact on the rate functions of general MFT models, as outlined in the present section. While our focus here is on diffusive systems, the measurement formalism we develop is, in principle, applicable to a broader class of models—including, for instance, systems governed by reaction-type dynamics involving multiple particle species, where current fluctuations can become non-Gaussian and the functional $\mathcal{U}(\rho,\mathbf{j})$ takes on a more intricate form (see~\cite{LecompteActiveLatticeGas}). Another interesting possible extension of our measurement formalism concerns systems exhibiting ballistic transport, for which a recently proposed formalism analogous to MFT, known as ballistic macroscopic fluctuation theory, has been developed~\cite{10.21468/SciPostPhys.15.4.136}.
Understanding how measurements influence such more complex systems remains an open direction for future investigation.

\subsection{Classical weak measurements}
\label{sec:MeasForm}
Having introduced, in Sec.~\ref{subsec:micro}, the microscopic dynamics and the observer's a-priori description $d\mathbb{P}_{\Lambda_L,[-t,0]}(\mathcal{C})$ of the unmonitored system, we now let the observer improve their knowledge of the system through measurements.
The central idea is to perform indirect measurements of the system’s configuration at discrete, sparsely distributed times separated by $\Delta t$, and to subsequently update the observer's probability distribution with the information extracted from these observations.
Clearly, if these measurements allow perfect knowledge of the state of the system, the posterior distribution of the observer will concentrate on the true trajectory of the system $\mathcal{C} =  \mathcal{C}^{(0)}$. More realistically, the information acquired is only partial. A mathematically effective approach is to consider the limit of \textit{weak continuous measurements} ($\Delta t \to 0$), where each individual measurement process acquires infinitesimal information about the system. This procedure is the classical analog~\cite{PhysRevB.110.L060202,PhysRevE.111.024135,gerbino2025measurement} of the one at the origin of the stochastic Schr\"odinger equation~\cite{Jacobs01092006, wiseman2009quantum, barchielli2009quantum}. As we will see, this mathematical procedure quantifies the effectiveness of the measurement process in terms of a rate $\widetilde{\gamma}$. In a more realistic setup where $\Delta t$ remains finite, we expect  the coarse-grained regime $t \gg \Delta t$ to be equivalent to $t \gg \widetilde{\gamma}^{-1}$ in our approach.

We model this by introducing a set of random variables $\{A_{k,t_n}\}$, representing the outcomes of measurements taken at discrete times $t_n = n \Delta t$, and at every point of the lattice $\Lambda_L$. Each variable $A_{k,t_n}$
provides partial information about an observable $\mathcal{N}_k(C_{t_n})$, that for the moment we leave as an unspecified functional of the configuration at time $t_n$. To model this, $A_{k, t_n}$ is sampled from a probability distribution $P_{\Delta t}(A_{k,t_n} \mid C_{t_n})$, conditioned on the configuration $C_{t_n}$ at time $t_n$ and satisfying
\begin{equation}
\label{eq:avevarA}
\mathbb{E}[A_{k,t_n} \mid C_{t_n}] = \mathcal{N}_k(C_{t_n}), \quad
\text{Var}[A_{k,t_n} \mid C_{t_n}] = \frac{1}{\widetilde{\gamma} \Delta t}\;.
\end{equation}
This distribution models the correlation between the microscopic state and the observed signal, with a larger $\widetilde{\gamma} $ corresponding to a smaller variance and thus more accurate measurements. $\widetilde{\gamma}$ is therefore a parameter characterizing the \textit{measurement strength}. 
Putting everything together, the joint distribution of measurement outcomes conditioned on the microscopic trajectory is given by $\mathbb{P}_{\Lambda_L,\widetilde{\gamma} ,[-t,0]}(\mathcal{A} \mid \mathcal{C}) = \prod_{k, n} P_{\Delta t}\left(A_{k,t_n} \mid C_{t_n}\right)$, where $\mathcal{A}=\{A_{k,t}\}_{k,t}$ and $\mathcal{C}=\{C_{k,t}\}_{k,t}$ denote the complete sets of measurement outcomes and configurations, respectively. In order to consider the continuous time limit $\Delta t \to 0$, it is convenient to encode the outcomes of the measurements into the measurement signal $S_{k,t_n} = \Delta t \sum_{n'<n} A_{k,t_{n'}}$.
In this limit, the details about the higher cumulants of $P_{\Delta t}(A_{k,t_n} \mid C_{t_n})$ become irrelevant and the measurement signal obeys the stochastic differential equation (SDE) $dS_{k}(t)=\mathcal N_{k}(C_t)\,dt+\widetilde\gamma^{-1/2}\,dW_{k}(t)$, with $W_{k}(t)$ independent Wiener processes; this is the classical analog of the measurement back-action term at the origin of the stochastic Schr\"odinger equation~\cite{wiseman}.  
Furthermore,  the joint distribution $\mathbb{P}_{\Lambda_L,\widetilde{\gamma} ,[-t,0]}(\mathcal{A} \mid \mathcal{C})$ turns into a functional measure over the $S_{k}(t)$, which reads
\begin{equation}\label{eq:measpathint}
    d\mathbb{P}_{\Lambda_L,\widetilde{\gamma},[-t,0]}(\mathcal{S}|\mathcal{C}) = \mathcal{D}[\mathcal{S}]
     \exp\left({-\int_{-t}^0 dt' \mathcal{M}_{t'}[\mathcal{S},\mathcal{C}] }\right),
\end{equation}
where $\mathcal{S}=\{S_{k}(t)\}_{k,t}$, $\mathcal{D}[\mathcal{S}]$ is meant as the $\Delta t \to 0$ limit of $\prod_{k, n} dS_{k, t_n}$ and
\begin{equation}
    \label{eq:Measweightfactor}
    \mathcal{M}_t[\mathcal{S},\mathcal{C}] = \frac{\widetilde{\gamma}}{2 } \sum_{i \in \Lambda_{L}} \left(\mathcal{N}_{i}(c_t) -  \partial_t S_{i}(t) \right)^2.
\end{equation}
By means of Bayes' theorem, Eq.~\eqref{eq:measpathint} can be used to express the distribution over the system's configuration inferred by the observer using the knowledge of the measurement outcomes $\mathcal{S}$
\begin{multline}
    d\mathbb{P}_{\Lambda_L,\widetilde{\gamma},[-t,0]}(\mathcal{C}|\mathcal{S}) = d\mathbb{P}_{\Lambda_L,[-t,0]}(\mathcal{C}) \frac{d\mathbb{P}_{\Lambda_L,\widetilde{\gamma},[-t,0]}(\mathcal{S}|\mathcal{C})}{d\mathbb{P}_{\Lambda_L,\widetilde{\gamma},[-t,0]}(\mathcal{S})} =\\= d\mathbb{P}_{\Lambda_L,[-t,0]}(\mathcal{C})\; \exp\left({-\int_{-t}^0 dt' \mathcal{M}_{t'}[\mathcal{S},\mathcal{C}] } - F[\mathcal{S}]\right),
\end{multline}
where $F[\mathcal{S}]$ enforces the normalization of the measure over the trajectories $\mathcal{C}$.

\subsection{Coarse-grained measurement and macroscopic limit}
We now consider the situation where the monitored observables are macroscopic and thus admit a coarse-grained description. We focus on Markov processes enforcing diffusive transport and 
we adopt from now on the diffusive scaling $x = i/L, \tau = \frac{t}{L^2}, \gamma=\widetilde{\gamma}L^2$.

With this scaling, the a-priori measure over microscopic trajectories can be replaced with the probability distribution over the empirical trajectories
\begin{equation}
    d\mathbb{P}_{\Lambda_L,[-\tau,0]}(\mathcal{C}) \quad \stackrel{\rm macroscopic}{\rightarrow} \quad
\mathcal{D}[\mathbf{j}] \exp\big\{-L^{d}\,\mathcal{I}_{[-\tau,0]}\left(\rho[\rho_{-\tau},\mathbf{j}],\mathbf{j}\right)\big\}    
\end{equation}
as expressed in Eq.~\eqref{eq:BigRate}. Then, we consider the continuous monitoring of the macroscopic density of particles, taking
\begin{equation}
\label{eq:Ndef}
 \mathcal{N}_k(C_{\tau_n}) := \frac{1}{l^d} \sum_{i \in \Lambda_L} C_{i,\tau_n} \, f\left(\frac{i - k}{l}\right).
\end{equation}
Here, $f(x)$ is a smooth, normalized kernel satisfying $\int dx \; f(x) = 1$ and decaying sufficiently fast as $\|x\| \to \infty$.  This function defines the spatial averaging used to approximate the density around site $k$.  With this choice, and for sufficiently well-behaved density profiles $\rho(x,\tau)$, it is clear that 
\begin{multline}
    \mathcal{N}_{k}(C_\tau) = \frac{1}{l^d} \sum_{i \in \Lambda_L} C_{i,\tau} f\left((i-k)/l\right) =\\= \left(\frac{L}{l}\right)^d \int_{\Lambda} dx'\; \rho(x',\tau) f\left(\frac{L}{l}\left( x' - \frac{k}{L}\right)\right) \xrightarrow[]{\rm macroscopic} \rho\left(x=\frac{k}{L},\tau\right).
\end{multline}
Similarly, the measurement signal scales as 
\begin{equation}
\label{eq:meassignal}
S_k(t)\to L^2 s(x,\tau).
\end{equation}
As a paradigmatic example, let us consider the calculation of the distribution of the time-averaged current. Conditioned to the measured signal $s$, 
the cumulant generating function in 
Eq.~\eqref{eq:BigCurrGenv} is modified to
\begin{equation}\label{eq:BigCurrGenvMeas}
    \lambda_\gamma(\mathbf{v}| s
    ) =  \lim_{\tau, L \to \infty} \frac{1}{L^d \tau} \log \int \mathcal{D}[\mathbf{j}] 
    \exp\left\{-L^{d} 
    \int_{-\tau}^0 d\tau' \left[ \mathcal{U}_{\gamma,\tau'}(\rho[\rho_{-\tau},\mathbf{j}],\mathbf{j} | s
    )    - \int_{\Lambda} dx\;\mathbf{v}\cdot \mathbf{j}\right]\right\},
\end{equation}
where we set 
\begin{equation}\label{eq:UtotalMeas}
\begin{aligned}
\mathcal{U}_{\gamma,\tau}(\rho,\mathbf{j} | 
s)&= \int_{\Lambda} dx\; \left(\frac{1}{2} [\mathbf{j}-\mathcal{J}(\rho)] \cdot \sigma(\rho)^{-1}[\mathbf{j}-\mathcal{J}(\rho)] +\frac{\gamma}{2} \left(\rho - \partial_\tau s\right)^2  \right)= \\ &=  \mathcal{U}_{\gamma=0,\tau}(\rho,\mathbf{j} | 
s)+ \frac{\gamma}{2}\int_{\Lambda} dx\;\left(\rho - \partial_\tau s\right)^2.
\end{aligned}
\end{equation}
Here, $\mathcal{U}_{\gamma=0,\tau}(\rho,\mathbf{j} | s)$ is given precisely by Eq.~(\ref{eq:BigRate}).
Formally, in the limit of large $L$, the functional integral reduces to a maximization over the empirical trajectories $(\rho(x,\tau), \mathbf{j}(x,\tau))$ and measurements provide an extra weight favoring trajectories close to the measured signal for the density $\rho \sim \partial_\tau s$.  
\subsubsection{Typical measurement signal}
We recall that the procedure we have in mind is that the observer is modeling the system through a probability distribution, which is updated by the measurement outcomes $\mathcal{S}$. We now show that in the diffusive scaling limit assumed above, for large $L$, the system will admit a coarse-grained description in terms of density and current $(\rho^{(0)}, \mathbf{j}^{(0)})$ which is peaked around typical profiles $(\rho^{(0)}_{\rm typ}, \mathbf{j}_{\rm typ}^{(0)})$. In turn, under the same hypothesis, the measurement outcomes $\mathcal{S}$ resulting from the measurements of coarse grained variables \eqref{eq:Ndef} have also a peaked distribution at large $L$ around $\partial_\tau s \sim \rho^{(0)}_{\rm typ}$.
To be more specific, consider a functional of the measurement outcomes $G[\mathcal{S}]$ and let us consider its average over different realizations of the protocol, involving both a realization of the system $\mathcal{C}^{(0)}$ and the corresponding set of measurement outcomes $\mathcal{S}$. Then, the average value of $G[\mathcal{S}]$ is
\begin{multline}
\label{eq:Gintsaddle}
    \overline{G[\mathcal{S}]} := 
    \int  d\mathbb{P}_{\Lambda_L,[-\tau,0]}(\mathcal{C}^{(0)}) 
d\mathbb{P}_{\Lambda_L,\widetilde{\gamma},[-t,0]}(\mathcal{S}|\mathcal{C}^{(0)}) 
     G[\mathcal{S}] 
\xrightarrow{\rm macroscopic}\\ 
\int \mathcal{D}[\rho^{(0)},\mathbf{j}^{(0)}] \mathcal{D}[s
] 
 \exp\big\{-L^{d} \int d\tau \; \mathcal{U}_{\gamma,\tau}(\rho^{(0)},\mathbf{j}^{(0)} | s)\big\}
 G[s]
\end{multline}
where with a slight abuse of notation we use $G[s]$ for the same functional computed on the macroscopic signal \eqref{eq:meassignal}.
We can see from this expression that at large $L$, the exponential factor concentrates the measure around the saddle points. For any functional $G[s]$ that does not give exponential weight to some realization of the measurement signal $s$ (e.g. any bounded function of the measured signal), we can then compute the integral by looking at the minima of $\int d\tau \; \mathcal{U}_{\gamma,\tau}(\rho^{(0)},\mathbf{j}^{(0)} | s)$ with respect to $(\rho^{(0)}, \mathbf{j}^{(0)})$ and $s$. Using Eq.~\eqref{eq:UtotalMeas}, we see that the effect of monitoring is simply to add a quadratic and positive contribution $\propto (\partial_\tau s - \rho^{(0)})^2$. Thus, the minimization can be easily done choosing $(\rho^{(0)} = \rho^{(0)}_{\rm typ}, \mathbf{j}^{(0)} = \mathbf{j}^{(0)}_{\rm typ})$, where $(\rho^{(0)}_{\rm typ},\mathbf{j}^{(0)}_{\rm typ})$ minimize the unmonitored rate function $\mathcal{I}(\rho^{(0)}, \mathbf{j}^{(0)})$ in Eq.~\eqref{eq:BigRate}; then one chooses $s_{\rm typ}$ such that $\partial_\tau s = \rho^{(0)}_{\rm typ}$. In practice, this amounts to solving the hydrodynamic equation
\begin{equation}
\label{eq:hydrotyp}
 \partial_\tau \rho_{\rm typ}^{(0)} + \nabla \cdot \mathbf{j}_{\rm typ}^{(0)} = 0 \;, \quad \mathbf{j}_{\rm typ}^{(0)} = \mathcal{J}(\rho_{\rm typ}^{(0)}) \;,   \quad \partial_\tau s_\text{typ} = \rho_{\rm typ}^{(0)}.    
\end{equation}

\section{Effect of Measurements on the Current statistics}\label{sec:stability}
In this section, we investigate the effect of monitoring on the statistical properties of the time-averaged (integrated) current, defined as $\mathbf{J}_{\tau}[\mathbf{j}](x) = \frac{1}{\tau} \int_{-\tau}^{0} d\tau'\, \mathbf{j}(x, \tau')$. 

Equivalently, we can look at the corresponding cumulant generating function introduced in Sec.~\ref{sec:LD&CGF} and Eq.~\eqref{eq:BigCurrGenv}.
Indeed, consider $G[s] = \lambda_{\gamma}(\mathbf{v} | s)$, the cumulant generating function in Eq.~\eqref{eq:BigCurrGenvMeas}. Then, from Eq.~\eqref{eq:Gintsaddle}, we have that
\begin{equation}
\label{eq:lambdaavevs}
    \overline{\lambda_{\gamma}(\mathbf{v} | s)} = 
    - \lim_{\tau \to \infty} \min_{\mathbf{j}} \left[\tau^{-1}\mathcal{I}_{\gamma,[-\tau,0]}(\rho[\rho_{-\tau},\mathbf{j}],\mathbf{j}) - \int_{\Lambda} dx \, \mathbf{v}(x)\cdot \mathbf{J}_{\tau}[\mathbf{j}](x)  \right]
\end{equation}
where we introduced the functional 
\begin{align}\label{eq:UtotalMeasTyp}
&\mathcal{I}_{\gamma,[-\tau,0]}(\rho,\mathbf{j}) = \int_{-\tau}^{0} d\tau'\;  \mathcal{U}_{\gamma,\tau'}(\rho(\tau'),\mathbf{j}(\tau')),\\ &\mathcal{U}_{\gamma,\tau}(\rho,\mathbf{j} )= \int_{\Lambda} dx\; \left(\frac{1}{2} [\mathbf{j}-\mathcal{J}(\rho)] \cdot \sigma(\rho)^{-1}[\mathbf{j}-\mathcal{J}(\rho)] +\frac{\gamma}{2} \left(\rho - \rho^{(0)}_{\rm typ}\right)^2  \right),
\end{align}
where we have included the subscript $\tau$ on $\mathcal{U}_{\gamma,\tau}(\rho,\mathbf{j})$ since it depends explicitly on time through $\rho^{(0)}_{\rm typ}(x,\tau)$.
It is straightforward to verify that $\mathcal{I}_{\gamma,[-\tau,0]}(\rho^{(0)}_{\rm typ},\mathbf{j}^{(0)}_{\rm typ}) = 0$ and thus Eq.~(\ref{eq:UtotalMeasTyp}) is properly normalized and consistently $\overline{\lambda_{\gamma}(\mathbf{v} = 0 | s)} = 0$. In the following, we remove the overline notation since, at large $L$, $\lambda_{\gamma}(\mathbf{v} | s) \xrightarrow{\text{in law}} \overline{\lambda_{\gamma}(\mathbf{v} | s)} = {\lambda_{\gamma}(\mathbf{v} | s_{\rm typ})}$ and, for simplicity, we denote $\lambda_{\gamma}(\mathbf{v})\equiv\lambda_{\gamma}(\mathbf{v} | s_{\rm typ})$. 

As in standard MFT, the functional \eqref{eq:UtotalMeasTyp} provides the natural probabilistic framework through which an observer monitoring the system can describe fluctuations of the density and current around the hydrodynamic solution in the mesoscopic regime. Indeed, the analysis of the large deviations of the density profile 
can be performed with similar methods through Eq.~(\ref{eq:ratefunctionofthedensity}) and Eq.~(\ref{eq:DensGenF}) (or more generally through Eq.~(\ref{eq:UtotalMeas})).

It is also convenient to consider the rate function of the integrated current first introduced in Eq.~\eqref{eq:DefPhiJ}, given by  
\begin{equation}\label{eq:CurrLDgamma}
   \phi_{\gamma}(\mathbf{J}) = \lim_{\tau \to \infty}\frac{1}{\tau}\min_{\mathbf{j}:\mathbf{J}=\mathbf{J}_{\tau}[\mathbf{j}]} \mathcal{I}_{\gamma,[-\tau,0]}(\rho[\rho_{-\tau},\mathbf{j}],\mathbf{j}). 
\end{equation}
and related to $ {\lambda_{\gamma}(\mathbf{v})} $ via the Legendre transform (see Eqs.~(\ref{eq:BigCurrGenv}, \ref{eq:legphilambda})).

Generally, the hydrodynamic equation \eqref{eq:hydrotyp} is expected to lead
to a stationary profile 
$\rho_{\infty}(x)=\lim_{\tau \to \infty} \rho^{(0)}(x,\tau)$ at large times.
Then, the contribution arising from any transient relaxation becomes subleading and therefore does not affect the rate function $\phi_{\gamma}(\mathbf{J})$. In practice, this means we can replace the full evolution of the hydrodynamic density profile $\rho_{\rm typ}^{(0)}(x,\tau)$ with its asymptotic limit $\rho_{\infty}(x)$ in Eq.~(\ref{eq:UtotalMeasTyp}) without altering the value of $\phi_{\gamma}(J)$ or $\lambda_{\gamma}(\mathbf{v})$. At stationarity, $\rho_{\infty}(x)$ satisfies the time-independent condition 
\begin{equation}\label{eq:hydroSS}
    \nabla_x \cdot  \mathcal{J}(\rho_{\infty}) = 0
    .
\end{equation}
This observation leads to the following simpler form of the relevant rate functional,
\begin{multline}\label{eq:UtotalMeasSS}
\mathcal{I}_{\gamma,[-\tau,0]}(\rho,\mathbf{j}) = \int_{-\tau}^{0} d\tau'\;  \mathcal{U}_\gamma(\rho(\tau'),\mathbf{j}(\tau')),\\ \mathcal{U}_{\gamma}(\rho,\mathbf{j} )= \int_{\Lambda} dx\; \left(\frac{1}{2} [\mathbf{j}-\mathcal{J}(\rho)] \cdot \sigma(\rho)^{-1}[\mathbf{j}-\mathcal{J}(\rho)] +\frac{\gamma}{2} \left(\rho - \rho_{\infty}\right)^2  \right).
\end{multline}
We stress that this expression for $\mathcal{I}_{\gamma,[-\tau,0]}(\rho,\mathbf{j})$ is meaningful only in the context of long-time, or steady-state, properties of the system. Throughout the remainder of this work, we shall always evaluate $\mathcal{I}_{\gamma,[-\tau,0]}(\rho,\mathbf{j})$ appearing in Eq.~(\ref{eq:CurrLDgamma}) using the definition given in Eq.~(\ref{eq:UtotalMeasSS}).

When the prescribed integrated current coincides with the typical one, i.e. the stationary solution of the hydrodynamic equations, $\mathbf{J}=\mathbf{j}_\infty\coloneqq\mathcal{J}(\rho_\infty)$, there exists a unique profile minimizing the action $\mathcal{I}_{\gamma,[-\tau,0]}(\rho[\rho_{-\tau},\mathbf{j}],\mathbf{j})$, namely $(\rho, \mathbf{j}) = (\rho_\infty,\mathbf{j}_\infty)$. This implies that the description of the observer correctly captures the typical properties of the system itself and in this case the measurements do not provide any relevant information. In contrast, the effect of measurements becomes visible when quantifying large deviations where $\mathbf{J} \neq \mathbf{j}_{\infty}$: in this case, we can already see from Eq.~\eqref{eq:UtotalMeasSS} that increasing $\gamma > 0$ keeps $\rho$ closer to $\rho_{\infty}$, thus suppressing fluctuations away from typicality. 

We emphasize that the annealed cumulant generating function, in which the average over measurement outcomes is taken before the logarithm,
\begin{equation}
\label{eq:annealedphi}
    \lambda_{\rm annealed}(\mathbf{v}) = \lim_{\tau, L \to \infty} \frac{1}{L^d \tau}\log \overline{\mathbb{E}^{(\mathbb{Q})}_{\Lambda_L,\gamma,\tau}\left[e^{\tau L^d \int_\Lambda dx\, \mathbf{v}(x)\cdot\mathbf{J}_\tau(x)} \,\middle|\, s\right]},
\end{equation}
is independent of $\gamma$ and equals the unmonitored one~\cite{gopalakrishnan2025_ClassicalMeasurements}. This is because averaging the conditional expectation over the measurement record simply returns the unconditional one, $\overline{\mathbb{E}[\,\cdot\,|\,s]}=\mathbb{E}[\,\cdot\,]$. The non-trivial effect of monitoring is instead seen in the quenched average \eqref{eq:lambdaavevs}, where the logarithm is taken before averaging over $s$.

In the following subsections, we shall identify the conditions under which, in the presence of measurements, the additivity principle remains valid or, more generally, under which no dynamical phase transition can occur. We shall also clarify the role played by measurements in shaping the onset and nature of dynamical phase transitions.

\subsection{Absence of Dynamical Phase Transitions and the Additivity Principle}\label{sec:covexcond}

As we discussed in the previous section, when $\mathbf{J} = \mathbf{j}_{\infty}$, the minimization in Eq.~\eqref{eq:CurrLDgamma}
is trivially solved by the unique stationary solution to the hydrodynamic equation \eqref{eq:hydrotyp}, i.e. $(\rho, \mathbf{j}) = (\rho_{\infty}, \mathbf{j}_{\infty})$.
One naturally expects a unique solution to persist for integrated current profiles $\mathbf{J}$ lying in a sufficiently small neighborhood of the stationary current $\mathbf{j}_\infty$.
However, for highly unlikely integrated current profiles, sufficiently far from the steady-state one, multiple minimizing solutions of $\mathcal{I}_{\gamma,[-\tau,0]}(\rho[\rho_{-\tau},\mathbf{j}],\mathbf{j})$ may emerge, generally through spontaneous symmetry breaking. This signals the onset of a dynamical phase transition (DPT), manifested as a non-analyticity in the rate functional $\phi_{\gamma}(\mathbf{J})$ or, equivalently, in the cumulant generating function $\lambda(\mathbf{v})$ defined in Eq.~(\ref{eq:BigCurrGenv}).

In the following subsection, we seek to establish conditions that rule out the existence of dynamical phase transitions in the formalism we established in the previous section, of a monitored diffusive system described by MFT. 
In Ref.~\cite{Bertini_2006}, it was established that if there exists a change of variables $f: \mathbb{R} \to \mathbb{R}$ for the local density $\rho \to f(\rho) := \omega$ such that the functional
\[
\mathcal{V}_{\gamma}(\omega,\mathbf{j}) \coloneqq \mathcal{U}_{\gamma}(f^{-1}(\omega),\mathbf{j}),
\]
is convex in both arguments, no DPT ensues. To see this, note that convexity of $\mathcal{V}_\gamma(\omega,\mathbf{j})$ implies that for any time-dependent profiles $\rho(\tau)$ and $\mathbf{j}(\tau)$,
\begin{multline}
    \tau^{-1}\mathcal{I}_{\gamma,[-\tau,0]}\left(\rho,\mathbf{j}\right) =\frac{1}{\tau} \int_{-\tau}^0 d\tau' \, \mathcal{U}_{\gamma}(\rho(\tau'), \mathbf{j}(\tau')) 
    = \frac{1}{\tau} \int_{-\tau}^0 d\tau' \, \mathcal{V}_{\gamma}(\omega(\tau'), \mathbf{j}(\tau')) \geq \\
    \geq \mathcal{V}_{\gamma} \left( \frac{1}{\tau} \int_{-\tau}^0 d\tau' \, \omega(\tau'), \frac{1}{\tau} \int_{-\tau}^0 d\tau' \, \mathbf{j}(\tau') \right) 
    \geq \inf_\omega \mathcal{V}_{\gamma}(\omega, \mathbf{J}_{\tau}[\mathbf{j}]) = \inf_\rho \mathcal{U}_{\gamma}(\rho, \mathbf{J}_{\tau}[\mathbf{j}]) = U_{\gamma}(\mathbf{J}_{\tau}[\mathbf{j}]),
\end{multline}
which shows that $U_\gamma(\mathbf{J}) \leq \phi_{\gamma}(\mathbf{J})$. 
The opposite inequality can always be proved by restricting the minimization in Eq.~\eqref{eq:CurrLDgamma} to functions $\mathbf{j}(x, \tau)$ that admit a large-$\tau$ limit $\mathbf{j}(x, \tau\to\infty) = \mathbf{J}$, see the discussion around Eq.~\eqref{eq:largetauItoU}.
Thus, convexity of $\mathcal{V}_{\gamma}$ ensures $U_\gamma(\mathbf{J}) = \phi_{\gamma}(\mathbf{J})$. Moreover, it implies that any saddle-point solution of $\mathcal{V}_{\gamma}(\omega,\mathbf{J})$ (and consequently of $\mathcal{U}_{\gamma}(\rho,\mathbf{J})$) is indeed a minimizer of these functionals. Additionally, if $\mathcal{V}_{\gamma}(\omega,\mathbf{J})$ is \emph{strictly} convex, there exists a unique minimal saddle-point solution $\rho_{\gamma,\mathbf{J}}$ of $\mathcal{U}_{\gamma}(\rho,\mathbf{J})$.
Since, for sufficiently regular mobility $\sigma(\rho)$ and diffusion coefficients $D(\rho)$, the saddle-point equations derived from $\mathcal{U}_{\gamma}(\rho,\mathbf{J})$ take the form of well-posed nonlinear second-order partial differential equations that depend smoothly on $\mathbf{J}$, one expects the minimizer $\rho_{\gamma,\mathbf{J}}$ to inherit the same smooth dependence on $\mathbf{J}$. This, in turn, implies smoothness of the current rate functional, $\phi_\gamma(\mathbf{J}) = \mathcal{U}_{\gamma}(\rho_{\gamma,\mathbf{J}},\mathbf{J})$. Moreover, as current rate functionals are always convex (see~\cite{Bertini_2006}), analyticity of $\phi_{\gamma}(\mathbf{J})$ guarantees that its Legendre transform, $\lambda_\gamma(\mathbf{v})$, is also smooth. It becomes thus clear that, more generally, convexity of $\mathcal{V}_{\gamma}(\omega,\mathbf{J})$ precludes the existence of dynamical phase transitions.

Of course, to make any meaningful progress, one must first identify a useful choice for the function $f(\rho)$. In the context of MFT without measurements ($\gamma=0$), with $\mathcal{J}(\rho) = -D(\rho) \boldsymbol\nabla_x \rho  + \sigma(\rho) \mathbf{E}$ (see Eq.~(\ref{eq:constrel})),
it was shown in~\cite{Bertini_2006} that choosing $f(\rho)= \int^{\rho} d\rho' \; D(\rho')$ renders the functional $\mathcal{V}_0(\omega(\tau), \mathbf{j}(\tau))$ convex provided that
\begin{subequations}
\label{eq:cond1}
\begin{align}
    D(\rho) \sigma''(\rho) \leq D'(\rho) \sigma'(\rho), \; \text{for any } \rho \; \text{, if } \mathbf{E}=0,
    \label{eq:cond1E0}
    \\
     D(\rho) \sigma''(\rho) = D'(\rho) \sigma'(\rho), \; \text{for any } \rho \; \text{, if } \mathbf{E} \neq 0.
     \label{eq:cond1Enz}
\end{align}
\end{subequations}
We stress that these are only sufficient conditions for convexity.
When measurements are included, an additional term appears in the expression for $\mathcal{V}_\gamma(\omega, \mathbf{j})$. Since convexity is preserved under addition, the entire rate functional is ensured to be convex if each term is convex. In particular, it suffices to require that $\left(f^{-1}(\omega)-\rho_{\infty}\right)^2$ is convex (see Eq.~(\ref{eq:UtotalMeasSS})), where $f$ is the map introduced before Eq.~(\ref{eq:cond1}). This leads to the additional condition
\begin{equation}\label{eq:cond3-a}
    \partial_\omega^2(f^{-1}(\omega)-\rho_\infty)^2=\tfrac{2}{D(\rho)^2}\big[1-(\rho-\rho_\infty)\tfrac{D'(\rho)}{D(\rho)}\big] \geq 0 \Longrightarrow \left(\rho-\rho_{\infty}\right) \frac{D'(\rho)}{D(\rho)} \leq 1 \;.
\end{equation}
Since we require convexity of each term of $\mathcal V_\gamma$ separately, rather than of their sum, Eqs.~(\ref{eq:cond1},\ref{eq:cond3-a}) are only sufficient conditions; nevertheless, they are enough to rule out any dynamical phase transition.
Note that Eqs.~(\ref{eq:cond1},\ref{eq:cond3-a}) must hold for every $x\in\Lambda$ and every admissible density $\rho$. 
For instance, both conditions hold for the SSEP (with transport coefficients given in Eq.~\eqref{eq:SSEPtranspcoeff}) and for non-interacting particles ($D=1$, $\sigma=\rho$): since $D'=0$, Eq.~\eqref{eq:cond3-a} is trivially satisfied, and Eq.~\eqref{eq:cond1E0} reduces to $\sigma''\leq0$. Hence, at zero external field, no dynamical phase transition can occur in either model, for any $\gamma$.

As it is clear from Eq.~(\ref{eq:cond1Enz}), the presence of an external field imposes a much stricter constraint on the type of models where convexity of $\mathcal{V}_{\gamma}(\omega,J)$ can be ensured from the outset. Indeed, substituting the SSEP transport coefficients in Eq.~\eqref{eq:SSEPtranspcoeff} shows that this equality does not hold and, as explained in~\cite{Lecomte_2017}, a DPT occurs for high enough external field $\mathbf{E}$ and no monitoring $(\gamma = 0)$. An interesting question is how measurements can affect this DPT, which we shall address in the next subsection.

Apart from cases in which the measurements may alter pre-existing transitions, a genuine measurement-induced phase transition, in this context, would correspond to a situation where no DPT is present at $\gamma=0$, but upon increasing the measurement strength $\gamma$ a threshold $\gamma_{\rm thresh}$ is crossed such that a DPT emerges as a function of $J$ only for $\gamma>\gamma_{\rm thresh}$. Clearly, if Eq.~(\ref{eq:cond1}) holds, such a scenario can only arise in models that do not satisfy Eq.~(\ref{eq:cond3-a}). A natural candidate is the zero-range process~\cite{SPITZER1970246,EvansHanney2005}, in which a site occupied by $k$ particles releases one of them at rate $g(k)$. Its transport coefficients are $D(\rho)=\Phi'(\rho)$ and $\sigma(\rho)=2\Phi(\rho)$, where the fugacity $\Phi(\rho)$ is fixed by the rates $g(k)$. Hence $D\sigma''=D'\sigma'$ holds identically, so that Eq.~(\ref{eq:cond1}) is satisfied for any external field, while Eq.~(\ref{eq:cond3-a}) reads $(\rho-\rho_\infty)\Phi''(\rho)/\Phi'(\rho)\leq1$ and fails for a sufficiently curved $\Phi$. For instance, for constant rates, $\Phi(\rho)=\rho/(1+\rho)$, it is violated for $\rho<(2\rho_\infty-1)/3$ whenever $\rho_\infty>1/2$. We leave a systematic investigation of this possibility as an open direction for future research.

We conclude this section with a remark on the additivity principle. Note that the action $\mathcal{I}_{\gamma,[-\tau,0]}(\rho[\rho_{-\tau},\mathbf{j}],\mathbf{j})$ is always characterized by invariance under time translations, as it contains no explicit dependencies on time. 
Breaking of time-translation symmetry leads to a manifold of time-dependent saddle-point solutions that minimize the action. Thus, restated in the terminology presented here, the additivity principle, introduced at the end of Sec.~\ref{sec:LD&CGF} can be viewed as the absence of a symmetry breaking of the invariance under time translations.
The conditions~(\ref{eq:cond1},\ref{eq:cond3-a}) identified in this section ensure the absence of any kind of symmetry breaking (including time translation) and can therefore be regarded as sufficient conditions for the validity of the additivity principle as well.
It is clear that the additivity principle provides a significant computational advantage in determining the rate functional $\phi_\gamma(\mathbf{J})$. Instead of minimizing the full action $\mathcal{I}_{\gamma,[-\tau,0]}(\rho,\mathbf{j})$ over density and current profiles that depend on both space and time, the problem is reduced in dimensionality: for each prescribed integrated current profile $\mathbf{J}(x)$, one only needs to minimize the functional $\mathcal{U}_\gamma\left(\rho,\mathbf{J}\right)$ over spatially dependent density profiles defined on the domain $\Lambda$. In the next section, we will assume that additivity principle holds so that this simplification can be used and the minimization reduces to
the one over the stationary profile $\rho(x)$ (see discussion around Eq.~\eqref{eq:largetauItoU})
\begin{equation}\label{eq:APdef}
    \phi_\gamma(\mathbf{J}) = U_\gamma(\mathbf{J}) \coloneqq \min_{\rho}  \mathcal{U}_\gamma\left(\rho,\mathbf{J}\right).
\end{equation}

\subsection{Effect of measurements on DPTs in one dimension and particle-hole symmetry}\label{subsec:DPTs}

We now focus on one-dimensional transport models under the influence of an external field and subject to continuous monitoring. 
As outlined in the introduction, we consider two reservoirs connected to the boundaries of the domain $\Lambda = [0,1]$, which impose fixed boundary densities: $\rho(0) = n_0$ and $\rho(1) = n_1$. 
Our goal is to compute $U_{\gamma}(J)$ in Eq.~\eqref{eq:APdef}, which serves as an upper bound to the rate function $\phi_{\gamma}(J)$, $\phi_{\gamma}(J) \leq U_{\gamma}(J)$. 
As, in one dimension, the current reduces to a scalar, we suppress the boldface notation. 
Since $\phi_{\gamma}(J)$ is known to be convex~\cite{Bertini_2006}, a tighter bound can be obtained by considering the convex envelope of $U_{\gamma}(J)$, denoted by $U_{\gamma}^{**}(J)$, which yields the improved inequality, $\phi_{\gamma}(J) \leq U_{\gamma}^{**}(J)$. As shown in the previous subsection, conditions Eq.~(\ref{eq:cond1}) and Eq.~(\ref{eq:cond3-a}) imply validity of the additivity principle, in which case this inequality becomes simply an equality $\phi_{\gamma}(J) = U_{\gamma}(J) = U_{\gamma}^{**}(J)$. 

More generally, for the sake of simplicity, we assume that a weaker form of the Additivity Principle (AP) is valid, namely $\phi_{\gamma}(J) = U_{\gamma}^{**}(J)$.
We emphasize that the assumption $\phi_\gamma(J)=U_\gamma^{**}(J)$ is not proven and underlies all results of this section. Should it fail, the discrepancy $U_\gamma^{**}(J)>\phi_\gamma(J)$ would necessarily signal a time-dependent minimizer, i.e. a spontaneous breaking of time-translation invariance. We regard the conclusions below as conditional on this weak additivity principle, whose violations, although documented in closed chains~(see, e.g.,\cite{DerridaKMP,ZarfatyKMP}), are rare and poorly understood in open boundary-driven systems~(e.g.,\cite{OhadShpielbergNumerical}).

Denoting the (constant in time) optimal solution that minimizes $\mathcal{U}_{\gamma}(\rho,J)$ by $\rho_{\gamma,J}(x)$, one obtains $U_{\gamma}(J) = \int_0^1 dx \; \mathcal{L}_{\gamma,J}\left[\rho_{\gamma,J}(x),x\right]$, for 
\begin{equation}\label{eq:LJgamma0}
    \mathcal{L}_{\gamma,J}\left[\rho_{\gamma,J}(x),x\right] =  \left(\frac{\left( J + D(\rho_{\gamma,J}) \partial_x \rho_{\gamma,J} - \sigma(\rho_{\gamma,J}) E\right)^2 }{2 \sigma(\rho_{\gamma,J})} + \frac{\gamma}{2}  \left( \rho_{\infty}(x) - \rho_{\gamma,J}(x) \right)^2 \right),
\end{equation}
where we have considered a constant external field along the chain and we remind that $\rho_{\infty}(x)$ is the steady state solution of the hydrodynamic equations.
Defining 
\begin{equation}
\label{eq:kappamudef}
\kappa = \int_{n_1}^{n_0}d\rho\; \frac{ D(\rho)}{\sigma(\rho)} \;, \quad \mu=\int_{n_1}^{n_0}d\rho\; D(\rho),
\end{equation}
we rewrite $U_{\gamma}(J)$ in a more convenient form, 
\begin{multline}\label{eq:phiJgamma}
    U_{\gamma}(J) =\int_0^1 dx \; \left(\frac{J^2 + \left(D(\rho_{\gamma,J}) \partial_x \rho_{\gamma,J}\right)^2 + \left(\sigma(\rho_{\gamma,J}) E\right)^2}{2 \sigma(\rho_{\gamma,J})} + \right. \\ \left. + \frac{\gamma}{2}  \left( \rho_{\infty} - \rho_{\gamma,J}\right) ^2 \right) +\mu E - \kappa J -JE,
\end{multline}
It is worth observing that this equation satisfies the current fluctuation relation, also known as Gallavotti-Cohen symmetry~\cite{gallavotti1995dynamical,LebowitzSpohn1999,Kurchan1998}  
\begin{equation}
    \label{eq:GKU}
    U_{\gamma}(-J) = U_{\gamma}(J) + 2 (\kappa + E) J, 
\end{equation}
irrespective of $\gamma$. Under the hypothesis of weak additivity principle stated above, the same parity relation holds for $\phi_{\gamma}(J)$, extending its validity beyond the unmonitored case~\cite{DerridaAdditivityPrinciple,PhysRevLett.94.030601,Derrida_2007}.
In principle, one might expect conditioning on the measurement outcomes to break the time-reversal invariance on which the Gallavotti-Cohen relation is based. However, in our case, we are monitoring the density (a time-reversal-invariant observable) while examining the large deviations of the current. For this reason, the relation remains valid even in the presence of conditioning. 
This statement concerns the typical/quenched generating function within the additivity-principle regime.

We remind that, in this subsection, our goal is to describe the effect that measurements can have on the nature of a pre-existent dynamical phase transition associated to the breaking of a symmetry of the system. As a concrete example, for the remainder of this section, we adopt the equilibrium setup introduced by Lecomte \textit{et al.}~\cite{Lecomte_2017}, in which the system satisfies a "particle-hole" symmetry around the stationary density $n_* \coloneqq n_0 = n_1$, implying that all odd-order derivatives of the mobility $\sigma(\rho)$ and the diffusion constant $D(\rho)$ vanish at $n_*$, i.e., $\sigma^{(k)}(n_*) = 0$ and $D^{(k)}(n_*) = 0$ for all odd $k$. We note that in this scenario, Lecomte \textit{et al.} showed via a perturbative analysis around $\rho \sim n_{*}$, that time-independent saddle-point solutions are stable, and thus the Additivity Principle (AP) cannot be locally broken. 
This argument can be straightforwardly extended to systems under continuous measurement to show that in this case also (local) stability of time-independent solutions persists for arbitrary values of $\gamma$. Note, however, that this result does not ensure the global validity of the AP. In particular, it remains possible for a dynamical phase transition to occur via time-dependent solutions that are not perturbatively connected to the symmetric profile near $n_*$. Such transitions, if they exist, would lie beyond the reach of the perturbative approach and the symmetry-based stability analysis.

In the following, we work out the consequences of the "weak" additivity principle assumption in one dimension and in the presence of "particle-hole" symmetry. We develop a systematic procedure to compute $U_{\gamma}(J)$ and then use it to analyze a few cases of interest.

\subsubsection{Implications of (Weak-)Additivity Principle in 1 dimension}\label{subsec:weakadditivity1d}

\begin{figure}[!htbp]
    \centering
    \includegraphics[width=0.78\linewidth]{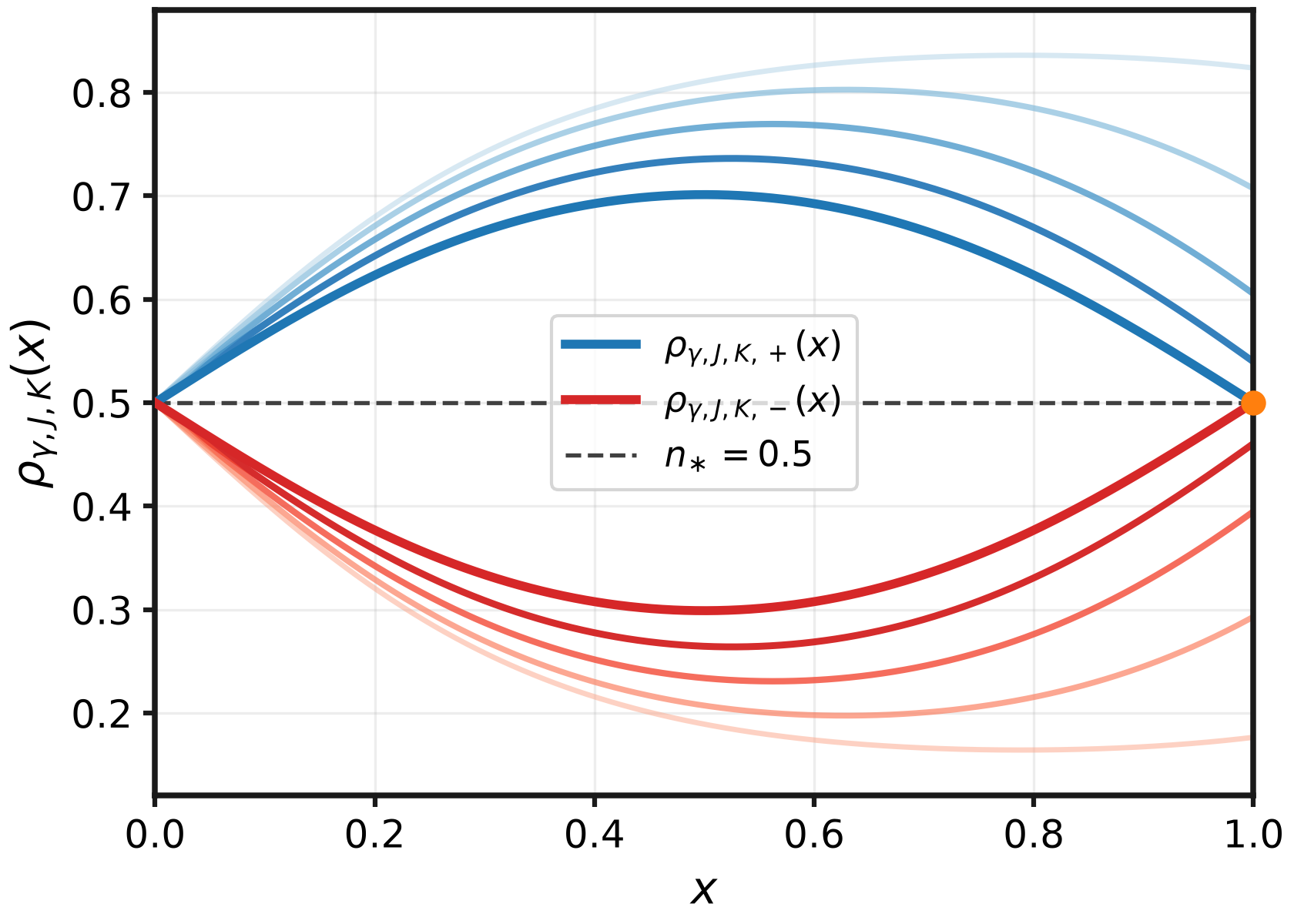}
    \caption{The two saddle-point branches generated by Eq.~(\ref{eq:KconservationT}) are plotted for several values of $K$ for SSEP at $E=4$, $J=1$, $\gamma=0$, and $n_*=1/2$. The line intensity decreases as $K$ increases.}
    \label{fig:KbranchesSketch}
\end{figure}

To analyze the particle-hole symmetric setup that we have just described, we build on the formalism developed in~\cite{DerridaAdditivityPrinciple,OhadShpielbergNumerical}, which offers a general and consistent framework for computing $U_{\gamma}(J)$. The Lagrangian in Eq.~(\ref{eq:LJgamma0}) has the structure of a one-dimensional mechanical problem: the coordinate is $\rho$, the spatial variable $x$ plays the role of time, and the saddle-point profile is a trajectory in this effective mechanics. In particular, this approach begins with the observation that when $\mathcal{L}_{\gamma,J}[\rho_{\gamma,J}(x), x]$ does not explicitly depend on $x$, the system exhibits a continuous symmetry under spatial translations, similar to conservation of the energy in time-independent Hamiltonian systems. By Noether's theorem, this symmetry implies the existence of a conserved quantity $K$, which is the effective mechanical energy. This conserved quantity is obtained from Eq.~(\ref{eq:LJgamma0}) via
\begin{equation}\label{eq:KconservationT}
    K:=V_{\gamma,J}(\rho_{\gamma,J})+\frac{\left(D(\rho_{\gamma,J})\partial_x \rho_{\gamma,J}\right)^2}{2 \sigma(\rho_{\gamma,J})} ,\;V_{\gamma,J}(\rho)\coloneqq-\frac{J^2+\left(\sigma(\rho)E\right)^2}{2 \sigma(\rho)}- \frac{\gamma}{2}\left(\rho-n_*\right)^2.
\end{equation} 
Here $V_{\gamma,J}(\rho)$ is the effective potential. Eq.~\eqref{eq:KconservationT} is useful as it states that the combination defining $K$ is independent of $x$. 
Note that this result holds for general MFT models (at $\gamma=0$), both in and out of equilibrium, and remains valid in the equilibrium setting even when measurements are included. In contrast, for $n_0 \neq n_1$, $\rho_\infty$ becomes space dependent, so that space translational symmetry is explicitly broken when $\gamma\neq 0$. 
Here, we focus on the equilibrium case. 
Note that in our case, the boundary conditions $\rho(x=0)=n_0$ and $\rho(x=1)=n_1$ are given. In the analogy with the mechanical system in which $K$ represents the energy, these are conditions that are non-local in time. For each value of $K$, Eq.~(\ref{eq:KconservationT}) admits exactly two solutions satisfying the initial condition $\rho_{\gamma,J,K,\pm}(0) = n_*$, denoted $\rho_{\gamma,J,K,\pm}(x)$, with $\rho_{\gamma,J,K,-}(x) \leq n_*$ and $\rho_{\gamma,J,K,+}(x) \geq n_*$. The origin of this twofold choice is the square root obtained when Eq.~(\ref{eq:KconservationT}) is solved for $\partial_x\rho$: once the trajectory leaves $n_*$, it can do so with either sign of the derivative. This construction is shown in Fig.~\ref{fig:KbranchesSketch} for the SSEP (see Eq.~\eqref{eq:SSEPtranspcoeff}) with $E=4$ and $\gamma=0$.

\begin{figure}[!htbp]
    \centering
    \begin{minipage}[t]{0.48\linewidth}
        \centering
        \includegraphics[width=\linewidth]{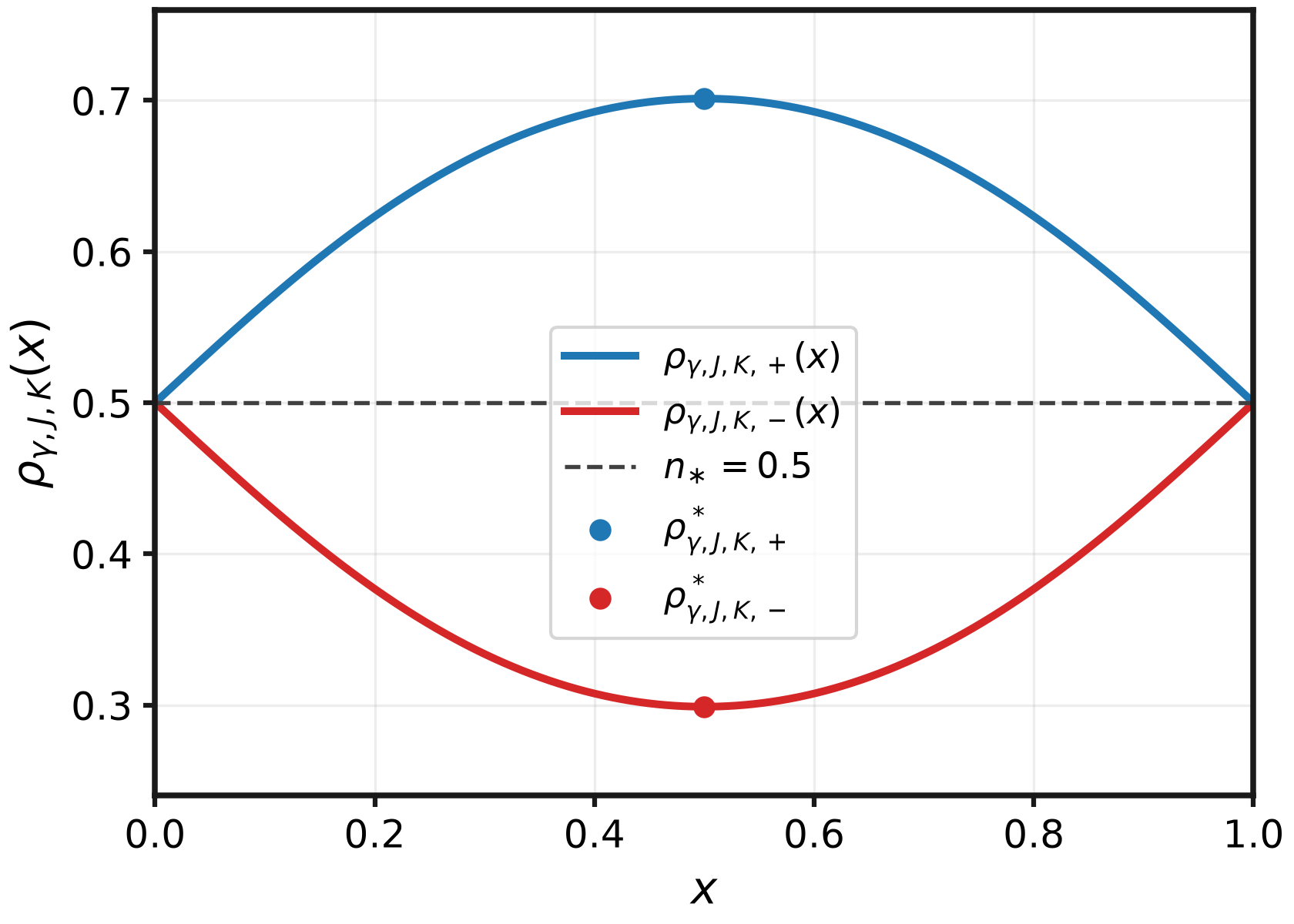}\\[-0.35em]
        {\small (a) Profile satisfying condition $\mathcal{C}$}
    \end{minipage}\hfill
    \begin{minipage}[t]{0.48\linewidth}
        \centering
        \includegraphics[width=\linewidth]{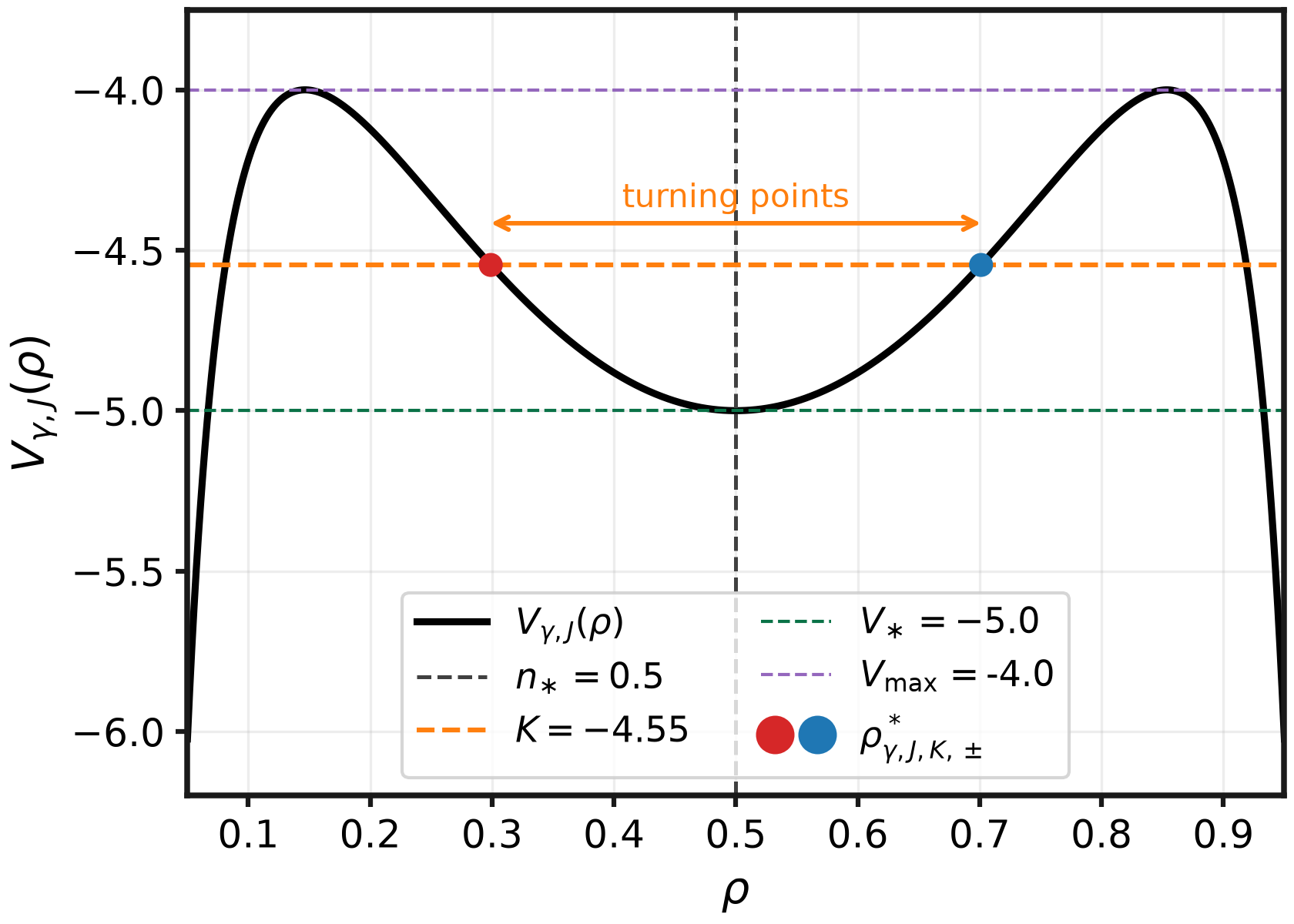}\\[-0.35em]
        {\small (b) $K$ fixes the excursion amplitude}
    \end{minipage}
    \caption{In panel (a), we show the two $\pm$ saddle-point profiles that satisfy condition $\mathcal{C}$ for SSEP at $E=4$, $J=1$, $\gamma=0$, and $n_*=1/2$. In panel (b), we plot the corresponding effective potential $V_{\gamma,J}(\rho)$, marking the value of $K$ that satisfies condition $\mathcal{C}$ in yellow. $K$ fixes the turning points of the saddle-point profiles, which we show in red and blue.}
    \label{fig:KconstraintSketch}
\end{figure}

\begin{figure}[!htbp]
    \centering
    \begin{tikzpicture}[x=1cm,y=1cm,>=Latex,thick]
        \begin{scope}
            \draw[->] (0,0) -- (3.25,0) node[right] {$\rho$};
            \draw[->] (0,-2.15) -- (0,0.35) node[above] {$V_{\gamma,J}(\rho)$};
            \draw[densely dashed] (1.55,0) -- (1.55,-0.62);
            \node[above=1pt] at (1.55,0) {$n_*$};
            \draw[black,very thick] plot[smooth] coordinates {(0.25,-1.85) (0.65,-1.35) (1.05,-0.9) (1.55,-0.62) (2.05,-0.9) (2.45,-1.35) (2.85,-1.85)};
            \draw[orange!80!black,densely dashed] (0,-0.62) -- (1.55,-0.62);
            \fill[red!75!black] (0,-0.62) circle (3.2pt);
            \node[orange!80!black,left=1pt] at (0,-0.62) {$V_*=V_{\text{max}}$};
            \draw[->] (0.35,-2.55) -- (2.85,-2.55) node[right] {$\rho^*$};
            \draw[orange!80!black,densely dashed] (1.55,-0.62) -- (1.55,-2.55);
            \fill[orange!80!black] (1.55,-2.55) circle (2.0pt);
            \draw (1.55,-2.48) -- (1.55,-2.62) node[above=2pt] {$n_*$};
            \node[font=\scriptsize] at (1.55,-3.18) {(a) Trivial domain};
        \end{scope}
        \begin{scope}[shift={(4.0,0)}]
            \draw[->] (0,0) -- (3.25,0) node[right] {$\rho$};
            \draw[->] (0,-2.15) -- (0,0.35) node[above] {$V_{\gamma,J}(\rho)$};
            \draw[densely dashed] (1.55,0) -- (1.55,-1.42);
            \node[above=1pt] at (1.55,0) {$n_*$};
            \draw[black,very thick] plot[smooth] coordinates {(0.25,-1.9) (0.65,-0.75) (1.1,-1.08) (1.55,-1.42) (2.0,-1.08) (2.45,-0.75) (2.9,-1.9)};
            \draw[red!75!black,line width=2.8pt] (-0.055,-1.42) -- (-0.055,-0.75);
            \draw[orange!80!black,densely dashed] (0,-0.75) -- (3.0,-0.75);
            \draw[orange!80!black,densely dashed] (0,-1.42) -- (1.65,-1.42);
            \node[orange!80!black,left=1pt] at (0,-0.75) {$V_{\text{max}}$};
            \node[orange!80!black,left=1pt] at (0,-1.42) {$V_*$};
            \draw[->] (0.35,-2.55) -- (2.85,-2.55) node[right] {$\rho^*$};
            \draw[orange!80!black,densely dashed] (0.65,-0.75) -- (0.65,-2.55);
            \draw[orange!80!black,densely dashed] (2.45,-0.75) -- (2.45,-2.55);
            \draw[orange!80!black,very thick] (0.65,-2.55) -- (2.45,-2.55);
            \draw (0.65,-2.48) -- (0.65,-2.62) node[above=2pt] {$\rho_-$};
            \draw (1.55,-2.48) -- (1.55,-2.62) node[above=2pt] {$n_*$};
            \draw (2.45,-2.48) -- (2.45,-2.62) node[above=2pt] {$\rho_+$};
            \node[font=\scriptsize] at (1.55,-3.18) {(b) Simple well};
        \end{scope}
        \begin{scope}[shift={(8.0,0)}]
            \draw[->] (0,0) -- (3.25,0) node[right] {$\rho$};
            \draw[->] (0,-2.15) -- (0,0.35) node[above] {$V_{\gamma,J}(\rho)$};
            \draw[densely dashed] (1.55,0) -- (1.55,-0.82);
            \node[above=1pt] at (1.55,0) {$n_*$};
            \draw[black,very thick] plot[smooth,tension=0.75] coordinates {(0.2,-1.95) (0.43,-1.38) (0.58,-0.82) (0.72,-0.34) (0.98,-0.82) (1.12,-1.28) (1.38,-1.02) (1.55,-0.82) (1.72,-1.02) (1.98,-1.28) (2.12,-0.82) (2.38,-0.34) (2.52,-0.82) (2.67,-1.38) (2.9,-1.95)};
            \draw[red!75!black,line width=2.8pt] (-0.055,-0.82) -- (-0.055,-0.42);
            \draw[orange!80!black,densely dashed] (0,-0.42) -- (2.48,-0.42);
            \draw[orange!80!black,densely dashed] (0,-0.82) -- (3.0,-0.82);
            \node[orange!80!black,left=1pt] at (0,-0.42) {$V_{\text{max}}$};
            \node[orange!80!black,left=1pt] at (0,-0.82) {$V_*$};
            \draw[->] (0.35,-2.55) -- (2.85,-2.55) node[right] {$\rho^*$};
            \draw[orange!80!black,densely dashed] (0.72,-0.34) -- (0.72,-2.55);
            \draw[orange!80!black,densely dashed] (0.98,-0.82) -- (0.98,-2.55);
            \draw[orange!80!black,densely dashed] (1.55,-0.82) -- (1.55,-2.55);
            \draw[orange!80!black,densely dashed] (2.12,-0.82) -- (2.12,-2.55);
            \draw[orange!80!black,densely dashed] (2.38,-0.34) -- (2.38,-2.55);
            \draw[orange!80!black,very thick] (0.72,-2.55) -- (0.98,-2.55);
            \draw[orange!80!black,very thick] (2.12,-2.55) -- (2.38,-2.55);
            \fill[orange!80!black] (1.55,-2.55) circle (2.0pt);
            \draw (1.55,-2.48) -- (1.55,-2.62) node[above=2pt] {$n_*$};
            \node[font=\scriptsize] at (1.55,-3.18) {(c) Nested wells};
        \end{scope}
    \end{tikzpicture}
    \caption{Generic particle-hole symmetric potential profiles. The red segment on the vertical axis marks the corresponding domain of $Q_{\gamma,J}(K)$, while the orange marks on the lower line indicate the possible turning values $\rho^*$.}
    \label{fig:QdomainSketch}
\end{figure}
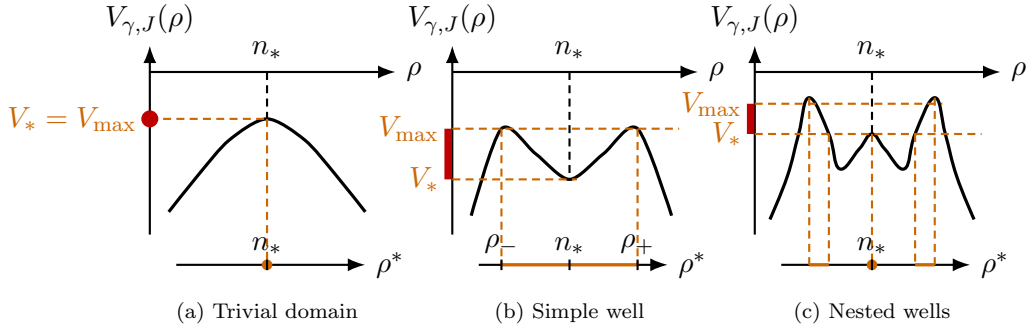

Due to the assumed particle-hole symmetry, both solutions contribute equally to the rate function and thus we may drop the $\pm$ index, defining, without loss of generality, $\rho_{\gamma,J,K}(x) \coloneqq \rho_{\gamma,J,K,+}(x)$. The remaining boundary condition, $\rho_{\gamma,J,K}(1)=n_*$, requires the trajectory to return to $n_*$ after traversing the full interval $[0,1]$. This requirement imposes the constraint
\begin{equation}\label{eq:timeconstraint}
     \mathcal{C}:G_{\gamma,J}(K)=2\int_{n_*}^{\rho_{\gamma,J,K}^*} d\rho\;\frac{D(\rho)}{\sqrt{\sigma(\rho)}} \frac{1}{\sqrt{2\left(K-V_{\gamma,J}(\rho)\right)}} - 1= 0,
\end{equation}
where $\rho_{\gamma,J,K}^*$ denotes the unique extremal value of the saddle-point profile $\rho_{\gamma,J,K}(x)$ (see appendix~\ref{sec:MethodForRF} for a proof that the minimal saddle-point profile must have at most a single extremal point). More rigorously, $\rho_{\gamma,J,K}^*$ satisfies $V_{\gamma,J}\left(\rho_{\gamma,J,K}^*\right)=K$ and corresponds precisely to the turning point of the particle's trajectory according to the classical mechanics analogy made. Condition $\mathcal{C}$ in Eq.~(\ref{eq:timeconstraint}) reduces the admissible values of $K$ to at most a discrete set. For SSEP, it turns out there is only one value of $K$ satisfying condition $\mathcal{C}$, for which we plot the corresponding saddle-point solutions in panel (a) of Fig.~\ref{fig:KconstraintSketch} (choosing $E=4$ and $\gamma=0$).
$U_{\gamma}(J)$ is then given by
\begin{multline}\label{eq:rfrhov0}
    U_{\gamma}(J)=\min_{K:\mathcal{C}} Q_{\gamma,J}(K) - J E, \\ \;  \text{with  } Q_{\gamma,J}(K) \coloneqq -K + 2 \int_{n_*}^{\rho_{\gamma,J,K}^*} d\rho\;\frac{D(\rho)}{\sqrt{\sigma(\rho)}} \sqrt{2\left(K-V_{\gamma,J}(\rho)\right)},
\end{multline}
and where the minimization over $K$ is constrained to condition $\mathcal{C}$. 
In Fig.~\ref{fig:KconstraintSketch}, we also plot the potential $V_{\gamma,J}(\rho)$ for SSEP with the same choice of parameters.
Since $V_{\gamma,J}(\rho)\leq 0$, it is bounded from above. As we are interested in bounded trajectories starting and ending at $\rho = n_*$, the "energy" $K$ must be chosen in $[V_*,V_{\text{max}}]$, where $V_* \coloneqq V_{\gamma,J}(n_*)$ and $V_{\rm max} = \max_{\rho} V_{\gamma,J}(\rho)$. Note that the functions $Q_{\gamma,J}$ and $G_{\gamma,J}$ are well-defined only for $K$ within this domain. In Fig.~\ref{fig:QdomainSketch} we exemplify some possible scenarios for the shape of the potential and the corresponding domain for $K$.

At this point, we make an observation not explicitly noted in either of the cited works: namely that $\partial_K Q_{\gamma,J}(K) = G_{\gamma,J}(K)$, which implies that the constraint $\mathcal{C}$ is automatically satisfied at the extremal points of $Q_{\gamma,J}(K)$ lying in the interior of its domain.
Indeed, differentiating $Q_{\gamma,J}(K)$ with respect to $K$, the term generated by the $K$-dependence of the upper limit $\rho_{\gamma,J,K}^*$ vanishes as $\left.\sqrt{2(K-V_{\gamma,J}(\rho))}\right|_{\rho=\rho_{\gamma,J,K}^*}=0$; hence $\partial_K Q_{\gamma,J}(K)=G_{\gamma,J}(K)$ and any interior extremum of $Q_{\gamma,J}$ automatically satisfies constraint $\mathcal C$. In principle, the global minimum of $Q_{\gamma,J}(K)$ could still occur at the boundaries without $\mathcal{C}$ being satisfied, but, with further analysis, one can show that this cannot be the case for $K=V_{\text{max}}$ (see appendix~\ref{sec:MethodForRF}) and that, at $K=V_*$, $\mathcal{C}$ is trivially fulfilled. This observation allows us to eliminate the constraint in the minimization appearing in Eq.~(\ref{eq:rfrhov0}), thus yielding
\begin{equation}\label{eq:rfrhov1}
    U_{\gamma}(J)=\min_{K} Q_{\gamma,J}(K) - J E.
\end{equation}
Although the functions $Q_{\gamma,J}(K)$ and $G_{\gamma,J}(K)$ can be analytically intractable and may not admit closed-form expressions, they nonetheless lead to a straightforward and numerically efficient method for computing the rate function using Eqs.~(\ref{eq:APdef},\ref{eq:rfrhov1}). Despite their complexity, it is still possible to detect the onset of a dynamical phase transition (DPT) as $J$ increases, by studying the behavior of $G_{\gamma,J}(K)$ perturbatively around $K = V_*$.

\subsubsection{Classification of DPTs and perturbative analysis}\label{subsec:classificationFPT}

\begin{figure}[!htbp]
    \centering
    \includegraphics[width=0.78\linewidth]{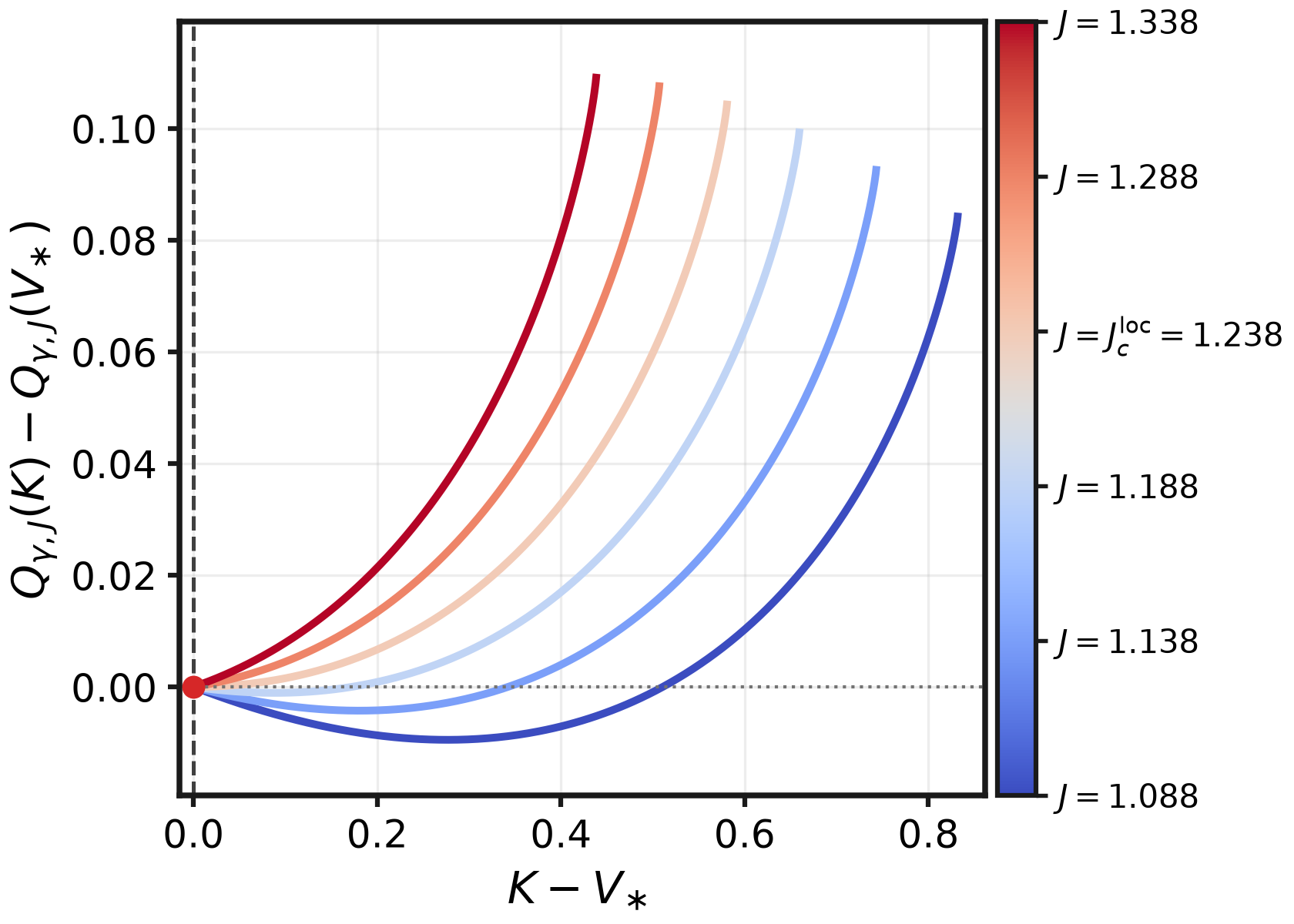}
    \caption{$Q_{\gamma,J}(K)-Q_{\gamma,J}(V_*)$ vs $K-V_*$, for different values of $J$, near the local critical current. At $J=J_c^{\rm loc}$ (light orange), the perturbative condition is $Q'_{\gamma,J_c^{\rm loc}}(V_*)=G_{\gamma,J_c^{\rm loc}}(V_*)=0$. For $J<J_c^{\rm loc}$, the global minimum of $Q_{\gamma,J}(K)$ is pinned to $V_*$ and for $J>J_c^{\rm loc}$ it becomes larger than $V_*$. This plot is obtained for SSEP at $E=4$ and $\gamma=0$.}
    \label{fig:QcriticalSketch}
\end{figure}

For currents sufficiently close to the average current $\overline{J} = \sigma(n_*) E$, the value of $K$ minimizing Eq.~\eqref{eq:rfrhov1} remains $K = V_*$ and the density profile remains pinned to the symmetric profile $\rho_{\gamma,J}(x)=n_*$.
However, changing $J$ sufficiently away from $\overline{J}$ may lead to a symmetry breaking transition. 
If the transition is continuous, it can be studied perturbatively by determining when $Q_{\gamma,J}(V_*)$ ceases to be a local minimum and new symmetry-breaking profiles $\rho_{\gamma,J}(x)$ that minimize the action appear in its vicinity. Since $V_{*}$ sits at the boundary of the accessible domain for $K$, the condition for $K=V_*$ to be a local minimum is precisely that $G_{\gamma,J}(V_*) = \partial_K Q_{\gamma,J}(K)\rvert_{K=V_*} > 0$, while it becomes a local maximum when $G_{\gamma,J}(V_*) < 0$. Perturbatively, one thus argues that the DPT occurs at the boundary between these two regimes, at a critical current $J_c^{\rm loc}$ for which $G_{\gamma,J_c^{\rm loc}}(V_*) = 0$, as shown in Fig.~\ref{fig:QcriticalSketch} for the SSEP with $E=4$ and $\gamma=0$. Here, the superscript "loc" emphasizes that this condition only probes the local stability of the symmetric profile: for a first-order transition, $J_c^{\rm loc}$ generally differs from the true critical current $J_c$, which is instead determined by an exchange of global minima.
Note that, in the previous discussion, we require $V_{\gamma,J_c^{\rm loc}}''(n_*) \geq 0$ (see Fig.~\ref{fig:QdomainSketch}(b)) so that  $\rho^*_{\gamma,J}(K \sim V_*) \sim n_*$ and $Q_{\gamma,J}(K)$ is smooth around $K = V_*$. In contrast, for $V_{\gamma,J_c^{\rm loc}}''(n_*) < 0$ (see Fig.~\ref{fig:QdomainSketch}(a,c)), the turning points $\rho_{\gamma,J,K}^*$ jump discontinuously when $K > V_\ast$ and consequently $Q_{\gamma,J}(K)$ is not a smooth function in this neighborhood.
Under the assumption of local convexity for $V_{\gamma,J}(\rho \sim n_*)$, we can expand $V_{\gamma,J_c^{\rm loc}}(\rho)$ up to second order around $n_*$, which allows one to express $J_c^{\rm loc}$ as a function of $\gamma$, $E$, and the model parameters,
\begin{equation}\label{eq:GeqCond1}
 G_{\gamma,J_c^{\rm loc}}(V_*) = 0\Longrightarrow \left(J_c^{\rm loc}\right)^2=\frac{\sigma_*}{\sigma_*''}\left(2 \pi^2 D_*^2 + 2 \gamma\,\sigma_* \right)+ \left(\sigma_*\,E\right)^2 >0,
\end{equation}
where we introduced the shorthand $D_* \coloneqq D(n_*)$ and $\sigma_* \coloneqq \sigma(n_*)$.

For $\gamma = 0$, this result is equivalent to that of Ref.~\cite{Lecomte_2017}: a DPT always occurs when $\sigma_*'' > 0$, and, conversely, if $\sigma_*'' < 0$, a DPT is expected only if the external field exceeds a threshold $E_\text{thresh}$. Turning to the monitored case $\gamma \neq 0$, we find that the qualitative picture remains unchanged: measurements cannot induce DPTs, but rather shift the critical values of the relevant parameters. Both $J_c^{\rm loc}$ and $E_\text{thresh}$ are quantitatively modified, while the underlying structure of the transition persists. Indeed, from Eq.~(\ref{eq:GeqCond1}) one finds that for $\sigma_*''>0$ a DPT always occurs, whereas for $\sigma_*''<0$,
\begin{equation}\label{eq:Ec?}
E_\text{thresh}^2 = \frac{2 \left(\pi^2 D_*^2 + \sigma_*\,\gamma\right)}{\sigma_* |\sigma_*''|}.
\end{equation}
For the SSEP, substituting the transport coefficients in Eq.~\eqref{eq:SSEPtranspcoeff} into Eq.~\eqref{eq:Ec?} gives $E_\text{thresh}^2=\pi^2+\gamma/2$, so the critical field grows monotonically with the measurement strength: monitoring \emph{suppresses} the dynamical phase transition, pushing it to larger driving. Note that, even though measurements cannot induce a DPT, they can still change the order of the DPT, as we shall address later.

As previously noted, this discussion is valid for continuous (second or higher-order) transitions. The study of first-order DPTs requires keeping track of changes of global minima of $Q_{\gamma,J}(K)$ that correspond to a discontinuous change in the profile $\rho_{\gamma,J}(x)$ with $J$.  Nonetheless, we can determine two conditions that generally ensure a second-order transition.

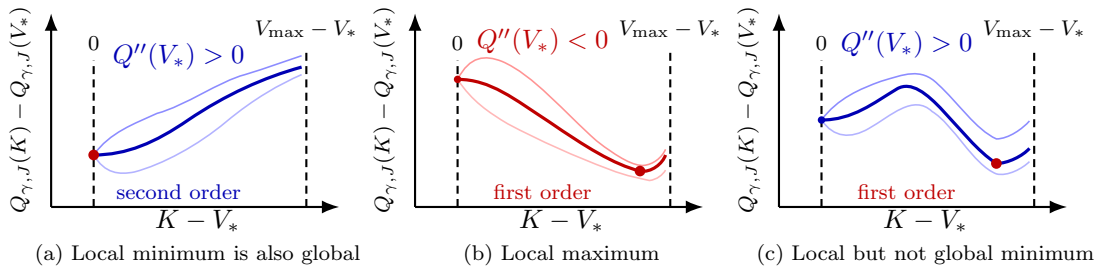
\begin{figure}[!htbp]
    \centering
    \begin{tikzpicture}[x=1.25cm,y=1.25cm,>=Latex,thick]
        \begin{scope}
            \draw[->] (0,0) -- (3.05,0);
            \node[font=\small] at (1.55,-0.15) {$K-V_*$};
            \draw[->] (0,0) -- (0,2.05);
            \node[rotate=90,font=\scriptsize] at (-0.32,1.03) {$Q_{\gamma,J}(K)-Q_{\gamma,J}(V_*)$};
            \draw[densely dashed] (0.45,0) -- (0.45,1.55) node[above,font=\scriptsize] {$0$};
            \draw[densely dashed] (2.7,0) -- (2.7,1.68) node[above,font=\scriptsize] {$V_{\max}-V_*$};
            \draw[blue!30,semithick] (0.45,0.55) .. controls (0.57,0.35) and (0.72,0.34) .. (0.88,0.37) .. controls (1.18,0.45) and (1.55,0.65) .. (1.85,0.9) .. controls (2.15,1.15) and (2.42,1.3) .. (2.65,1.4);
            \draw[blue!45,semithick] (0.45,0.55) .. controls (0.62,0.79) and (0.88,0.88) .. (1.12,0.99) .. controls (1.42,1.07) and (1.78,1.33) .. (2.05,1.4) .. controls (2.28,1.47) and (2.5,1.55) .. (2.65,1.6);
            \draw[blue!70!black,very thick] (0.45,0.55) .. controls (0.8,0.55) and (1.1,0.68) .. (1.45,0.92) .. controls (1.85,1.19) and (2.25,1.37) .. (2.65,1.48);
            \fill[blue!70!black] (0.45,0.55) circle (1.4pt);
            \fill[red!75!black] (0.45,0.55) circle (2.0pt);
            \node[blue!70!black,font=\small,fill=white,inner sep=0.8pt] at (1.35,1.59) {$Q''(V_*)>0$};
            \node[blue!70!black,font=\scriptsize,fill=white,inner sep=0.8pt] at (1.35,0.16) {second order};
            \node[align=center,font=\scriptsize] at (1.55,-0.5) {(a) Local minimum is also global};
        \end{scope}
        \begin{scope}[shift={(3.85,0)}]
            \draw[->] (0,0) -- (3.05,0);
            \node[font=\small] at (1.55,-0.15) {$K-V_*$};
            \draw[->] (0,0) -- (0,2.05);
            \node[rotate=90,font=\scriptsize] at (-0.32,1.03) {$Q_{\gamma,J}(K)-Q_{\gamma,J}(V_*)$};
            \draw[densely dashed] (0.45,0) -- (0.45,1.55) node[above,font=\scriptsize] {$0$};
            \draw[densely dashed] (2.7,0) -- (2.7,1.68) node[above,font=\scriptsize] {$V_{\max}-V_*$};
            \draw[red!28,semithick] (0.45,1.35) .. controls (0.6,1.12) and (0.78,1.0) .. (0.98,0.88) .. controls (1.32,0.7) and (1.7,0.53) .. (2.05,0.4) .. controls (2.25,0.32) and (2.4,0.3) .. (2.48,0.28) .. controls (2.56,0.29) and (2.62,0.34) .. (2.65,0.39);
            \draw[red!42,semithick] (0.45,1.35) .. controls (0.58,1.6) and (0.74,1.61) .. (0.92,1.54) .. controls (1.25,1.4) and (1.58,1.0) .. (1.92,0.7) .. controls (2.15,0.52) and (2.32,0.43) .. (2.42,0.44) .. controls (2.52,0.45) and (2.61,0.52) .. (2.65,0.61);
            \draw[red!75!black,very thick] (0.45,1.35) .. controls (0.8,1.35) and (1.05,1.14) .. (1.42,0.9) .. controls (1.8,0.64) and (2.1,0.45) .. (2.38,0.38) .. controls (2.5,0.36) and (2.62,0.46) .. (2.65,0.55);
            \fill[red!75!black] (0.45,1.35) circle (1.4pt);
            \fill[red!75!black] (2.38,0.38) circle (2.0pt);
            \node[red!75!black,font=\small,fill=white,inner sep=0.8pt] at (1.35,1.75) {$Q''(V_*)<0$};
            \node[red!75!black,font=\scriptsize,fill=white,inner sep=0.8pt] at (1.35,0.16) {first order};
            \node[align=center,font=\scriptsize] at (1.55,-0.5) {(b) Local maximum};
        \end{scope}
        \begin{scope}[shift={(7.7,0)}]
            \draw[->] (0,0) -- (3.05,0);
            \node[font=\small] at (1.55,-0.15) {$K-V_*$};
            \draw[->] (0,0) -- (0,2.05);
            \node[rotate=90,font=\scriptsize] at (-0.32,1.03) {$Q_{\gamma,J}(K)-Q_{\gamma,J}(V_*)$};
            \draw[densely dashed] (0.45,0) -- (0.45,1.55) node[above,font=\scriptsize] {$0$};
            \draw[densely dashed] (2.7,0) -- (2.7,1.68) node[above,font=\scriptsize] {$V_{\max}-V_*$};
            \draw[blue!30,semithick] (0.45,0.92) .. controls (0.58,0.73) and (0.76,0.72) .. (0.95,0.83) .. controls (1.18,0.96) and (1.42,1.2) .. (1.62,1.01) .. controls (1.85,0.87) and (2.05,0.45) .. (2.3,0.35) .. controls (2.42,0.32) and (2.58,0.41) .. (2.65,0.48);
            \draw[blue!45,semithick] (0.45,0.92) .. controls (0.62,1.15) and (0.86,1.25) .. (1.08,1.33) .. controls (1.32,1.4) and (1.5,1.48) .. (1.68,1.3) .. controls (1.9,1.1) and (2.08,0.82) .. (2.3,0.72) .. controls (2.43,0.72) and (2.58,0.83) .. (2.65,0.91);
            \draw[blue!70!black,very thick] (0.45,0.92) .. controls (0.78,0.92) and (1.0,1.1) .. (1.25,1.25) .. controls (1.55,1.42) and (1.9,0.72) .. (2.3,0.46) .. controls (2.42,0.46) and (2.58,0.54) .. (2.65,0.62);
            \fill[blue!70!black] (0.45,0.92) circle (1.4pt);
            \fill[red!75!black] (2.3,0.46) circle (2.0pt);
            \node[blue!70!black,font=\small,fill=white,inner sep=0.8pt] at (1.35,1.72) {$Q''(V_*)>0$};
            \node[red!75!black,font=\scriptsize,fill=white,inner sep=0.8pt] at (1.35,0.16) {first order};
            \node[align=center,font=\scriptsize] at (1.55,-0.5) {(c) Local but not global minimum};
        \end{scope}
    \end{tikzpicture}
    \caption{Possible scenarios for the behavior of $Q_{\gamma,J}(K)-Q_{\gamma,J}(V_*)$ as a function of $K-V_*$ at $J=J_c^{\rm loc}$, where $Q'(V_*)=G(V_*)=0$. The lighter curves sketch the profiles immediately before and after the transition. Panels (a) and (b) show cases when the local criterion (sign of $Q''(V_*)$) is sufficient to detect the order of the transition; panel (c) shows a case when a distant minimum can still make the transition first order even though $Q''(V_*)>0$; this clarifies why $Q''(V_*)>0$ is a necessary but not sufficient condition for a putative second order transition.}
    \label{fig:QperturbativeSketch}
\end{figure}

First, a necessary condition for a second-order DPT is $G'_{\gamma,J_c^{\rm loc}}(V_*) = Q''_{\gamma,J_c^{\rm loc}}(V_*) > 0$, as illustrated in Fig.~\ref{fig:QperturbativeSketch}. Indeed, if $Q''_{\gamma,J_c^{\rm loc}}(V_*) < 0$ [Fig.~\ref{fig:QperturbativeSketch}(b)], then $Q(V_*)$ becomes a local maximum, forcing the global minimum to lie elsewhere. As shown in Appendix~\ref{app:gencond}, this condition can be written as $A_*(E,\gamma)\coloneqq E^2+A_\gamma \,\gamma + A_0 >0$, where the model-dependent coefficients $A_0$ and $A_\gamma$ are defined in Eq.~\eqref{eq:AcoeffsStability}.
A second condition follows from our use of the additivity principle: $U_{\gamma}(J)$ must remain convex in a neighborhood of $J_c^{\rm loc}$. Otherwise, its convex envelope $U_{\gamma}^{**}(J)$ develops a linear segment, signaling a first-order transition that cannot be detected through the local analysis of $Q_{\gamma,J}(K)$. In Appendix~\ref{app:gencond}, this convexity condition is written as $B_*(E,\gamma)/A_*(E,\gamma)>0$, where $B_*(E,\gamma)\coloneqq E^2+B_\gamma \,\gamma + B_0$ and the coefficients $B_0$ and $B_\gamma$ are defined in Eq.~\eqref{eq:BcoeffsStability}. Thus, a second-order DPT requires both $A_*(E,\gamma)>0$ and $B_*(E,\gamma)>0$. Within the region where a local DPT can occur, however, $B_*(E,\gamma)>0$ implies $A_*(E,\gamma)>0$, so the two requirements reduce to the single condition $B_*(E,\gamma)>0$.

In the following, we will use this condition to detect situations where, by tuning the measurement strength $\gamma$, one can induce a change in the order of the phase transition, from second order to first order.
We now turn to providing concrete examples of particle-hole symmetric models in which these behaviors can be observed.

\subsubsection{Analysis of specific models}

We now apply the local accessibility criterion, Eq.~\eqref{eq:GeqCond1}, and the transition-order condition $B_*(E,\gamma)>0$ discussed at the end of the previous subsection to the SSEP and Katz--Lebowitz--Spohn (KLS) models. 
For the SSEP, the local conditions place the model in case (a) of Fig.~\ref{fig:SketchStability}: whenever the DPT is locally accessible, it is predicted to be second order, and monitoring only shifts its threshold. In Appendix~\ref{sec:SSEPextensiveanalysis}, we solve the full stationary problem analytically and rule out a loss of convexity of $U_\gamma(J)$. This excludes first-order transitions within the weak-additivity description and confirms the local analysis.

The KLS model~\cite{KLS1984} features both on-site exclusion (implying $0\leq\rho\leq 1$) and nearest-neighbor interactions and has a richer phenomenology that captures the change of the order of the DPT with the increase of the measurement strength. This model is governed by two parameters, $\delta$ and $\epsilon$, which modulate the hopping dynamics through the following transition rates:
\begin{align*}
  0100 \xrightarrow{1+\delta} 0010, \quad 1101 \xrightarrow{1-\delta} 1011, \\
  1100 \xrightarrow{1+\epsilon} 1010, \quad 0101 \xrightarrow{1-\epsilon} 0011,  
\end{align*}
with transitions between spatially inverted (left/right) configurations having the same rates. We restrict our attention to particle-hole symmetric cases where $\delta = 0$, $n_* = 1/2$. The KLS model admits a macroscopic fluctuation theory (MFT) description~\cite{Hager2001MinimalCP,Krapivsky_2013,Lecomte_2017}, under which the transport coefficients take the form
\begin{equation}\label{eq:KLStranspcoeff}
    D(\rho)= \frac{
4 \left(1 - 4 \epsilon \rho (1 - \rho)  + \sqrt{1-\frac{ 8 \epsilon \rho (1 - \rho) }{1 + \epsilon}} \right)
}{
\sqrt{1-\frac{ 8 \epsilon \rho (1 - \rho) }{1 + \epsilon}}
\left(1 + \sqrt{1-\frac{ 8 \epsilon \rho (1 - \rho) }{1 + \epsilon}} \right)^3
}, \;  \frac{\sigma(\rho)}{D(\rho)} = 2 \rho  (1 - \rho) \sqrt{1-\frac{ 8 \epsilon \rho (1 - \rho) }{1 + \epsilon}}. 
\end{equation}

The local analysis of Appendix~\ref{app:gencond} shows that the KLS model realizes all four cases of Fig.~\ref{fig:SketchStability} as $\epsilon$ is varied (see Table~\ref{tab:coefficients}). We focus on $\epsilon=0.6$, which lies in case (d), the only one in which both first- and second-order DPTs occur for every value of $\gamma$, and explore this structure by varying $E$ and $\gamma$.

\begin{figure}[!htbp]
    \centering
    \includegraphics[width=1\linewidth]{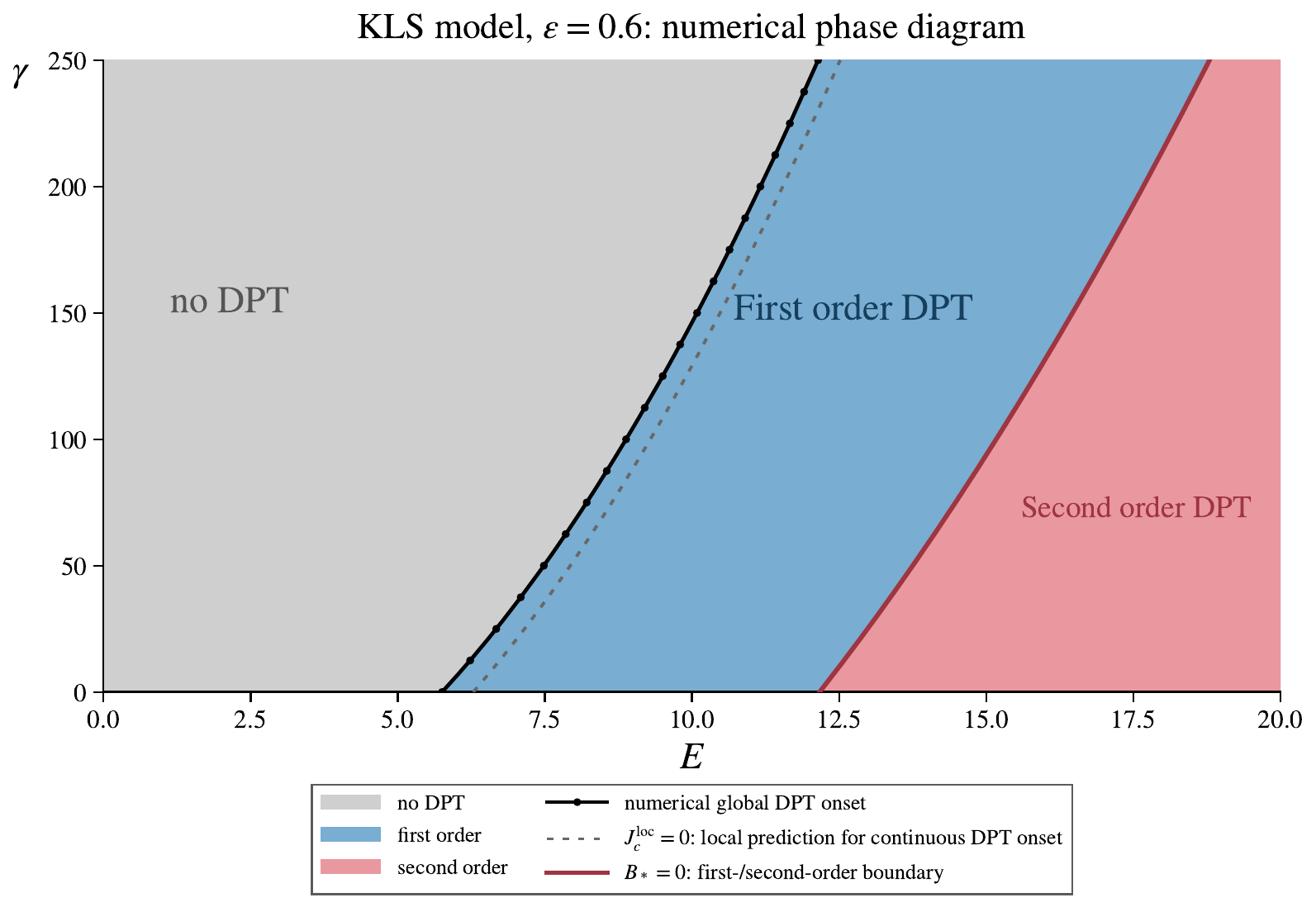}
    \caption{
    Nature of the putative critical point ($J=J_c$) for the particle-hole-symmetric KLS model at $\epsilon=0.6$ and $\delta=0$. Gray denotes the region with no DPT, while blue and red indicate first- and second-order DPTs, respectively. The solid black curve corresponds to $J_c=0$. This curve is distinct from the gray dotted line, $J_c^{\rm loc}=0$, which gives the local condition for the onset of a DPT. The solid red curve represents our derived condition $B_*(E,\gamma)=0$ for the boundary between first- and second-order transitions, in agreement with the numerical results.}
    \label{fig:KLSEps060PhaseDiagram}
\end{figure}

\begin{figure}[!htbp]
    \centering
    \includegraphics[width=\linewidth]{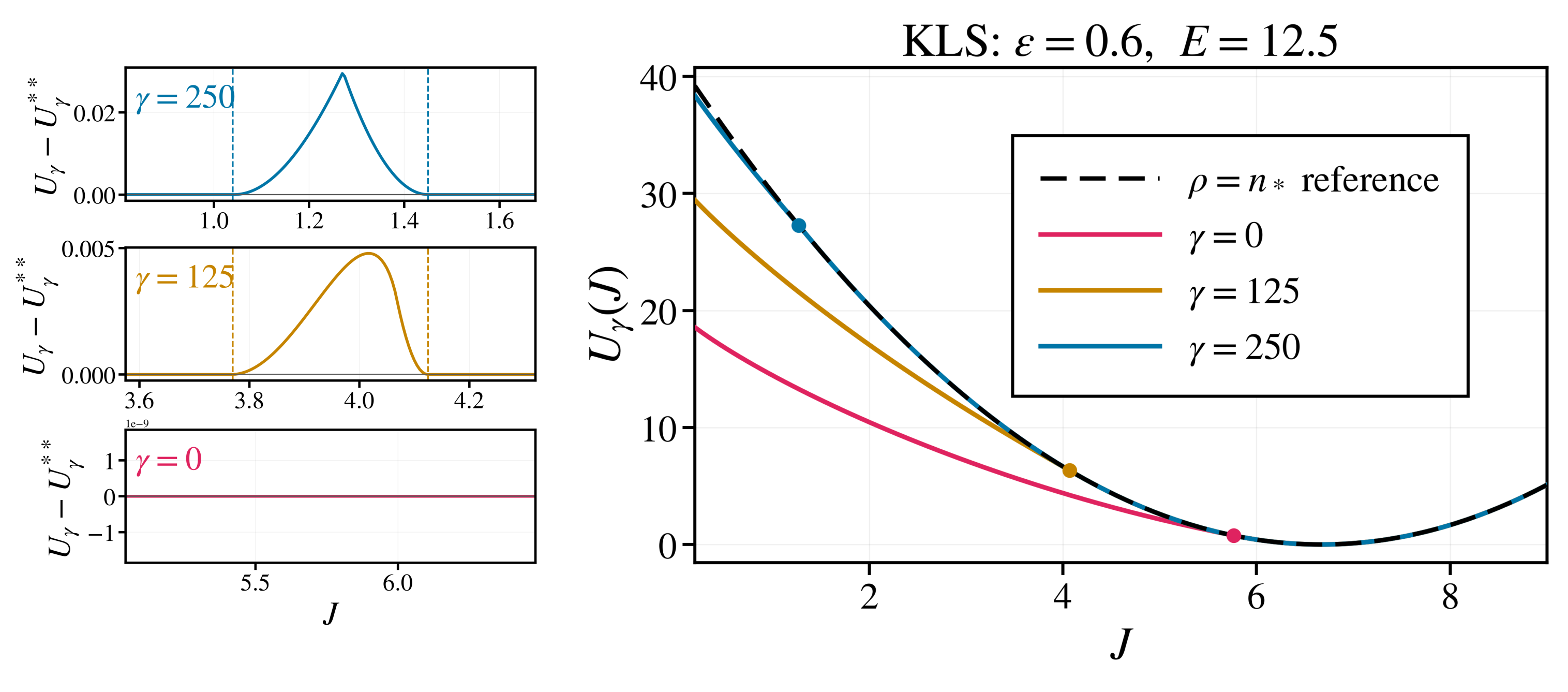}
    \caption{Measurement-driven change of the DPT in the rate function of the current of KLS at $\epsilon=0.6$, $\delta=0$, and $E=12.5$. Right: the function $U_\gamma(J)$ for $\gamma=0,125,250$, together with the homogeneous-profile reference $\mathcal{U}_{\gamma}(\rho\equiv 1/2,J)$ (black dashed line). 
    For sufficiently large $\gamma$, no DPT is present (see Fig.~\ref{fig:KLSEps060PhaseDiagram}) and $U_\gamma(J) = \mathcal{U}_\gamma(1/2, J)$. 
    Left: the convexification gap $U_\gamma(J)-U_\gamma^{**}(J)$. The transition is second order at $\gamma=0$ as $U_\gamma(J)-U_\gamma^{**}(J)=0$ and first order at $\gamma=125$ and $\gamma=250$.}
    \label{fig:DPT_Meas}
\end{figure}

Figure~\ref{fig:KLSEps060PhaseDiagram} shows the resulting numerical phase diagram. Besides the region in which no DPT is detected, it contains a first-order region and a second-order region, in agreement with the local analysis. After global minimization, the contour $B_*(E,\gamma)=0$ remains accurate, as it coincides exactly with the boundary between first- and second-order DPTs. By contrast, the local condition $J_c^{\rm loc}=0$ in Eq.~(\ref{eq:GeqCond1}) does not give the global onset of the DPT, because a finite-amplitude first-order transition extends into the locally inaccessible region represented by $\left(J_c^{\rm loc}\right)^2<0$. Appendix~\ref{app:KLSBoundaryTests} presents the full stationary analysis, including the underlying density maps and independent high-resolution tests of the local predictions.
Interestingly, along a vertical cut at fixed $E$, increasing $\gamma$ can drive the system from the second-order region into the first-order region, demonstrating that monitoring can change the order of the DPT.
Figure~\ref{fig:DPT_Meas} illustrates this along a cut at $E=12.5$. At $\gamma=0$, $U_\gamma(J)$ is convex and coincides with its convex envelope, so the transition is second order. At $\gamma=125$ and $\gamma=250$, instead, $U_\gamma(J)$ differs from its convex envelope, which produces a coexistence interval and makes the transition first order.

We emphasize that, for first order transitions, we cannot assert with certainty that the weak additivity principle holds, i.e. $\phi_{\gamma}(J)=U_{\gamma}^{**}(J)$. 
The cases where it does not hold correspond to a DPT to a time-translation symmetry broken phase.

Near second order transitions, one can check perturbatively, by a straightforward extension of the results of Ref.~\cite{Lecomte_2017}, that the time-independent solutions are always stable. 
This means that, upon increasing $J$, no breaking of the time-translation symmetry appears, but the system transitions from a uniform density profile to a static particle-hole symmetry broken phase. 

Having established how monitoring can impact the stability of saddle-point solutions—and thereby influence dynamical phase transitions—we now turn, in the next section, to the strong monitoring limit in general MFT models that satisfy the convexity conditions discussed in Sec.~\ref{sec:covexcond}.

\section{Non-equilibrium setups and strong-measurement expansion}\label{sec:LMS}

In this section, we provide a characterization of the current statistics in non-equilibrium setups. 
We restrict our analysis to one-dimensional transport systems with fixed boundary densities $\rho(0) = n_0$ and $\rho(1) = n_1$ and no external field,  $E = 0$. 
Note that the method introduced in the previous section assumes translation invariance of the stationary profile in space $x$, a hypothesis that breaks for $n_0 \neq n_1$, so we follow a different route. 
For simplicity, we assume the diffusion constant does not depend on the density, $D(\rho)=1$. 
This assumption can be relaxed without qualitatively changing the conclusions of this section.\footnote{With $\omega = f(\rho) = \int^{\rho} d\rho'\, D(\rho')$, the rate function becomes $\phi_{\gamma}(J) = \int_0^1 dx\,\Big(\frac{(J + \partial_x \omega_{\gamma,J})^2}{2 \widetilde{\sigma}(\omega_{\gamma,J})} + \frac{\gamma}{2}\big(f^{-1}(\omega_\infty) - f^{-1}(\omega_{\gamma,J})\big)^2\Big)$, where $\omega_{\gamma,J}=f(\rho_{\gamma,J})$, $\widetilde{\sigma}(\omega)=\sigma(f^{-1}(\omega))$ and $\omega_\infty= f(n_0)+\left( f(n_1)-f(n_0) \right) x$. Compared with the case $D=1$, the only change is in the measurement term, which near $\omega_\infty$ acts as a position-dependent measurement strength $\gamma/D(\rho_\infty(x))^2$. This modifies some prefactors in the results of this section, but neither the methods nor the conclusions.}
Furthermore, we assume strict convexity of the functional $\mathcal{U}_{\gamma}(\rho,J)$, which not only implies that the additivity principle holds, and thus the rate function of the current is given by Eq.~(\ref{eq:APdef}), but also that there is a single minimal density profile $\rho_{\gamma,J}(x)$ minimizing $\mathcal{U}_{\gamma}(\rho,J)$ for each $J$ (and so no DPT). 
In particular, these assumptions are fully verified by the SSEP (see Eq.~\eqref{eq:SSEPtranspcoeff}), making it a prototypical example for the current discussion. 

We emphasize once again that in the regime we are considering here, no MIPT appears, in particular, no form of charge-sharpening transition~\cite{nahum2025_ClassicalMeasurements, gopalakrishnan2025_ClassicalMeasurements}: the reason is that we are looking at the coarse-grained version of the measurement process, which therefore measures the density rather than the number of onsite particles.

From the assumption $D(\rho) = 1$, one concludes that $\rho_{\infty}(x) = n_0 + (n_1 - n_0) x$; moreover,  from Eqs.~(\ref{eq:APdef},\ref{eq:phiJgamma}), we can write \begin{equation}\label{eq:phiJgamma1}
    \phi_{\gamma}(J) = \int_0^1 dx \; \left( \frac{J^2 + \left( \partial_x \rho_{\gamma,J} \right)^2}{2 \sigma(\rho_{\gamma,J})} + \frac{\gamma}{2} \left( n_0 + (n_1 - n_0) x - \rho_{\gamma,J} \right)^2 \right) - \kappa J,
\end{equation}
with $\kappa$ defined in Eq.~\eqref{eq:kappamudef}. In the case of SSEP, 
\begin{equation}\label{eq:kappaforSSEP}
    \kappa_{\rm SSEP} = \int_{n_1}^{n_0}  \frac{d\rho}{\sigma(\rho)} = \frac{1}{2} \log\left(\frac{n_0 (1 - n_1)}{n_1 (1 - n_0)}\right)
\end{equation}

From this, one can straightforwardly derive the Euler-Lagrange equation for the unique saddle-point profile $\rho_{\gamma,J}(x)$. This yields
\begin{equation}\label{eq:rhof}
    \gamma \left(n_0 + (n_1 - n_0)x - \rho_{\gamma,J} \right)
    + c'(\rho_{\gamma,J}) \left( \left( \partial_x \rho_{\gamma,J} \right)^2 - J^2 \right)
    + 2 c(\rho_{\gamma,J}) \, \partial_x^2 \rho_{\gamma,J} = 0,
\end{equation}
where $c(\rho) \equiv \left(2 \sigma(\rho)\right)^{-1}$, together with boundary conditions $\rho_{\gamma,J}(0) = n_0$ and $\rho_{\gamma,J}(1) = n_1$. Solving Eq.~(\ref{eq:rhof}) allows us to directly compute the rate function $\phi_{\gamma}(J)$ and indirectly the cumulant generating function,
\begin{equation}
    \lambda_{\gamma}(v) = v J(v) - \phi_{\gamma}(J(v)),
\end{equation}
with the relation $J(v)$ established implicitly through
\begin{equation}
    v = \int_0^1 dx \; \frac{\left( J + \partial_x \rho_{\gamma,J}\right)}{ \sigma(\rho_{\gamma,J})}.
\end{equation}

Eq.~(\ref{eq:rhof}) can be efficiently solved numerically using standard techniques for ordinary differential equations (ODEs), providing a direct method to compute $\phi_{\gamma}(J)$ and $\lambda_{\gamma}(v)$ for each specific model. 
For large $\gamma$, the numerical solution reveals a quick variation of $\rho_{\gamma,J}(x)$ at the edges $x=0$ and $x=1$: to fully capture this behavior, we employ an adaptive algorithm, starting with an initial number of grid points \( N_x \) and increasing the number of points according to the estimated gradient of the function. In the following, we explore analytically the regime of strong measurements ($\gamma \to \infty$).
Before going into detail, we anticipate the behavior of the large $\gamma$ limit. Let us first consider the minimization of the function $\mathcal{U}_{\gamma}(\rho,J)$ but with free boundary conditions, i.e. without fixing the values $\rho_{\gamma,J}(0)$ and $\rho_{\gamma,J}(1)$. This unconstrained minimization of $\mathcal{U}_{\gamma}(\rho,J)$ provides an overall minimal solution that we denote by $\rho_{\gamma,J}^{\rm free}$ and clearly $\mathcal{U}_{\gamma}(\rho_{\gamma,J}^{\rm free},J) \leq \mathcal{U}_{\gamma}(\rho_{\gamma,J},J)$. As a consequence, the values of $\rho_{\gamma,J}^{\rm free}$ at the boundaries are entirely determined from the form of the functional Eq.~(\ref{eq:phiJgamma1}) and in general $\rho_{\gamma,J}^{\rm free}(0) \neq n_0$ and $\rho_{\gamma,J}^{\rm free}(1) \neq n_1$, although the functional form of $\rho^{\rm free}_{\gamma, J}$ still depends on $n_0, n_1$ as they enter explicitly the functional $\phi_{\gamma}(J)$ in Eq.~(\ref{eq:phiJgamma1}). 
We will see that the large-$\gamma$ expansion of $\rho_{\gamma, J}(x)$ splits into a \emph{regular} part in powers of $1/\gamma$, which coincides precisely with $\rho_{\gamma,J}^{\rm free}(x)$ in the bulk, and a non-analytic part which is understood as the matching of the correct boundary conditions $n_0/n_1$ with the behavior of $\rho_{\gamma,J}^{\rm free}(x)$ at the boundaries.

\subsection{Near-equilibrium expansion with particle-hole symmetry
\label{sec:neareqparticlehole}
}
To understand the behavior at large $\gamma$, it is useful to consider a regime where we can have explicit analytic expressions. Thus, we start by tackling the problem of solving the saddle-point equation, Eq.~(\ref{eq:rhof}), by considering a particle-hole symmetric model in the vicinity of its symmetric point $n_*$, i.e.\ $n_0=n_* + b_0 \delta$ and $n_1=n_* + b_1 \delta$, for small $\delta$. At $n_0=n_1=n_*$ exactly, $c'(n_*)=0$ and thus, as discussed in the previous section, the saddle-point profile is pinned to $n_*$, i.e.\ $\rho_{\gamma,J}=n_*$, for any $J$, provided no DPT occurs, which we assume throughout this section. 
This implies that $\rho_{\gamma,J} = n_*+\rho_{\gamma,J}^{(1)} \delta + \mathcal{O}(\delta^2)$ for small $\delta$.
At first order in $\delta$, Eq.~(\ref{eq:rhof}), becomes
\begin{equation}
    \gamma (b_0 + (b_1 - b_0) x - \rho_{\gamma,J}^{(1)}) -c''(n_*) J^2 \rho_{\gamma,J}^{(1)} + 2 c(n_*) \frac{d^2 \rho_{\gamma,J}^{(1)}}{dx^2} = 0.
\end{equation}
Subjected to the boundary conditions $\rho_{\gamma,J}^{(1)}(0)=b_0$ and $\rho_{\gamma,J}^{(1)}(1)=b_1$, the exact solution of this equation is 
\begin{equation}\label{eq:rhofpertdens}
    \rho_{\gamma,J}^{(1)}(x) = \frac{\gamma \left( b_0 + (b_1-b_0)x \right) + c''(n_*) J^2 \left(b_0 \frac{\sinh{\left(\eta (1-x)\right)}}{\sinh{\eta}} + b_1 \frac{\sinh{\left(\eta x\right)}}{\sinh{\eta}}\right)}{\gamma+c''(n_*) J^2} 
\end{equation}
where $\eta=\sqrt{\frac{\gamma+c''(n_*) J^2}{2 c(n_*)}}$.

The behavior of $\rho_{\gamma,J}^{(1)}$ reveals the key feature of the $\gamma \to \infty$ limit.
In fact, when increasing $\gamma$ to very large values, for $x$ at $O(1)$ distance from the boundaries, 
\begin{equation}
    \sinh(x \eta) / \sinh \eta \sim e^{-\sqrt{\frac{\gamma}{2 c(n_*)}}(1-x)},\;\sinh((1-x) \eta) / \sinh \eta \sim e^{-\sqrt{\frac{\gamma}{2 c(n_*)}}\,x},
\end{equation}
implying that these terms do not contribute to a perturbative expansion in $\gamma^{-1}$. Expanding Eq.~\eqref{eq:rhofpertdens} in $\gamma^{-1}$ at fixed $x$ in the bulk yields 
\begin{equation}
\label{eq:rho1largegamma}
\rho_{\gamma,J}^{(1)}(x) \stackrel{\gamma \to \infty}{\sim}(b_0 + (b_1 - b_0) x)\left(1- \gamma^{-1} c''(n_*) J^2 +\mathcal{O}\left(\gamma^{-2}\right)\right).
\end{equation}
We see that the $1/\gamma$ correction in this expansion leads to a violation of the boundary conditions, as the right hand side of Eq.~\eqref{eq:rho1largegamma} differs from $b_0/b_1$ when $x \to 0 / x\to 1$ respectively.
This signals the non-commutativity of the $\gamma \to \infty$ and $x \to 0,1$ limits. Quantitatively, one can see that 
\begin{equation}
\rho_{\gamma,J}^{(1)}(x = y/\sqrt{\gamma}) = b_0 + \left( b_1 - b_0 \right) \frac{y}{\sqrt{\gamma}} - \frac{b_0}{\gamma} c''(n_*) J^2\left(1-e^{-y/\sqrt{2c(n_*)}} \right) +\mathcal{O}\left(\gamma^{-3/2}\right),
\end{equation}
which indeed correctly interpolates between the right value at the boundary $b_0$ as $y \to 0$ ($x \ll \gamma^{-1/2}$) and the value that arises in the bulk (i.e. the limit as $x \to 0$ of the right-hand side of Eq.~\eqref{eq:rho1largegamma}). This is, in fact, the general mechanism: the optimal profile splits into a bulk part---which, at large $\gamma$, reduces to the solution of the unconstrained minimization---and matching functions living on the scales $x \sim \gamma^{-1/2}$ and $1-x \sim \gamma^{-1/2}$, which connect the bulk value to the imposed boundary conditions.

In the next section, we generalize this discussion to general values of $n_0$ and $n_1$.

\subsection{Method for irregular expansion at large measurement strength}\label{sec:asynptmethod}
\subsubsection{Expansion in the bulk and $\rho_{\gamma,J}^{\rm free}(x)$
\label{sec:bulkexp}
}
\begin{figure}[ht]
    \centering    \includegraphics[width=1\linewidth]{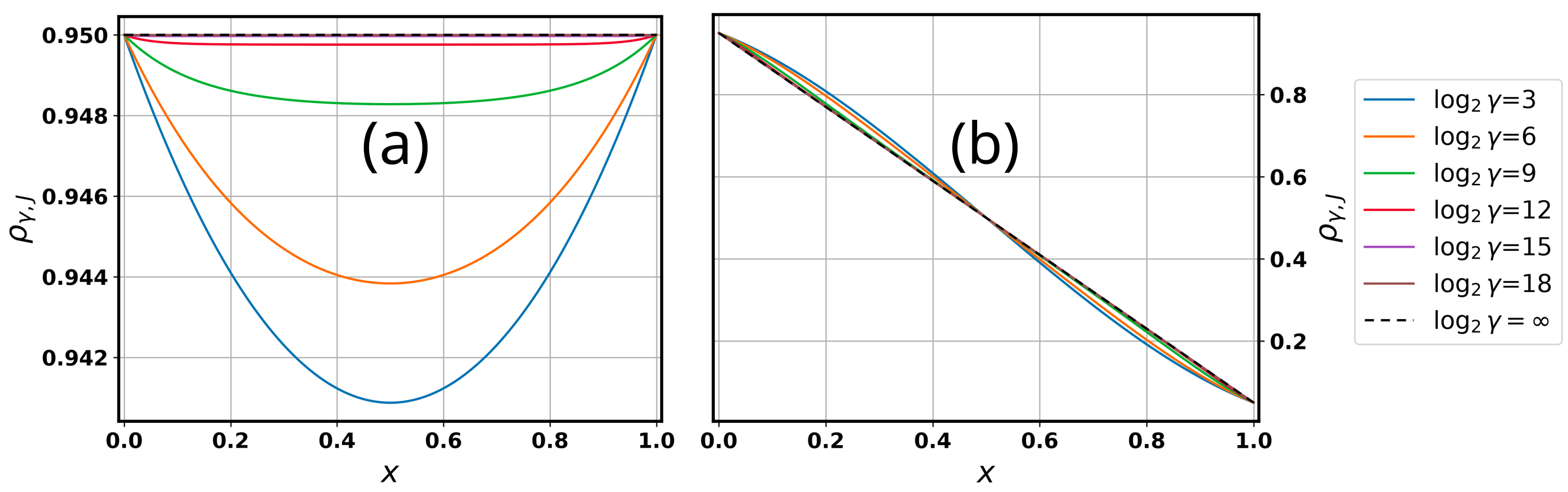}
    \caption{We consider monitored SSEP. We plot the typical saddle-point density profile as $\gamma$ is increased up to infinity. In (a), we set $J = 0.1$ and the boundary conditions to $n_0=n_1=0.95$ (equilibrium) and, in (b),  $J = 1.0$ and $n_0=1-n_1=0.95$ (out of equilibrium).}
    \label{fig:DensityView}
\end{figure}

In order to perform a large-$\gamma$ expansion of the rate/cumulant generating function, it is necessary to understand the behavior of $\rho_{\gamma,J}(x)$ in the limit $\gamma \to \infty$. We thus start by considering a Laurent expansion at each $x$, that we denote by 
\begin{equation}
\label{eq:laurentrho}
\rho_{\gamma,J}^{(\mathcal{B})}(x) = \sum_{n \in \mathbb{N}_0} \gamma^{-n} \rho_{J}^{(\mathcal{B}; n)}(x)
\;.
\end{equation} 
Because of the shape of Eq.~(\ref{eq:rhof}), one obtains that, at all orders $n$, $\rho_{J}^{(\mathcal{B}; n)}(x)$ satisfies an equation of the form
\begin{equation}
    \rho_{J}^{(\mathcal{B}; n)}(x) = F_{n}(x; J,\rho_{J}^{(\mathcal{B}; 0)},...,\rho_{J}^{(\mathcal{B}; n-1)}),
    \label{eq:rhonF}
\end{equation}
for a set of functionals $F_n$ determined from Eq.~(\ref{eq:rhof}). This provides a straightforward iterative method for computing the functions $\rho_{J}^{(\mathcal{B}; n)}$ at all orders. For instance, for $n=0$, $F_{0} = n_0-(n_0-n_1) x$ and thus $\rho_{J}^{(\mathcal{B}; 0)} = n_0-(n_0-n_1) x = \rho_\infty(x)$, as expected in the $\gamma \to \infty$ limit of the rate function, Eq.~(\ref{eq:phiJgamma1}) (see Fig.~\ref{fig:DensityView} for the case of SSEP).
This zero order solution already saturates the boundary conditions $\rho_{J}^{(\mathcal{B}; 0)}(0) = n_0$, $\rho_{J}^{(\mathcal{B}; 0)}(1) = n_1$.
Crucially, Eq.~\eqref{eq:rhonF} uniquely determines the $n$-order function $\rho_{J}^{(\mathcal{B}; n)}(x)$ in terms of the previous ones, without any freedom to account for the boundary conditions, which would, in principle, require the entire sum to satisfy $\rho_{\gamma,J}^{(\mathcal{B})}(x = 0) = n_0$ and $\rho_{\gamma,J}^{(\mathcal{B})}(x = 1) = n_1$. Indeed, immediately at first order, one can show that
\begin{equation}\label{eq:rhopert1}
    \rho_{J}^{(\mathcal{B}; 1)} = - c'(n_0 + (n_1 - n_0) x) \times \left( J^2 - \left(n_0-n_1\right)^2\right).
\end{equation}
Generally $c'(\rho)$ is non-vanishing and thus, unless $J^2 = \left(n_0-n_1\right)^2$, the boundary conditions are not satisfied. 
This suggests that complete minimization of Eq.~\eqref{eq:phiJgamma1} is not compatible with the Laurent expansion we used in $\rho_{\gamma,J}^{(\mathcal{B})}(x)$, which does not capture exponential terms (see Eq.~\eqref{eq:rhofpertdens} for the near-equilibrium case). More specifically, we see that the solution to the hierarchy in Eq.~\eqref{eq:rhonF}, since it does not depend on the boundary conditions, reduces to the unconstrained minimization, i.e. $\rho_{\gamma,J}^{(\mathcal{B})} = \rho^{\rm free}_{\gamma,J}(x)$, as anticipated at the beginning of this section.

\begin{figure}[ht]
    \centering
    \includegraphics[width=1\linewidth]{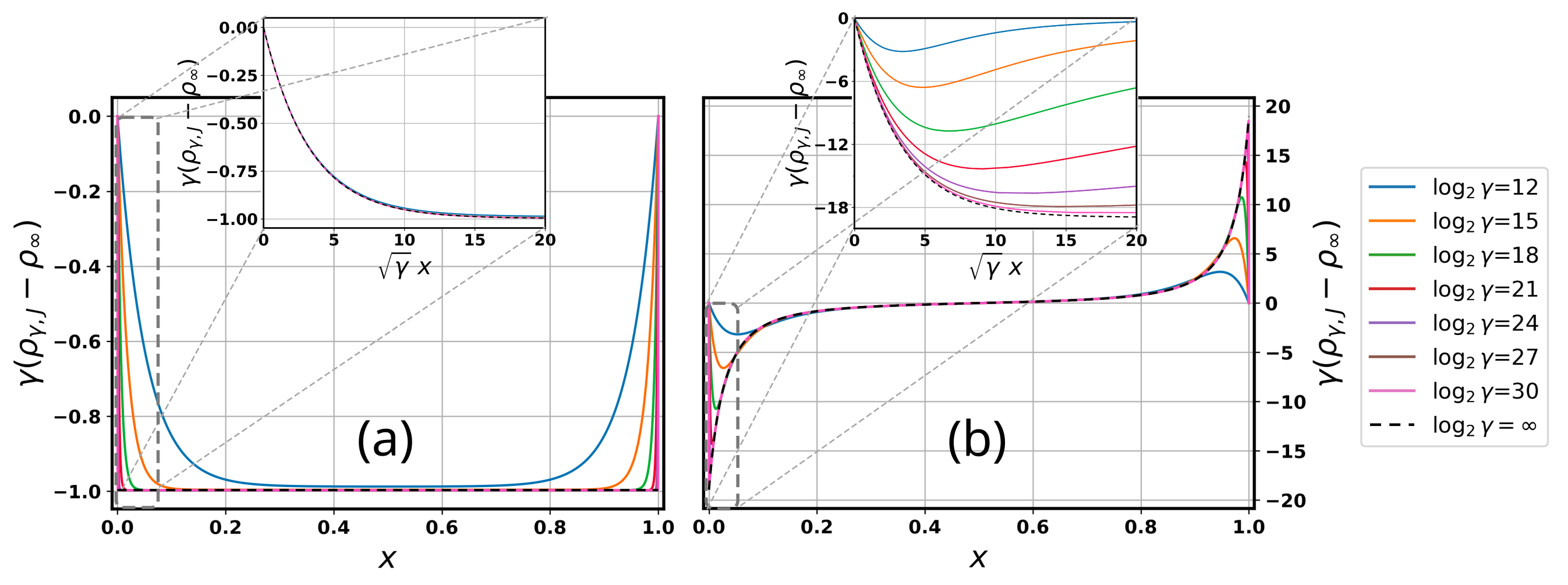}
    \caption{We consider monitored SSEP. We plot the difference of the typical saddle-point density to $\rho_{\infty}$ as $\gamma$ is increased up to infinity. In (a), we set $J = 0.1$ and boundary conditions $n_0=n_1=0.95$ and, in (b),  $J = 1.0$ and $n_0=1-n_1=0.95$ . In the insets, we show the convergence of the boundary density to $\rho_{J}^{(\text{L}; 1)}$, after the appropriate space rescaling $x \to \gamma^{1/2}x$. }
    \label{fig:DensityNS}
\end{figure}
As an example, in the equilibrium case $n_0 = n_1 = n_*$, $\rho_{\gamma,J}^{\rm free}(x) = \rho_{\gamma,J}^{\rm free} $ is a constant and can be determined solving Eq.~\eqref{eq:rhof}, which in this case reduces to
\begin{equation}
    J^2 c'(\rho_{\gamma,J}^{(\mathcal{B})}) + \gamma \left(\rho_{\gamma,J}^{(\mathcal{B})}-n_*\right)=0
\end{equation}
Comparing to Sec.~\ref{sec:neareqparticlehole}, in the presence of particle-hole symmetry, $c'(n_*) = 0$ and this equation is always solved by $\rho_{\gamma,J}^{(\mathcal{B})} = \rho_{\gamma, J}^{\rm free} = n_*$, even though in general it has a non-trivial dependence on $\gamma, J$.

\subsubsection{Matching functions at the boundary
\label{sec:matchingbound}
}
Assuming that $\sigma(\rho) \neq 0$ in the interval $\rho \in [n_1,n_0]$, from the structure of Eq.~(\ref{eq:rhof}), the correct way to extract the matching function from the boundaries to the bulk is to set
\[
    \rho_{\gamma,J}^{(\theta)}(y)=\rho_{\gamma,J}\left(x=\beta_{\gamma}^{(\theta)}(y)\right), \qquad
    \beta_{\gamma}^{(\theta)}(y)=\iota_\theta+s_\theta\frac{y}{\sqrt{\gamma}},
\]
where $\theta\in\{\text{L},\text{R}\}$, $\iota_{\text{L}}=0$, $\iota_{\text{R}}=1$, and $s_\theta=(-1)^{\iota_\theta}$. Thus $\beta_{\gamma}^{(\text{L})}(y)=y/\sqrt{\gamma}$ and $\beta_{\gamma}^{(\text{R})}(y)=1-y/\sqrt{\gamma}$. With this convention, increasing $\gamma$ simultaneously zooms in at the boundaries $x=0$ and $x=1$. Injecting this definition into Eq.~\eqref{eq:rhof}, one obtains
\begin{equation}\label{eq:rofResc}
n_0 + (n_1 - n_0) \beta_{\gamma}^{(\theta)}(y) - \rho_{\gamma,J}^{(\theta)}
     +c'(\rho_{\gamma,J}^{(\theta)}) \left( \left(\partial_y\rho_{\gamma,J}^{(\theta)}\right)^2 - \frac{J^2}{\gamma}\right) + 2 \;c(\rho_{\gamma,J}^{(\theta)}) \partial_y^2\rho_{\gamma,J}^{(\theta)} = 0 \;.
\end{equation}
This differential equation is complemented with the boundary conditions matching the value at the physical boundary with the bulk one,
\begin{equation}
\label{eq:boundaryrhoL}
    \rho_{\gamma,J}^{(\theta)}(0) = n_{\iota_\theta} \;, \qquad \lim_{y \to \infty} \left(\rho_{\gamma,J}^{(\theta)}(y)-\rho_{\gamma,J}^{(\mathcal{B})}\big(\beta_{\gamma}^{(\theta)}(y)\big)\right) = 0.
\end{equation}
It is also useful to consider the series expansion at large $\gamma$ of $\rho_{\gamma,J}^{(\theta)}(y)$. Due to the presence of the $\gamma^{-1/2}$ factor, one is forced to include half-integer powers of $\gamma^{-1}$,
\begin{equation}
    \rho_{\gamma,J}^{(\theta)}(y) = \sum_{n \in \mathbb{N}_0/2} \gamma^{-n} \rho_{J}^{(\theta; n)}(y).
\end{equation}
Each $\rho_{J}^{(\theta; n)}(y)$ is determined from a perturbative expansion of Eq.~(\ref{eq:rofResc}), which can be written in the form
\begin{equation}\label{eq:rofRescPert}
    \partial_y^2\rho_{J}^{(\theta; n)}(y) - \frac{\rho_{J}^{(\theta; n)}(y)}{2 c(n_{\iota_\theta})}   = F^{(\theta)}_{n}\left(\rho_{J}^{(\theta; 0)},...,\rho_{J}^{(\theta; n-1/2)}\right),
\end{equation}
for some sequence of functions $F^{(\theta)}_{n}$ determined from Eq.~(\ref{eq:rofResc}). The general solution of Eq.~\eqref{eq:rofRescPert} at each order takes the form
\begin{equation}
\label{eq:rhoLexp}
    \rho_{J}^{(\theta; n)} = A^{(\theta)}_n e^{- \frac{y}{\sqrt{2 c(n_{\iota_\theta})}}}+ B^{(\theta)}_n e^{\frac{y}{\sqrt{2 c(n_{\iota_\theta})}}} + C^{(\theta)}_n(y),
\end{equation}
where $C^{(\theta)}_n(y)$ is fixed by the inhomogeneous terms in Eq.~(\ref{eq:rofRescPert}). 
This hierarchy is complemented with boundary conditions obtained by expanding Eq.~\eqref{eq:boundaryrhoL} order by order at large $\gamma$, which impose that $B^{(\theta)}_n = 0$ for any $n$.
In Fig.~\ref{fig:DensityNS}, we present some examples of numerical solutions, clarifying the bulk/boundary convergence at large $\gamma$.

For the first few orders, one can easily check that  
\begin{subequations}
\begin{align}
 \rho_{\gamma,J}^{(\theta)}(y) &= n_{\iota_\theta} + \frac{n_{1-\iota_\theta}-n_{\iota_\theta}}{\sqrt{\gamma}} y+ \frac{\rho_{J}^{(\theta; 1)}(y) }{\gamma} + \mathcal{O}\left(\gamma^{-3/2} \right), \\ 
   \rho_{J}^{(\theta; 1)}(y) &= - c'(n_{\iota_\theta}) \left(J^2 - \left(n_0-n_1\right)^2\right) \left(1-e^{-\frac{y}{\sqrt{2 c(n_{\iota_\theta})}}}\right).
\end{align} 
\end{subequations}

\subsubsection{Large-$\gamma$ expansion of the rate function}
Having established an iterative procedure to determine the asymptotic expansion of $\rho_{\gamma,J}$, we are now ready to describe how it can be used to compute the rate function perturbatively. We start by noting that $\phi_{\gamma}(J)$ can be rewritten as 
\begin{equation}
\label{eq:phisplit}
    \phi_{\gamma}(J) = \int_0^1 dx\; \mathcal{L}_{\gamma,J}\left[\rho_{\gamma,J}^{(\mathcal{B})}(x),x\right] + \int_0^1 dx\; \left(\mathcal{L}_{\gamma,J}\left[\rho_{\gamma,J}(x),x\right]-\mathcal{L}_{\gamma,J}\left[\rho_{\gamma,J}^{(\mathcal{B})}(x),x\right]\right).
\end{equation}
Consider a small $\epsilon > 0$. We can bound the error made in the bulk by replacing $\rho_{\gamma, J}(x)$ with $\rho_{\gamma,J}^{(\mathcal{B})}(x)$ precisely by the exponential decay of the boundary term in Eq.~\eqref{eq:rhoLexp}, so that accounting for the left/right boundaries we have
\begin{equation}
    \max_{x \in [\epsilon, 1 - \epsilon]} \left|\mathcal{L}_{\gamma,J}\left[\rho_{\gamma,J}(x),x\right]-\mathcal{L}_{\gamma,J}[\rho_{\gamma,J}^{(\mathcal{B})}(x),x]\right| = \max[\mathcal{O}(e^{- \sqrt{\gamma/(2\,c(n_0))}\,\epsilon}), \mathcal{O}(e^{- \sqrt{\gamma/(2\,c(n_1))} \epsilon})]
\end{equation}
Thus, we can see that choosing $\epsilon = \mathcal{O}(\gamma^{-\nu})$, with $0<\nu<1/2$, the inner part of the second integral in Eq.~\eqref{eq:phisplit} goes to zero faster than any power. We are thus left with 
\begin{multline}\label{eq:phiexp}
        \phi_{\gamma}(J) \simeq \int_0^{1} dx\; \mathcal{L}_{\gamma,J}\left[\rho_{\gamma,J}^{(\mathcal{B})}(x),x\right] + \\ + \sum_{\theta\in\{\text{L},\text{R}\}} \int_0^{\epsilon(\gamma) \gamma^{1/2}} \frac{dy}{\sqrt{\gamma}}\; \left(\mathcal{L}_{\gamma,J}\left[\rho_{\gamma,J}^{(\theta)}(y), \beta_{\gamma}^{(\theta)}(y)\right]-\mathcal{L}_{\gamma,J}\left[\rho_{\gamma,J}^{(\mathcal{B})}(\beta_{\gamma}^{(\theta)}(y)), \beta_{\gamma}^{(\theta)}(y)\right]\right),
\end{multline}
where the equality holds up to terms decaying faster than any power of $1/\gamma$, and $\rho_{\gamma,J}^{(\text{L})}$ and $\rho_{\gamma,J}^{(\text{R})}$ are the inner expansions generated by $\beta_{\gamma}^{(\text{L})}$ and $\beta_{\gamma}^{(\text{R})}$, respectively.
One concludes that the integrand of the second term in Eq.~(\ref{eq:phiexp}) decays faster to $0$ than any power of $y^{-1}$ and so the integration can be extended to infinity without affecting the perturbative expansion:
\begin{multline}\label{eq:phiexpF}
    \phi_{\gamma}(J) \simeq \int_0^{1} dx\; \mathcal{L}_{\gamma,J}\left[\rho_{\gamma,J}^{(\mathcal{B})}(x),x\right] + \\ + \sum_{\theta\in\{\text{L},\text{R}\}}  \int_0^{\infty} \frac{dy}{\sqrt{\gamma}}\; \left(\mathcal{L}_{\gamma,J}\left[\rho_{\gamma,J}^{(\theta)}(y), \beta_{\gamma}^{(\theta)}(y)\right]-\mathcal{L}_{\gamma,J}\left[\rho_{\gamma,J}^{(\mathcal{B})}(\beta_{\gamma}^{(\theta)}(y)), \beta_{\gamma}^{(\theta)}(y)\right]\right),
\end{multline}
where once again the equality holds up to terms decaying faster than any power in $1/\gamma$. 
Eq.~(\ref{eq:phiexpF}) concludes this section, as it provides a clear way to compute the asymptotic expansion of the rate function $\phi_{\gamma}(J)$ from knowledge of the previously described inner and outer expansions, $\rho_{\gamma,J}^{(\text{L}/\text{R})}$ and $\rho_{\gamma,J}^{(\mathcal{B})}$, respectively. We stress again that all the results obtained here were heavily dependent on the assumption that $\sigma(\rho) \neq 0$ in the interval $\rho \in [n_1,n_0]$, as it would otherwise imply a singularity in the pointwise expansion (see Eq.~(\ref{eq:rhopert1})) and even a different scaling with $\gamma$ near the boundaries. We shall address this singular case in subsection \ref{sec:singular}.

\subsection{Infinite measurement strength}

\begin{figure}[ht]
    \centering  \includegraphics[width=1\linewidth]{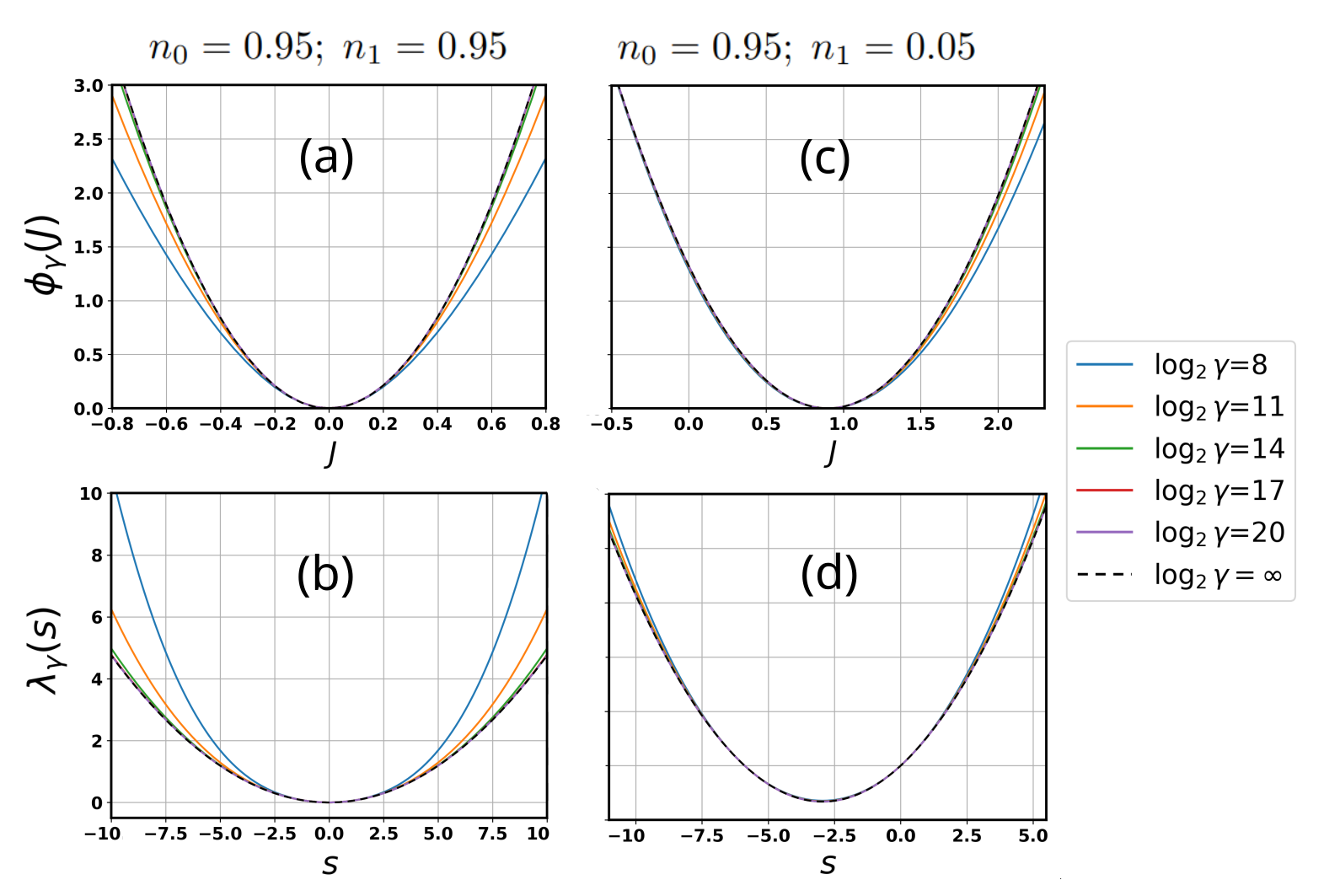}
    \caption{We consider monitored SSEP. Rate function, (a) and (c), and Cumulant Generating Function, (b) and (d), of the current for different values of $\gamma$, showing convergence to predicted $\gamma \to \infty$ functions. Figures (a) and (b) were obtained fixing the boundary densities to $n_0=n_1=0.95$, while $n_0=1-n_1=0.95$ in figures (c) and (d). These figures were obtained with a starting number of grid points $N_x=400$ (it was checked that the results did not change for $N_x=200$ and $N_x=800$).}
    \label{fig:RF_CGF_NS}
\end{figure}

The simplest and most straightforward application of the method described in the previous section is to determine the exact limit $\gamma \to \infty$. In fact, it is evident from Eq.~(\ref{eq:phiexpF}) that the boundary  expansion cannot contribute due to the overall factor of $\gamma^{-1/2}$. From the bulk expansion only $\rho_{J}^{(\mathcal{B}; 0)} = n_0 - (n_0-n_1) x$ contributes, as all higher-order corrections, when inserted in Eq.~(\ref{eq:phiexpF}), vanish identically in the limit $\gamma \to \infty$.
We thus obtain 
\begin{equation}\label{eq:InfGammaRF}
    \lim_{\gamma \to \infty} \phi_{\gamma}(J)= \int_0^1 dx\; \frac{\left(J-\left(n_0-n_1\right)\right)^2}{2 \sigma\left(n_0-\left(n_0-n_1\right) x\right)} = \frac{\left(J-\left(n_0-n_1\right)\right)^2}{2 \kappa_2},
\end{equation}
for $\kappa_2^{-1} = \int_0^1 dx \; \left(\sigma\left(n_0-\left(n_0-n_1\right) x \right)\right)^{-1} = \frac{1}{n_0-n_1}\int_{n_1}^{n_0} \frac{d\rho}{\sigma(\rho)} = \frac{\kappa}{n_0-n_1}$. From the Legendre-transform of this result, we also obtain
\begin{equation}\label{eq:cgfgammainf}
    \lim_{\gamma \to \infty} \lambda_{\gamma}\left(v\right) = \left(n_0-n_1\right) v + \frac{\kappa_2}{2} v^2 =  \left(n_0-n_1\right)\left( v + \frac{v^2}{2\kappa} \right).
\end{equation}
We thus conclude that in the limit $\gamma \to \infty$ the distribution of the current $J$ becomes precisely Gaussian with average $n_0-n_1$ and variance $\kappa_2$. 
For SSEP, for which $\kappa$ takes the explicit form in Eq.~(\ref{eq:kappaforSSEP}),
we complement this analysis with the numerical results shown in Fig.~\ref{fig:RF_CGF_NS}.

The same limiting result can be understood directly from the underlying MFT formula for monitored systems, Eq.~(\ref{eq:UtotalMeasTyp}).
In the strong-measurement limit, the observer has access to the full time evolution of the coarse-grained density $\rho(x,\tau)$. The conservation law $\partial_x j =-\partial_{\tau} \rho$ then fixes the current profile up to a spatially independent function $C(\tau)$, $j(x,\tau) = C(\tau)-\int_0^x dx' \partial_\tau \rho(x',\tau)$.
Substituting this representation into Eq.~(\ref{eq:UtotalMeasTyp}), one finds that $C(\tau)$ has the large-deviation form
\begin{equation}\label{eq:Cdistr}
    \begin{split}
     d\mathbb{P}_{\Lambda_L,\gamma,[-\tau,0]}\left(C\right) \asymp
     \exp\Bigg(&- L \int_{-\tau}^0 d\tau'\; \int_0^1 dx\\
     &\times \frac{\left(C(\tau')-\int^{x} \partial_{\tau'} \rho^{(0)}_{\rm typ}(x,\tau') - \mathcal{J}(\rho^{(0)}_{\rm typ}(x,\tau'))\right)^2}{2\sigma(\rho^{(0)}_{\rm typ}(x,\tau'))}\Bigg),
    \end{split}
\end{equation}
where we remind that $\rho^{(0)}_{\rm typ}$ corresponds to the solution of the hydrodynamic equations.
For a sufficiently long observation time, transient effects can be neglected and $\rho^{(0)}_{\rm typ}$ can be replaced by the steady profile in Eq.~(\ref{eq:hydroSS}). For $D(\rho)=1$, this profile is $\rho_\infty(x) = n_0 + \left(n_1-n_0\right) x$.
Additionally, we have for the CGF (see Eq.~\eqref{eq:BigCurrGenv})
\begin{equation}
 \lambda_{\gamma}(v) = 
\lim_{\tau, L \to \infty} \frac{1}{\tau\,L}\log\mathbb{E}_{\Lambda_L, \tau}^{(\mathbb{Q})}[e^{ L v \int_{-\tau}^0 d\tau'\int_0^1 dx j(x, \tau')}] \stackrel{\gamma \to \infty}{=} \lim_{\tau, L \to \infty}\frac{1}{\tau\,L} \log\mathbb{E}_{C} [e^{ L v \int_{-\tau}^0 d\tau' C(\tau') }],
\end{equation}
where in the second equality we used $\int d\tau' j(x,\tau') = \int d\tau' C(\tau') + O(1)$ and we denoted the expectation with respect to Eq.~\eqref{eq:Cdistr} by $\mathbb{E}_C[\ldots]$.
Thus, we have
\begin{equation}
    \lim_{\gamma \to \infty}\lambda_\gamma(v) = \lim_{\tau, L \to \infty}\frac{1}{L\tau}\log\left(\int D[C]\exp\left(L\int_{-\tau}^0 d\tau'\; \left(v\, C(\tau')-\frac{\left( C(\tau')+n_1-n_0\right)^2}{2\kappa_2}\right)\right)\right),
\end{equation}
where $D[C]$ denotes functional integration over the scalar history $C$. Since the exponent is local in time and contains no coupling between different values of $C(\tau)$, the Gaussian functional integral factorizes and can be performed explicitly. This yields precisely Eq.~(\ref{eq:cgfgammainf}), providing both a check of the strong-measurement expansion and a direct explanation of the limiting Gaussian current statistics.
Before concluding this section, we remark that had we not considered coarse-grained measurements, but rather single-site monitoring, the rate function would have become peaked around $J=n_0-n_1$ since the observer would know exactly the motion of every particle, thus eliminating all statistical fluctuations. It is the fact that, in our setup, the observer can only access the average density on a microscopically large region that prevents them from knowing exactly the net flow of particles across each boundary of such interval, only the total flow out of/into the interval.

\subsection{Scaling of cumulants in the non-singular case}\label{sec:cumulantscaling}
In the previous section, we learned that in the limit $\gamma \to \infty$ the statistics of the current converges exactly to a Gaussian, implying that all cumulants $\kappa_n$ for $n>2$ vanish identically in this limit. In this section, we address how fast these cumulants converge to $0$ with $\gamma$.

We start by noticing that there is a non-trivial double-scaling limit, where one considers $\gamma,J \to \infty$ keeping $\tilde J = J / \sqrt{\gamma}$ finite. Indeed, we set
\begin{equation}\label{eq:RFresum}
F_{0}(\tilde J^2 = J^2/\gamma) := \lim_{\gamma \to \infty}     \frac{\phi_{\gamma}(\tilde J \sqrt{\gamma})  + \kappa \tilde J \sqrt{\gamma}}{\gamma},
\end{equation}
To determine this function, we inspect Eqs.~(\ref{eq:phiJgamma1},\ref{eq:rhof}). Setting $\tilde \rho_{\tilde J} = \lim_{\gamma \to \infty} \rho_{\gamma, J = \sqrt{\gamma} \tilde J}$, we see from Eq.~\eqref{eq:rhof}, that it satisfies
\begin{equation}\label{eq:rhofdoublescallim}
    \left(n_0 + (n_1 - n_0)x - \tilde\rho_{\tilde J} \right)
    - c'(\tilde{\rho}_{\tilde J}) \tilde J^2 
     = 0,
\end{equation}
and using this expression, one can evaluate Eq.~\eqref{eq:RFresum} from Eq.~\eqref{eq:phiJgamma1}, leading to
\begin{equation}\label{eq:F0}
    F_0(\tilde J^2) = \int_0^1 dx \; \left( \frac{\tilde J^2}{2 \sigma(\tilde\rho_{\tilde J})} + \frac{1}{2} \left( n_0 + (n_1 - n_0) x - \tilde\rho_{\tilde J} \right)^2 \right) \;,
\end{equation}
Note that this equation coincides with the double-scaling limit of the bulk contribution discussed in Sec.~\ref{sec:bulkexp}, i.e. only the first integral in Eq.~\eqref{eq:phiexpF}; indeed, the boundary contributions become irrelevant in the double-scaling limit \eqref{eq:RFresum}, because of the overall $\gamma^{-1/2}$ factor in front of the second term in Eq.~\eqref{eq:phiexpF}.

This expansion for the rate function implies that the CGF takes the form
\begin{equation}\label{eq:G0definition}
    G_0(\tilde v) = \lim_{\gamma \to \infty} \gamma^{-1}\lambda_\gamma (v = \tilde v \sqrt{\gamma} - \kappa)= \max_{\widetilde{J}} \left(\widetilde{v} \widetilde{J} - F_0(\widetilde{J}^2) \right),
\end{equation}
with $\widetilde{v} = \left(v+\kappa\right)/\sqrt{\gamma}$.
The invariance under $J \to -J$ of $F_0(\tilde J^2)$ implies $G_0(\widetilde{v})=G_0(-\widetilde{v})$, once again a manifestation of the Gallavotti-Cohen symmetry \eqref{eq:GKU} in the large-$\gamma$ regime. 
By using Eq.~\eqref{eq:G0definition}, one can connect the Taylor expansion of $F_0(\tilde J^2)$ at small $\tilde J$ with the Taylor expansion $G_0(\tilde v)$ at small $\tilde v$,
\begin{equation}\label{eq:GExpansion}
    G_0(\widetilde{v}) = \sum_{k \geq 1} \frac{g_{0,2k}}{\left(2 k\right)!} \widetilde{v}^{2 k}.
\end{equation}
In fact, under a closer look, one observes that $G_0(\widetilde{v})$ does not actually obey one of the defining properties of a cumulant generating function: at $v=0$, $G_0(\kappa/\sqrt{\gamma}) \neq 0$. However, this does not constitute a problem as all non-zero terms in an expansion of $G_0(\kappa/\sqrt{\gamma})$ are subsequently canceled by the higher order contributions that were discarded in the $\gamma\to \infty$ limit of Eq~(\ref{eq:G0definition}).

\begin{figure}[ht]
    \centering
    \includegraphics[width=1\linewidth]{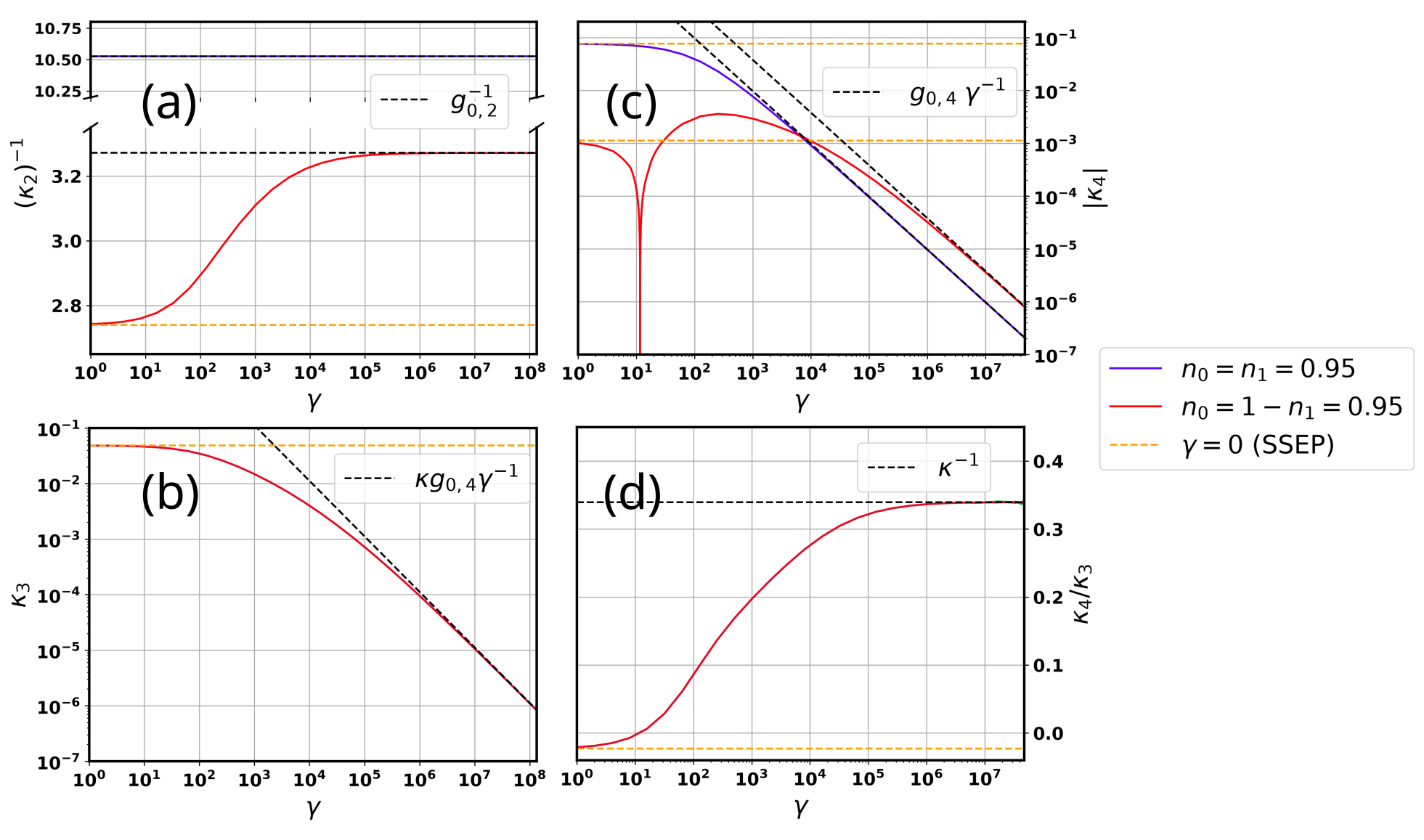}
    \caption{We consider monitored SSEP. Cumulants of the current as a function of $\gamma$ for different boundary conditions. In (a), we plot the (inverse of the) second cumulant $\kappa_2^{-1}$, in (b) the third cumulant $\kappa_3$ for $n_0=1-n_1=0.95$ (when $n_1=n_0$, $\kappa_3=0$ as discussed) and in (c) the absolute value of the fourth cumulant $|\kappa_4|$ (which changes sign from negative to positive as $\gamma$ increases).  Finally, in (d), we plot the ratio $\kappa_4/\kappa_3$, confirming the predicted convergence to $\kappa^{-1}$ as $\gamma \to \infty$. These figures were obtained with a starting number of grid points $N_x=400$ (it was checked that the results did not change for $N_x=200$ and $N_x=800$).} 
    \label{fig:Cumulants_NS}
\end{figure}

\begin{figure}[ht]
    \centering
    \includegraphics[width=1\linewidth]{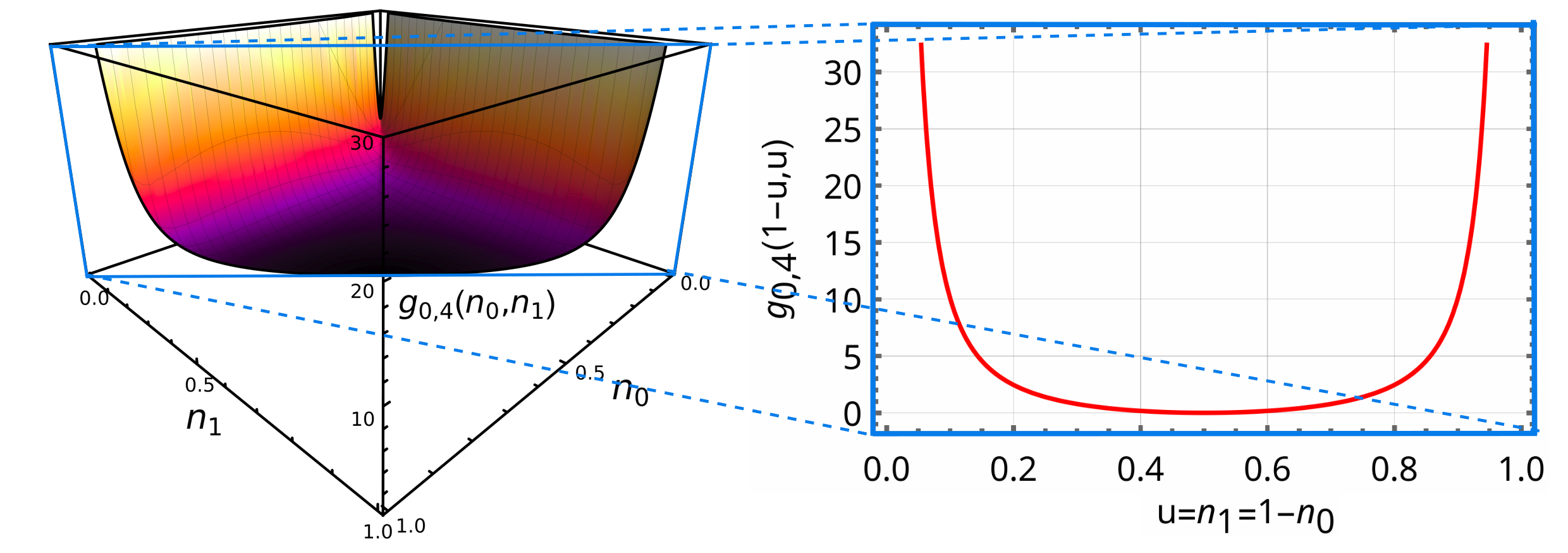}
    \caption{Coefficient $g_{0,4}$ in the Taylor expansion of the double-scaling CGF $G_0(\widetilde v)$ (Eq.~(\ref{eq:GExpansion})) for monitored SSEP, plotted as a function of the boundary densities $n_0$ and $n_1$. Its growth near the edges of the density domain signals the loss of regularity of the expansion when the mobility vanishes at the boundary densities.} 
    \label{fig:g04plot}
\end{figure}

In the absence of fine-tuning, one expects all $g_{0,2k} \neq 0$ for $k \geq 1$; then, $G_0(\widetilde{v})$ entirely provides the leading order behavior of all the cumulants.
Retaining only the leading order contributions from $G_0(\widetilde{v})$ to each power $v^{n}$, one writes
\begin{equation}
    \lambda_\gamma(v) \sim \gamma \sum_{n \geq 1} \frac{g_{0,2n}}{\left(2 n\right)!} \gamma^{-n} \left(v^{2 n} + 2 n \kappa  v^{2 n -1} \right).
\end{equation}
It is clear from this expression that the general large-$\gamma$ behavior of the cumulants is given by $\kappa_n \sim \gamma^{1-\lceil n/2 \rceil}$, as explicitly shown in Fig.~\ref{fig:Cumulants_NS} for the first four cumulants in SSEP.
Explicitly, $\kappa_2=\mathcal O(1)$, $\kappa_3,\kappa_4=\mathcal O(\gamma^{-1})$, $\kappa_5,\kappa_6=\mathcal O(\gamma^{-2})$, and so on: the odd cumulant $\kappa_{2n-1}$ and the following even cumulant $\kappa_{2n}$ decay at the same rate $\gamma^{1-n}$ and  their ratio tends to the finite constant, $\lim_{\gamma \to \infty} \kappa_3/\kappa_4=\kappa$ (see  Fig.~\ref{fig:Cumulants_NS}(d)).
We emphasize the logic of this derivation: the leading large-$\gamma$ behavior of each cumulant --- a property of the CGF at small $v$, or equivalently of the rate function near its minimum --- has been extracted from the double-scaling function $F_0$, which is defined in the large-current regime $J\sim\sqrt{\gamma}$. At first sight, this relates two different limits. The connection is justified by the analyticity of $F_0(\tilde J^2)$ at $\tilde J=0$, which follows from the regularity of the bulk and boundary expansions discussed in Sec.~\ref{sec:asynptmethod}. A complementary way to see the same point is to note that a joint expansion in $J$ and $\gamma^{-1}$ contains only terms of the form $J^{2n}/\gamma^{n+m}$, with $m\geq0$. Therefore, at each fixed power of $J$, the leading large-$\gamma$ contribution is the term with $m=0$; this is precisely the contribution that survives in the double-scaling regime $J\sim\sqrt{\gamma}$. As a consequence, under reasonable convergence assumptions, the leading-in-$\gamma$ part of the small-current expansion is smoothly connected to the double-scaling function. As we shall see in Sec.~\ref{sec:singular}, this connection breaks down precisely when $\sigma(\rho)$ vanishes at the boundary densities, because the required analyticity is then lost. 
This can be illustrated explicitly in SSEP, where $g_{0,2}=\frac{n_0-n_1}{\kappa}$, with $\kappa$ given in Eq.~(\ref{eq:kappaforSSEP}), and $g_{0,4}$ is shown in Fig.~\ref{fig:g04plot}.
As $n_0\to1$ and $n_1\to0$, the coefficient $g_{0,2}$ vanishes while $g_{0,4}$ diverges, indicating that a different scaling analysis is required.
This concludes our analysis of the current cumulants' behavior for large measurement strength on a generic diffusive model.

\subsection{Singular case}\label{sec:singular}
In this section, we address the singular behavior that is expected when $\sigma(\rho)=0$ for some value of the density $\rho \in [n_1,n_0]$, as discussed in subsection \ref{sec:asynptmethod}. For simplicity, we shall address the case where $\sigma(\rho)=0$ vanishes at both boundaries but not in the bulk, i.e. $\sigma(n_0)=0$ and $\sigma(n_1)=0$. We also assume $\sigma'(n_{0,1}) \neq 0$.
Note that choosing $n_0=1$ and $n_1=0$ in the SSEP corresponds precisely to this scenario.

\begin{figure}[ht]
    \centering
    \includegraphics[width=1\linewidth]{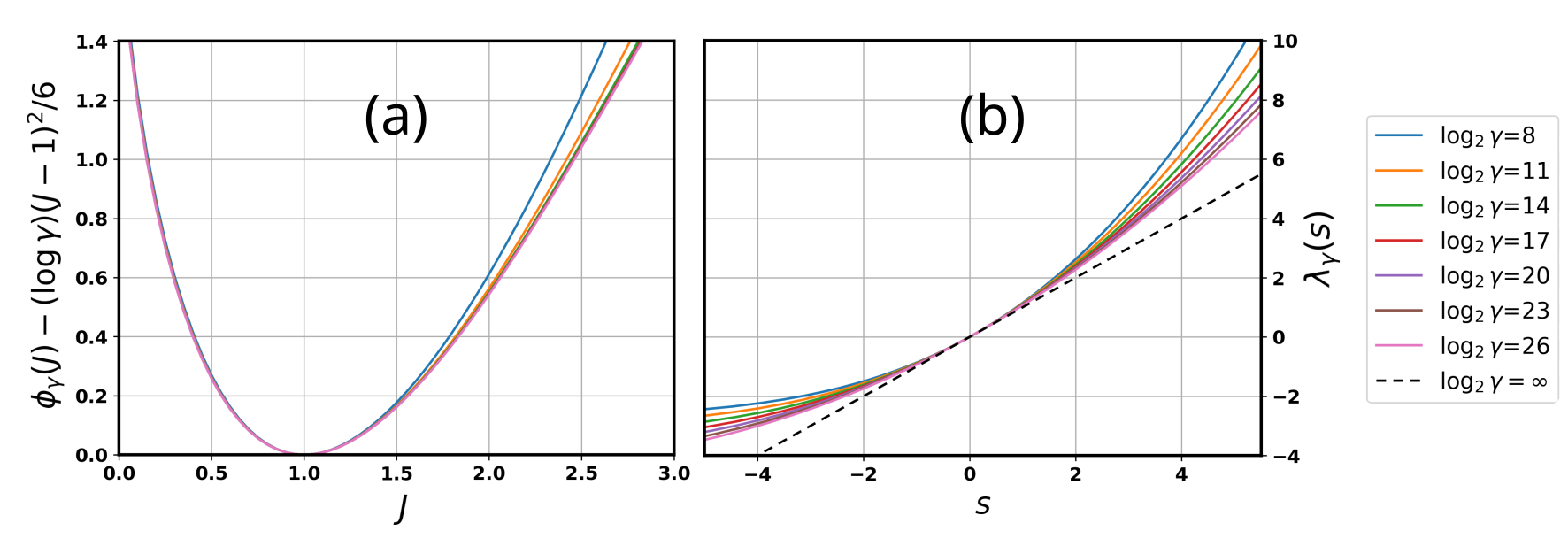}
    \caption{We consider monitored SSEP. We plot the rate function, (a), and Cumulant Generating Function, (b), of the current for different values of $\gamma$ and singular boundary conditions $n_0=1-n_1=1$, showing convergence to predicted $\gamma \to \infty$ functions. These figures were obtained with a starting number of grid points $N_x=400$ (it was checked that the results did not change for $N_x=200$ and $N_x=800$).}
    \label{fig:RF_CGF_S}
\end{figure}

\begin{figure}[ht]
    \centering
    \includegraphics[width=1\linewidth]{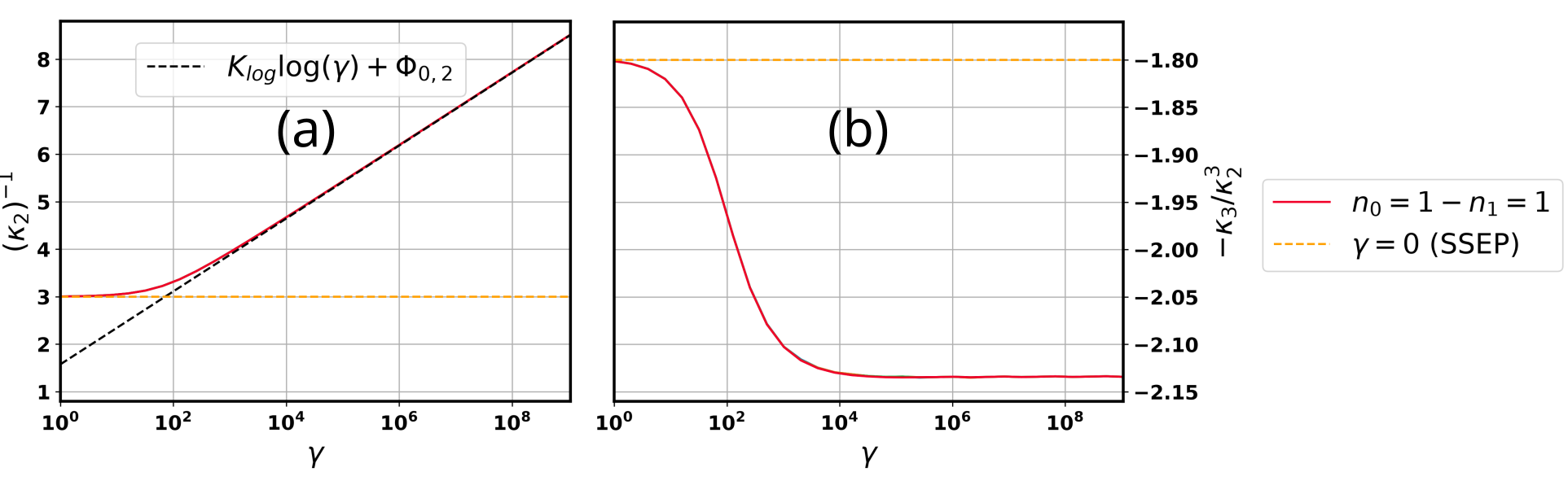}
    \caption{We consider monitored SSEP. Cumulants of the current as a function of $\gamma$ for boundary conditions $n_0=1-n_1=1$. We plot the (inverse of the) second cumulant, $\kappa_2^{-1}$, in (a) and in (b) the third derivative of the rate function at $J=1$, $\left.\partial_J^3\phi_{\gamma}(J)\right|_{J=1} = -\kappa_3/\kappa_2^3$ , which converges to $\Phi_{0,3}$ in the limit $\gamma \to \infty$. These figures were obtained with a starting number of grid points $N_x=400$ (it was checked that the results did not change for $N_x=200$ and $N_x=800$).}
    \label{fig:Cumulants_S}
\end{figure}

It is instructive to first examine what survives of the strategy of Sec.~\ref{sec:cumulantscaling}. The double-scaling limit in Eq.~\eqref{eq:RFresum} ($\kappa$ formally diverges but $\phi_{\gamma}(J) + \kappa J$ is well-defined) remains perfectly well defined, and is still given by the bulk formula, Eq.~\eqref{eq:F0}. However, for $\tilde J \to 0$, $\tilde\rho_{\tilde J}(x = 0,1) = n_0,n_1$ where the function $c'(\rho) = \partial_\rho (2 \sigma(\rho))^{-1}$ diverges. This signals a non-analytic behavior in the neighborhood of $\tilde J = 0$. Indeed, one can see that the optimal profile solving Eq.~\eqref{eq:rhofdoublescallim} \emph{detaches} from the singular boundary values $n_0, n_1$ by an amount $\sim |\tilde{J}|^{2/3}$ (for SSEP, $1-\tilde\rho_{\tilde J}(0) \simeq (\tilde{J}^2/4)^{1/3}$ at $x=0$), so that $\sigma$ never vanishes along the profile and $F_0(\tilde{J}^2)$ is finite and smooth for all $\tilde{J} \neq 0$, but \emph{not analytic at the origin}. To see this explicitly, we note that one can approximate $\tilde\rho_{\tilde J}(x) = \rho_{\infty}(x) $ for $|\tilde J|^{2/3} \ll x \ll 1 - |\tilde J|^{2/3}$ ($\tilde\rho_{\tilde J}(x)$ then approaches $\tilde\rho_{\tilde J}(0)$ on the scale $x \sim |\tilde J|^{2/3}$). Then, from Eq.~\eqref{eq:F0}, we extract the logarithmic divergence of $\int_{|\tilde J|^{2/3}}^{1-|\tilde{J}|^{2/3}} dx/\sigma(\rho_\infty(x))$, which produces the small-$\tilde{J}$ behavior
\begin{equation}\label{eq:F0singlog}
    F_0(\tilde{J}^2) = -K_{\rm log}\, \tilde{J}^{2} \log|\tilde{J}| + \mathcal{O}(\tilde{J}^{2}), \qquad K_{\rm log}=\frac{1}{3 \left(n_1-n_0\right)}\left(\frac{1}{\sigma'(n_0)}-\frac{1}{\sigma'(n_1)} \right).
\end{equation}
As a consequence, the two operations that commuted in the non-singular case no longer commute here: expanding at small $J$ and then taking the scaling limit does not give the same result as taking the double-scaling limit first and then expanding at small $\tilde J$. This failure stems from the non-analyticity of $F_0(\tilde J^2)$ at the origin. Although Eq.~\eqref{eq:F0singlog} does not fully determine the cumulants, one can still use it to extract their scaling with $\gamma$. Indeed, Eq.~\eqref{eq:F0singlog} dictates the leading large-$\gamma$ behavior of the rate function at fixed $J$,
\begin{equation}\label{eq:RFleadinglog}
    \phi_{\gamma}(J) = \frac{K_{\rm log}}{2} \left(J+n_1-n_0\right)^2 \log \gamma + \Phi_0(J) + o(1).
\end{equation}
To see this, we note that the coefficient of $\log\gamma$ must be at most quadratic in $J$, as higher powers of $J$ would be incompatible with the finite double-scaling limit ($J\sim\sqrt{\gamma}$) in Eq.~\eqref{eq:RFresum}. Additionally, the linear and constant terms follow from requiring that the rate function vanishes, together with its derivative, at the average current $J=n_0-n_1$ for every $\gamma$.
The next subleading term is independent of $\gamma$ and is represented by $\Phi_0(J)$ in Eq.~\eqref{eq:RFleadinglog}. It cannot, however, be obtained from the double-scaling treatment because the two limits do not commute. Determining $\Phi_0$ requires properly taking the large-$\gamma$ limit at fixed $J$, for which the boundary layers must be resolved explicitly. We carry out this calculation in Appendix~\ref{app:singularlog}; here we retain Eq.~\eqref{eq:RFleadinglog} and use it to derive the current statistics. In particular, Eq.~\eqref{eq:RFleadinglog} shows that the curvature at the minimum $J=n_0-n_1$ diverges logarithmically with $\gamma$, whereas all higher derivatives approach the corresponding finite derivatives of $\Phi_0$.
Substituting the SSEP mobility in Eq.~\eqref{eq:SSEPtranspcoeff} into Eq.~\eqref{eq:F0singlog} gives $K_{\rm log}=1/3$. We therefore plot the non-$\gamma$-divergent part of the SSEP rate function in Fig.~\ref{fig:RF_CGF_S}, showing convergence to a non-trivial function.
In parallel with section \ref{sec:cumulantscaling}, we now seek to describe the behavior of the current cumulants for large $\gamma$, introducing the scaled variable $t = K_{\rm log} \left(J + n_1 - n_0\right)\,\log(\gamma)$ that allows us to write
\begin{equation}
    \lambda_\gamma(v) = \left(n_0-n_1\right)v + \max_{t}\left(\frac{v\,t- \frac{1}{2} t^2}{K_{\rm log} \log\left(\gamma\right)} - \Phi_0\left(n_0-n_1+\frac{t}{K_{\rm log} \log\left(\gamma\right)}\right)+o(1)\right).
\end{equation}
The Legendre transform is obtained by first determining the implicit relation between $t$ and $v$,
\begin{equation}\label{eq:t(v)rel}
    v = t+\Phi_0'\left(n_0-n_1+\frac{t}{K_{\rm log} \log\left(\gamma\right)}\right)\Longrightarrow t(v) =v+\mathcal{O}\left(\left(\log\gamma\right)^{-1}\right).
\end{equation}
The implication in Eq.~(\ref{eq:t(v)rel}) follows because $J=n_0-n_1$ (i.e. $t=0$) is a stationary point of $\Phi_0$, so $\Phi_0'(n_0-n_1)=0$ and therefore $\Phi_0'\left(n_0-n_1+\frac{t}{K_{\rm log}\log(\gamma)}\right)=\mathcal{O}\left((\log\gamma)^{-1}\right)$. Thus, if one keeps only the leading contribution in $\gamma$ at each power $v^n$, the subleading corrections to $t(v)$ in powers of $(\log\gamma)^{-1}$ can be dropped, obtaining
\begin{equation}
    \lambda_\gamma(v) \sim \left(n_0-n_1\right)v + \frac{v^2}{2K_{\rm log}\log(\gamma)} - \Phi_0\left(n_0-n_1+\frac{v}{K_{\rm log}\log\left(\gamma\right)}\right).
\end{equation}
We emphasize that this expression for $\lambda_\gamma(v)$ should be understood in precisely this leading-order sense. The large-$\gamma$ behavior of the cumulants then follows from the Taylor expansion of $\Phi_0$ around $J=n_0-n_1$, $\Phi_0(n_0-n_1+u)=\sum_{k=2}^{\infty} \frac{\Phi_{0,k}}{k!} u^k$. Since the coefficients $\Phi_{0,k}$ are generically non-zero, the leading cumulant scalings are
\begin{equation}
    \kappa_1=n_0-n_1,\,\kappa_2 \sim \left(K_{\rm log}\log\gamma\right)^{-1}\left(1-\frac{\Phi_{0,2}}{K_{\rm log}\log\gamma} \right),\, \kappa_n \sim - \frac{\Phi_{0,n}}{\left(K_{\rm log}\log\gamma\right)^{n} }   \text{ for } n \geq 3.
\end{equation}
Thus, singular boundary conditions replace the algebraic suppression in $\gamma$ found in Sec.~\ref{sec:cumulantscaling} by a logarithmic one: the variance decays as $(\log\gamma)^{-1}$, while the higher cumulants decay as successive inverse powers of $\log\gamma$.
We test these predictions for monitored SSEP in Fig.~\ref{fig:Cumulants_S}, finding excellent agreement for the second and third current cumulants. The analytical prediction in Fig.~\ref{fig:Cumulants_S}(a) uses $\Phi_{0,2}$, computed from the expression for $\Phi_0$ derived in Appendix~\ref{app:singularlog}.

Overall, in both the regular and singular cases the distribution of the integrated current becomes Gaussian at large measurement strength. The distinction is that the limiting variance remains finite in the regular case, whereas it vanishes in the singular case. Correspondingly, the higher cumulants also decay in qualitatively different ways: in the regular case analyzed in Sec.~\ref{sec:cumulantscaling} they vanish as powers of $\gamma$, while in the singular case their decay is only logarithmic. This contrast can be traced to the behavior of the saddle-point profile near the boundaries, which in the regular case gives only a subleading contribution to the rate function, whereas in the singular case it controls the leading behavior through the boundary layers.

\section{Conclusion}

We have developed a macroscopic fluctuation theory for diffusive classical lattice gases subject to continuous, coarse-grained monitoring.  Starting from a microscopic Bayesian description of weak measurements, we showed that the hydrodynamic limit concentrates the measurement record around its typical value.  The observer's conditioned description to this typical measurement record is then governed by the usual MFT action supplemented by a positive quadratic term that confines the density to the stationary hydrodynamic profile---this corresponds to the quenched, rather than annealed, statistics.  This formulation captures how monitoring changes the statistics inferred by an observer without altering the underlying classical dynamics.  Coarse graining is essential here: even at arbitrarily strong monitoring, it leaves a residual uncertainty in the current that would be absent under site-resolved observation.

We used this action to study current large deviations.  Extending the convexity construction of Ref.~\cite{Bertini_2006}, we found an additional sufficient condition under which monitoring preserves convexity, excludes dynamical phase transitions, and guarantees the additivity principle. This also identifies a natural setting in which a genuinely measurement-induced transition could occur: a model that satisfies the unmonitored convexity conditions but violates the new one, such as the zero-range process~\cite{SPITZER1970246,EvansHanney2005}.

For boundary-driven one-dimensional systems with particle-hole symmetry, we analyzed how monitoring modifies pre-existing dynamical phase transitions.  Measurements raise the critical driving required for a continuous transition, thereby suppressing the transition in models such as the SSEP.  They can nevertheless have a qualitative effect on its character.  We derived criteria for a change in the order of the transition and exhibited this scenario in the KLS model, where increasing the measurement strength changes a second-order transition into a first-order one.  These conclusions rely on the weak additivity assumption $\phi_\gamma(J)=U_\gamma^{**}(J)$.  A failure of this assumption would instead reveal a time-dependent optimal history and hence a breaking of time-translation invariance, an interesting possibility that remains to be investigated directly.

On a different note, we also obtained a detailed strong-measurement expansion for one-dimensional open systems satisfying the additivity principle.  In the regular case, where the mobility does not vanish at any point in the system, the current rate function approaches an exactly quadratic form: the current becomes Gaussian, with mean $n_0-n_1$ and a variance fixed by the harmonic average of the mobility along the stationary profile.  The approach to this limit is controlled by bulk and boundary-layer expansions, and the higher cumulants obey $\kappa_n\sim\gamma^{1-\lceil n/2\rceil}$.  When the mobility vanishes at the boundary densities, this expansion becomes singular.  Boundary layers then have a different scaling and produce logarithmic rather than algebraic suppression: the variance vanishes as $1/\log\gamma$, while higher cumulants decay as inverse powers of $\log\gamma$.  Thus, strong monitoring still yields Gaussian current statistics, but with a sharply distinct route to the limit in the singular case.

Natural next steps are to determine the monitored large-deviation functional of the steady-state density, to treat multispecies and reaction-diffusion systems, and to extend the measurement construction to systems with ballistic transport.  These directions should clarify how broadly the interplay between partial observation, transport fluctuations, and dynamical phase transitions extends beyond diffusive single-species gases.

\section*{Acknowledgements}
We thank the Laboratoire de Physique Théorique et Modèles Statistiques (LPTMS) for hosting J.C. during most of this work. A.D.L. also thanks Adam Nahum, Pierre Le Doussal and Tridib Sadhu for stimulating discussions on this topic.
J.C. acknowledges support from Fundação para a Ciência e Tecnologia (FCT) through grant No. 2022.11940.BD. 
J.C. and P.R. acknowledge support by FCT through Grant No. UID/04540/2025 (DOI: 10.54499/UID/04540/2025) to the I\&D unit Centro de Física e Engenharia de Materiais Avançados (CeFEMA). P.R. further acknowledges support by FCT through project SCALE-QLT (DOI: 10.54499/2024.16192.PEX). 
P.R. acknowledges further support by the CY Initiative of Excellence (grant “Investissements d’Avenir” ANR-16-IDEX-0008) and part of it was developed during his stay at CY Advanced Studies.
This work has also benefited from discussions and networking activities within COST Action CA24109, Many-body Open Quantum Systems (QOpen), supported by COST (European Cooperation in Science and Technology).

\begin{appendix}

\section{Method to compute the rate function}\label{sec:MethodForRF}

In this section, we present an efficient method for computing the current rate function $\phi_{\gamma}(J)$ and, via a Legendre transform, the cumulant generating function $\lambda_{\gamma}(v)$. Our approach builds on the frameworks developed in~\cite{DerridaAdditivityPrinciple, OhadShpielbergNumerical}. Throughout, we assume the validity of the Additivity Principle (AP), under which $\phi_{\gamma}(J) = U_{\gamma}(J)$ (see the discussion preceding Eq.~(\ref{eq:APdef})). Accordingly, for each fixed $J$, we seek the density profile that globally minimizes the action $\mathcal U_\gamma(\rho,J)$ subject to the boundary conditions. Any smooth minimizer must, in particular, be a saddle-point solution of the action; we therefore first characterize all admissible saddle-point profiles and then compare their action values.

Although calculating these saddle-point profiles and subsequently computing the rate function is generally challenging, significant simplifications arise when the action $\mathcal{U}_{\gamma}(\rho, J)$ is invariant under spatial translations (or, more generally, exhibits continuous symmetries). In the monitored setting, such symmetry occurs only at equilibrium, defined by $n_* = n_0 = n_1$. Consequently, in this appendix, we restrict our analysis to the equilibrium case—possibly with an external field—and closely follow the methods of~\cite{DerridaAdditivityPrinciple, OhadShpielbergNumerical} to derive our main results. We stress that these results are valid for any system satisfying these conditions and are not restricted to the particle-hole-symmetric case analyzed in the main text.

In this setting, the Lagrangian in Eq.~(\ref{eq:LJgamma0}) can be viewed as that of a one-dimensional mechanical problem, with $\rho$ as the coordinate and $x$ as time. Since it has no explicit $x$ dependence, spatial translations imply a conserved quantity $K$. We denote a saddle-point trajectory with conserved value $K$ by $\rho_{\gamma,J,K}(x)$.
Applying Noether's theorem to the Lagrangian described in Eq.~(\ref{eq:LJgamma0}) gives
\begin{equation}\label{eq:Kconservation}
    K=-\frac{J^2+\left(\sigma(\rho_{\gamma,J,K})E\right)^2}{2 \sigma(\rho_{\gamma,J,K})}+\frac{\left(D(\rho_{\gamma,J,K})\partial_x \rho_{\gamma,J,K}\right)^2}{2 \sigma(\rho_{\gamma,J,K})} - \frac{\gamma}{2}\left(\rho_{\gamma,J,K}-n_*\right)^2.
\end{equation} 
Rewriting this equation for \(\partial_x \rho_{\gamma,J,K}\), we arrive at a first-order differential equation for \(\rho_{\gamma,J,K}\),
\begin{equation}\label{eq:1ODE}
    \partial_x \rho_{\gamma,J,K} = \pm \frac{1}{D(\rho_{\gamma,J,K})}  \sqrt{ J^2 +\left(\sigma(\rho_{\gamma,J,K})E\right)^2 + \sigma(\rho_{\gamma,J,K}) \left( 2 K + \gamma \left(\rho_{\gamma,J,K}-n_*\right)^2\right)},
\end{equation}
where the overall sign must be chosen appropriately. To account for possible changes in this sign, we consider the set of points $\{x_k\}_{1\leq k < n}$ at which the derivative $\partial_x \rho_{\gamma,J,K}$ changes sign. These points partition the unit interval $[0,1]$ into $n$ contiguous subintervals, denoted by $\mathcal{A}_k = [x_{k-1}, x_k]$, where we set $x_0 = 0$ and $x_n = 1$. We define the density at each of these extremal points as $\rho_k \coloneqq \rho_{\gamma,J,K}(x_k)$ ($\rho_0=\rho_n=n_*$). Since Eq.~(\ref{eq:1ODE}) uniquely determines the saddle-point solution $\rho_{\gamma,J,K}$ given the set of extremal values $\{\rho_k\}_k$, we will henceforth identify $\rho_{\gamma,J,K}$ by its corresponding set of extrema $\{\rho_k\}_k$. 

Defining the potential $V_{\gamma,J}(\rho) =-\frac{J^2+\left(\sigma(\rho)E\right)^2}{2\sigma(\rho)}-\frac{\gamma}{2}\left(\rho-n_* \right)^2$, we note that, at the points $x_k$ (for $1\leq k <n$), $\partial_x \rho_{\gamma,J,K}=0$ and thus the potential $V_{\gamma,J}(\rho)$ must assume the same value: $K=V_{\gamma,J}(\rho_1)=\cdots=V_{\gamma,J}(\rho_{n-1})$ (for $n=1$, $K=V_{\gamma,J}(n_*)$). We shall denote this restriction on the set of possible extrema of $\rho_{\gamma,J,K}$ as $\mathbf{C1}$. With these considerations, Eq.~(\ref{eq:1ODE}) can be simplified to
\begin{equation}\label{eq:1ODEbest}
    \partial_x \rho_{\gamma,J,K} = \pm \frac{\sqrt{\sigma(\rho_{\gamma,J,K})}}{D(\rho_{\gamma,J,K})}  \sqrt{2\left(K-V_{\gamma,J}(\rho_{\gamma,J,K})\right)}.
\end{equation}
On each interval $\mathcal{A}_k$, the saddle-point solution $\rho_{\gamma,J,K}$ is monotonic by definition, which implies that any integral of the form $\int_{\mathcal{A}_k} dx\; F[\rho_{\gamma,J,K}]$ can be simplified to 
\begin{equation}\label{eq:xtorhointegrals}
    \int_{\mathcal{A}_k} dx\; F[\rho_{\gamma,J,K}] = \int_{\mathcal{A}_k} dx\; \partial_x \rho_{\gamma,J,K}\frac{F[\rho_{\gamma,J,K}]}{\partial_x \rho_{\gamma,J,K}} = \int_{\rho_{k-1}}^{\rho_k} |d\rho|\;\frac{D(\rho)}{\sqrt{\sigma(\rho)}} \frac{F[\rho]}{\sqrt{2\left(K-V_{\gamma,J}(\rho)\right)}} ,
\end{equation}
where the symbol $|d\rho|$ means that $\int_a^b |d\rho| = \int_b^a |d\rho|$. Applying this to $F[\rho]=1$, one obtains the condition $\mathbf{C2}$:
\begin{equation}\label{eq:cond2_GH}
    \mathbf{C2}: \;\sum_{k=1}^n T_{\gamma,J}^{(\rho_{k-1},\rho_k)}[K] = 1, \; \text{with}\; \; T_{\gamma,J}^{(\rho_{k-1},\rho_k)}[V]\coloneqq \int_{\rho_{k-1}}^{\rho_k} |d\rho|\;\frac{D(\rho)}{\sqrt{\sigma(\rho)}} \frac{1}{\sqrt{2\left(V-V_{\gamma,J}(\rho)\right)}}.
\end{equation}
In essence, condition $\mathbf{C2}$ expresses both the boundary conditions $\rho_{\gamma,J,K}(0)=\rho_{\gamma,J,K}(1)=n_*$ and the requirement that the saddle-point solution traverse an interval of total length one.
Additionally, choosing $F[\rho]=\mathcal{L}_{\gamma,J}\left[\rho\right]$ in Eq.~(\ref{eq:xtorhointegrals}) and using that $\mu=\kappa=0$ at equilibrium, Eq.~(\ref{eq:phiJgamma}) reduces to
\begin{equation}\label{eq:UwithH}
    U_{\gamma}(J)=-K+\sum_{k=1}^n H_{\gamma,J}^{(\rho_{k-1},\rho_k)}[K] -JE, \text{with}\; H_{\gamma,J}^{(a,b)}[V]\coloneqq \int_{a}^{b} |d\rho|\;\frac{D(\rho)}{\sqrt{\sigma(\rho)}} \sqrt{2\left(V-V_{\gamma,J}(\rho)\right)}.
\end{equation}
Note that, for the function \( T_{\gamma,J}^{(\rho_{k-1}, \rho_k)}[K] \) to be well-defined, each \(\rho_k\) must lie in one of the following two domains:
\begin{align*}
    \mathcal{D}_+ = \left\{ \rho \geq n_* \mid \forall \varrho \in [n_*, \rho),\ V_{\gamma,J}(\varrho) < V_{\gamma,J}(\rho) \right\}, \quad\\
    \mathcal{D}_- = \left\{ \rho \leq n_* \mid \forall \varrho \in (\rho,n_*],\ V_{\gamma,J}(\varrho) < V_{\gamma,J}(\rho) \right\}.
\end{align*}
Accordingly, we define the two branches of \( V_{\gamma,J} \) by $V_{\gamma,J,s}(\rho) = V_{\gamma,J}(\rho)$, for $\rho \in \mathcal{D}_s,\ s \in \{-,+\}$. Since each $V_{\gamma,J,s}(\rho)$ is injective, condition $\mathbf{C1}$ implies that, for a given $K$, there is at most one possible extremum on each branch. When they exist, we define $\rho_{\gamma,J,-}^*(K)=V_{\gamma,J,-}^{-1}(K)$
and $\rho_{\gamma,J,+}^*(K)=V_{\gamma,J,+}^{-1}(K)$, satisfying $\rho_{\gamma,J,-}^*(K)<n_*<\rho_{\gamma,J,+}^*(K)$.

Between two consecutive extrema, $\rho_{\gamma,J,K}$ must cross $n_*$. If we partition the unit interval at these points (where \( \rho_{\gamma,J,K} = n_* \)), we obtain two types of subintervals, denoted by \( \mathcal{B}_+ \) and \( \mathcal{B}_- \), depending on whether \( \rho_{\gamma,J,K} \) reaches the value \( \rho_{\gamma,J,+}^*(K) \) or \( \rho_{\gamma,J,-}^*(K) \), respectively. The value of $U_\gamma(J)$ depends only on the multiset of excursions $\{\mathcal{B}_\pm\}$, not on their order along $[0,1]$, since the action is a sum of identical contributions over the subintervals.

Hence two excursions of the same type can be brought adjacent without changing $U_\gamma(J)$; but two consecutive $\mathcal{B}_+$ (or $\mathcal{B}_-$) excursions meet at $\rho=n_*$ with non-vanishing, opposite-sign derivative, producing a corner where $\partial_x\rho$ is discontinuous. Such a profile is not a smooth solution of the Euler--Lagrange equation and therefore cannot be the global minimizer of $\mathcal U_\gamma(\rho,J)$. It follows that a minimizing profile can contain at most one excursion of each type. It remains to rule out a profile containing precisely one $\mathcal{B}_+$ excursion and one $\mathcal{B}_-$ excursion.

Let $\ell_\pm$ be the lengths of the two excursions (with $\ell_+ + \ell_-=1$), and denote the contribution of $\mathcal{B}_s$ to the action by
\begin{align*}
    \mathcal U_{\gamma,J}^{(s)}[\rho_{\gamma,J,K}]
    &\coloneqq \mathcal U_{\gamma,J}^{(s,0)}[\rho_{\gamma,J,K}]
    +\mathcal U_{\gamma,J}^{(s,\nabla)}[\rho_{\gamma,J,K}],\;
    \mathcal U_{\gamma,J}^{(s,\nabla)}[\rho]
    \coloneqq\int_{\mathcal{B}_s}dx\;\frac{\left(D(\rho)\partial_x\rho\right)^2}{2\sigma(\rho)}
    \geq 0,\\
    \mathcal U_{\gamma,J}^{(s,0)}[\rho]&
    \coloneqq\int_{\mathcal{B}_s}dx\;\left(\frac{J^2+\left(\sigma(\rho)\,E\right)^2}{2\sigma(\rho)} +\frac{\gamma}{2}\left(\rho-n_*\right)^2\right) -J E \ell_s
    \geq 0.
\end{align*}
We now stretch this excursion to the full interval by setting $\widetilde\rho_{\gamma,J,K}^{(s)}(x)\coloneqq\rho_{\gamma,J,K}(\inf\mathcal{B}_s+\ell_s x)$. The non-gradient and gradient contributions then scale as $\ell_s^{-1}$ and $\ell_s$, respectively, so that
\begin{equation}\label{eq:stretchedprofileinequality}
    \mathcal U_\gamma\left(\widetilde\rho_{\gamma,J,K}^{(s)},J\right)
    =\frac{\mathcal U_{\gamma,J}^{(s,0)}[\rho_{\gamma,J,K}]}{\ell_s}
    +\ell_s\mathcal U_{\gamma,J}^{(s,\nabla)}[\rho_{\gamma,J,K}]
    <\frac{\mathcal U_{\gamma,J}^{(s)}[\rho_{\gamma,J,K}]}{\ell_s}.
\end{equation}
On the other hand, $\mathcal U_\gamma(\rho_{\gamma,J,K},J)=\sum_{s\in\{-,+\}}\ell_s \mathcal U_{\gamma,J}^{(s)}[\rho_{\gamma,J,K}]/\ell_s$,
and therefore at least one excursion satisfies $\mathcal U_{\gamma,J}^{(s)}[\rho_{\gamma,J,K}]/\ell_s \leq \mathcal U_\gamma(\rho_{\gamma,J,K},J)$.
Together with the strict inequality in Eq.~(\ref{eq:stretchedprofileinequality}), this shows that stretching this excursion produces an admissible single-excursion profile with strictly smaller action,
\begin{equation*}
    \mathcal U_\gamma\left(\widetilde\rho_{\gamma,J,K}^{(s)},J\right)
    <\mathcal U_\gamma(\rho_{\gamma,J,K},J).
\end{equation*}
Hence a profile containing one excursion of each type cannot be the global minimizer.

These considerations bring a significant simplification to Eq.~(\ref{eq:cond2_GH}), which we can now rewrite as
\begin{equation}\label{eq:cond2v_GH}
    \mathbf{C2}: \; T_{\gamma,J}^{(n_*,\rho_{\gamma,J,s}^*(K))}[K] = \frac{1}{2}, \qquad s\in\{-,+\}.
\end{equation}
Accordingly, the expression for the rate function becomes
\begin{equation}\label{eq:RFmindens}
    U_{\gamma}(J) = \min_{s\in\{-,+\}}\min_{K:\mathbf{C2}} Q_{\gamma,J}^{(n_*,\rho_{\gamma,J,s}^*(K))}(K)-JE,\;
    Q_{\gamma,J}^{(a,b)}(K)\coloneqq -K+2H_{\gamma,J}^{(a,b)}[K].
\end{equation}
We now show that condition $\mathbf{C2}$ need not be imposed explicitly in the minimization of $Q_{\gamma,J}^{(n_*,\rho_{\gamma,J,s}^*(K))}(K)$ in Eq.~(\ref{eq:RFmindens}).
To this end, we use the identity
\begin{equation}\label{eq:GderH}
    \partial_K H_{\gamma,J}^{(n_*,\rho_{\gamma,J,s}^*(K))}[K] = T_{\gamma,J}^{(n_*,\rho_{\gamma,J,s}^*(K))}[K],
\end{equation}
which implies that condition $\mathbf{C2}$ is automatically satisfied at the extrema of $Q_{\gamma,J}^{(n_*,\rho_{\gamma,J,s}^*(K))}(K)$ that lie in smooth regions. This function is, however, only piecewise smooth, since $\rho_{\gamma,J,s}^*(K)$ can jump between disconnected components of $\mathcal D_s$. It thus remains to examine its discontinuities and the boundaries of its domain.

The lower boundary is $K=V_*\coloneqq V_{\gamma,J}(n_*)$. If $V_{\gamma,J}'(n_*)=0$, it corresponds to the constant saddle-point profile $\rho_{\gamma,J,K}=n_*$, for which condition $\mathbf{C2}$ is automatic. If $V_{\gamma,J}'(n_*)\neq0$, one branch leaves $n_*$ continuously, whereas the other, when it exists, is reached after an outward jump and therefore has a larger value of $Q_{\gamma,J}^{(n_*,\rho_{\gamma,J,s}^*(K))}(K)$. On the continuous branch, $\rho_{\gamma,J,s}^*(K)\to n_*$ and hence $T_{\gamma,J}^{(n_*,\rho_{\gamma,J,s}^*(K))}[K]\to0$ as $K\to V_*^+$. It follows from Eq.~(\ref{eq:GderH}) that the derivative of the function being minimized tends to $-1$, so $K=V_*$ cannot be a minimum.

At a discontinuity $K_d$, since the distance $|\rho_{\gamma,J,s}^*(K)-n_*|$ always increases with $K$, the turning point jumps outwards from $\rho_{\rm in}$ to $\rho_{\rm out}$ and the function jumps upwards by $2H_{\gamma,J}^{(\rho_{\rm in},\rho_{\rm out})}[K_d]>0$. The point $\rho_{\rm in}$ must be a local maximum of $V_{\gamma,J}$, so $V_{\gamma,J}'(\rho_{\rm in})=0$; expanding the potential around this point therefore gives $\lim_{K\to K_d^-}T_{\gamma,J}^{(n_*,\rho_{\gamma,J,s}^*(K))}[K]=+\infty$. It thus follows that $Q_{\gamma,J}^{(n_*,\rho_{\gamma,J,s}^*(K))}(K)$ increases immediately before the discontinuity and then jumps upwards. A discontinuity therefore cannot contain the global minimum.

It remains to consider the terminal value $V_{\gamma,J,s}^{\rm sup}\coloneqq\sup_{\rho\in\mathcal D_s}V_{\gamma,J}(\rho)$.
Since physical densities are non-negative, $\mathcal D_-$ is bounded and, by continuity, the turning point $\rho_{\gamma,J,-}^*(K)$ converges to the finite terminal point $\min\mathcal D_-$ as $K\to V_{\gamma,J,-}^{\rm sup}$. Similarly, whenever $\rho_{\gamma,J,+}^*(K)$ remains bounded, it converges to $\max\mathcal D_+$ as $K\to V_{\gamma,J,+}^{\rm sup}$. By the definition of $\mathcal D_s$, the potential at each terminal point is a local maximum of $V_{\gamma,J}$ and thus satisfies $V_{\gamma,J}'=0$. Expanding the potential around it then gives $\lim_{K\to(V_{\gamma,J,s}^{\rm sup})^-}T_{\gamma,J}^{(n_*,\rho_{\gamma,J,s}^*(K))}[K]=+\infty$\footnote{One might worry that a terminal point may coincide with a boundary of the physical density domain (such as $\rho=0$), in which case being a local maximum of $V_{\gamma,J}$ need not imply $V_{\gamma,J}'=0$. For physical particle systems, however, $\sigma(\rho)\to0$ as $\rho\to0$ because no particles remain to carry current fluctuations. Thus, $V_{\gamma,J}(\rho)\to-\infty$ in this limit, so $\rho=0\notin\mathcal D_-$ and cannot be a terminal point. Likewise, if the density has a finite upper bound (as $\rho=1$ in exclusion processes), $\sigma(\rho)$ vanishes there, and the corresponding endpoint is also excluded from $\mathcal D_+$.}.
The only remaining possibility is $\rho_{\gamma,J,+}^*(K)\to+\infty$ as $K\to(V_{\gamma,J,+}^{\rm sup})^-$, which can occur only for $\gamma=0$. Provided that $\lim_{\rho\to\infty}D(\rho)\neq0$, the same divergence follows, i.e., $\lim_{K\to(V_{\gamma,J,+}^{\rm sup})^-} T_{\gamma,J}^{(n_*,\rho_{\gamma,J,+}^*(K))}[K]=+\infty$.
These results show that every admissible global minimum satisfies condition $\mathbf{C2}$ and thus this constraint can be removed from the minimization in Eq.~(\ref{eq:RFmindens}). We are now ready to write Eq.~(\ref{eq:RFmindens}) in its final form,
\begin{equation}\label{eq:RFmindensFinal}
        U_{\gamma}(J) = \min_{K,s} Q_{\gamma,J}^{(n_*,\rho_{\gamma,J,s}^*(K))}(K)-JE .
\end{equation}

Even though Eq.~(\ref{eq:RFmindensFinal}) is generally extremely complicated to solve analytically, it provides a compact and numerically much more efficient way (analogous to that in~\cite{DerridaAdditivityPrinciple}) to compute the rate function and CGF (through a Legendre transform) in equilibrium.

To conclude this section, we remark that this method can also be extended to unmonitored ($\gamma=0$) out-of-equilibrium systems, i.e. $n_0>n_1$. Since $\gamma=0$ throughout this discussion, we suppress it from the notation and write $U(J)$, $V_J$, $\rho_{J,K}$, $T_J$, $H_J$, and $Q_J$. These objects have the same definitions as in the equilibrium discussion above, evaluated at $\gamma=0$. In these cases, it is not guaranteed that the derivative of $\rho_{J,K}$ vanishes inside the unit interval $[0,1]$. Nevertheless, by an analysis similar to the one carried out in this appendix, one can show that the rate function is given by
\begin{multline}\label{eq:RENEq}
    U(J) = \frac{1}{2}\min \left( \min_{\substack{s\in\{-,+\}\\ K_{\rm mon}<K\leq V_{\rm max}}}\sum_{k=0}^{1}Q_J^{(n_k,\rho_{J,s}^*(K))}(K) ,\right.\\
    \left. -K_{\rm mon} +Q_J^{(n_0,n_1)}(K_{\rm mon})\right)+\mu E-\kappa J-JE .
\end{multline}
Here $\rho_{J,+}^*(K)>n_0>n_1$ and $\rho_{J,-}^*(K)<n_1<n_0$ denote, respectively, the accessible upper and lower turning points satisfying $V_J(\rho_{J,s}^*(K))=K$.
Additionally, $V_{\rm max}\equiv\max_\rho V_J(\rho)$, as in the main text, while
\begin{equation}
    V_{\rm int}^{\max}\coloneqq\max_{\rho\in[n_1,n_0]}V_J(\rho),
    \qquad
    K_{\rm mon}\coloneqq\operatorname*{arg\,max}_{K\geq V_{\rm int}^{\max}}
    \left(-K+Q_J^{(n_0,n_1)}(K)\right).
\end{equation}
The first and second arguments of the minimum in Eq.~(\ref{eq:RENEq}) describe profiles with a turning point and monotonic profiles, respectively. For the monotonic branch, $\partial_K[-K+Q_J^{(n_0,n_1)}(K)]=2(T_J^{(n_0,n_1)}[K]-1)$; since $T_J^{(n_0,n_1)}[K]$ decreases with $K$, $K_{\rm mon}$ is the unique value selected by the corresponding length constraint. A turning profile, on the other hand, contains the monotonic crossing from $n_0$ to $n_1$ plus an additional segment. Its unit-length constraint therefore requires $T_J^{(n_0,n_1)}[K]<1$ and hence $K>K_{\rm mon}$.

\section{Conditions for First-order vs Continuous Dynamical Phase Transitions}\label{app:ConditionsFist}

\subsection{Explicit general conditions
\label{app:gencond}}
In Sec.~\ref{subsec:DPTs}, we specialized our analysis of dynamical phase transitions to the one dimensional particle-hole-symmetric setting considered in Ref.~\cite{Lecomte_2017}. There, we extended to the monitored setting the condition derived in Ref.~\cite{Lecomte_2017} under which a spatially inhomogeneous, symmetry-breaking solution can emerge continuously from the homogeneous profile. This condition determines the candidate critical current $J_c^{\rm loc}$ given in Eq.~\eqref{eq:GeqCond1}.

\begin{figure}[!htbp]
    \centering
    \includegraphics[width=1\linewidth]{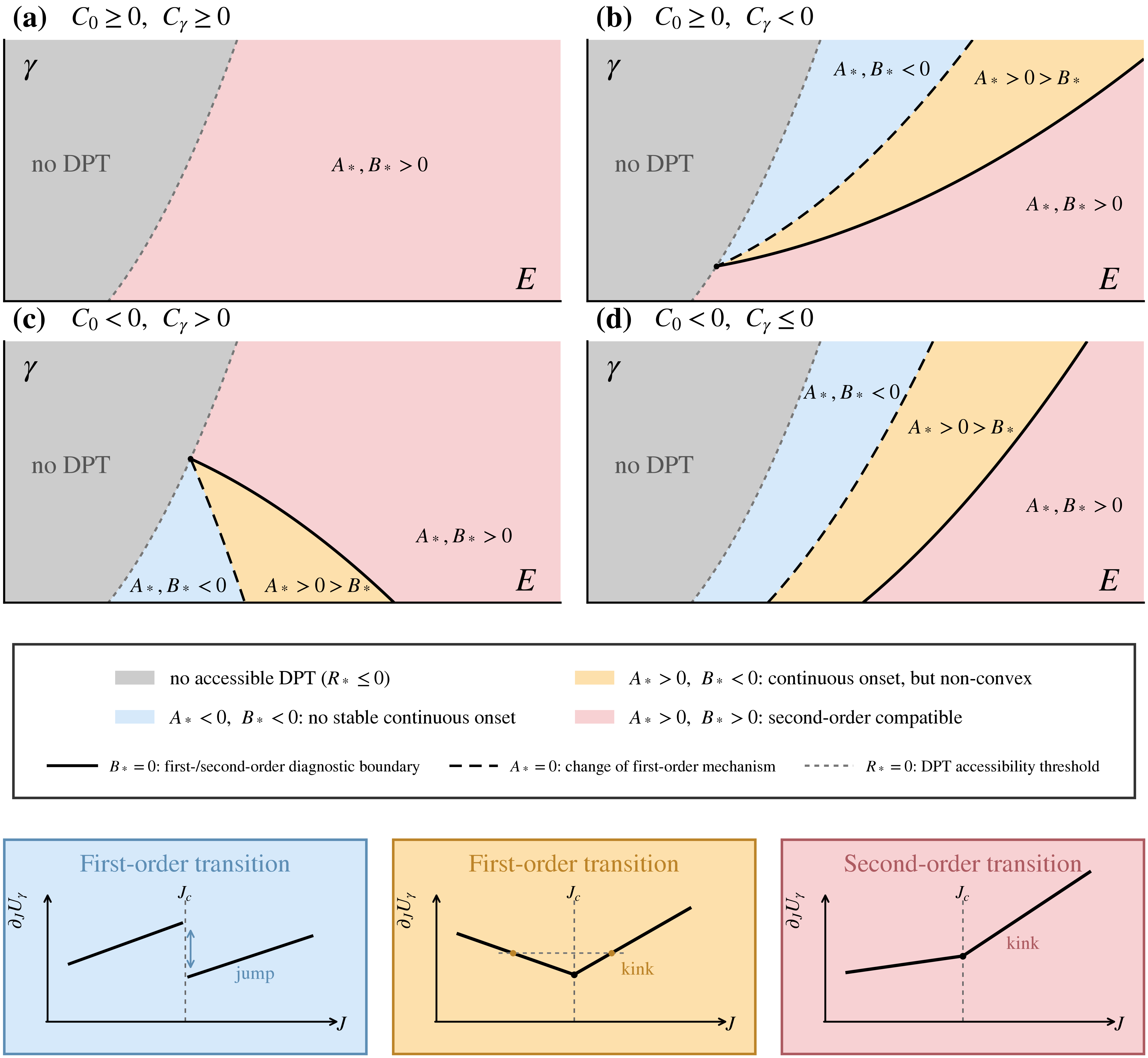}
    \caption{Schematic classification of the DPT nature in terms of $C_*(\gamma)=C_0+\gamma C_\gamma$. The four panels form the exhaustive partition (a) $C_0\geq0$, $C_\gamma\geq0$; (b) $C_0\geq0$, $C_\gamma<0$; (c) $C_0<0$, $C_\gamma>0$; and (d) $C_0<0$, $C_\gamma\leq0$. Gray denotes the inaccessible region $R_*(E,\gamma)\leq0$. In the accessible domain, blue has $A_*(E,\gamma)<0$ and $B_*(E,\gamma)<0$, so there is no stable continuous onset; yellow has $A_*(E,\gamma)>0$ and $B_*(E,\gamma)<0$, so the onset can be continuous but is non-convex; and red has $A_*(E,\gamma)>0$ and $B_*(E,\gamma)>0$, as required for a second-order transition. The solid $B_*(E,\gamma)=0$ line separates the first-order diagnostic from the second-order-compatible region, while the dashed $A_*(E,\gamma)=0$ line separates the two first-order mechanisms. The lower sketches show the corresponding behavior of $\partial_JU_\gamma(J)$ in the blue, yellow, and red regions: a jump, a continuous but non-injective kink, and a monotone kink, respectively.}
    \label{fig:SketchStability}
\end{figure}

Here we go beyond this existence condition and derive two additional local criteria that constrain the order of the resulting transition. The first condition is $G'_{\gamma,J_c^{\rm loc}}(V_*)=Q''_{\gamma,J_c^{\rm loc}}(V_*)\geq0$, which is necessary for the homogeneous branch to remain a local minimum at the critical point, as required for a second-order transition. The functions $G_{\gamma,J}(K)$ and $Q_{\gamma,J}(K)$, the conserved quantity $K$, and the boundary value $V_*=V_{\gamma,J}(n_*)$ are defined in Eqs.~\eqref{eq:KconservationT}--\eqref{eq:rfrhov1}. As in Sec.~\ref{subsec:DPTs}, we use the shorthand $D_*\coloneqq D(n_*)$ and $\sigma_*\coloneqq\sigma(n_*)$, with primes denoting derivatives evaluated at $n_*$. Expanding $V_{\gamma,J_c^{\rm loc}}(\rho)$ to fourth order and eliminating $J_c^{\rm loc}$ using Eq.~\eqref{eq:GeqCond1}, we obtain
\begin{equation}\label{eq:cond1stRF}
    G_{\gamma,J_c^{\rm loc}}'(V_*) \propto  A_*(E,\gamma)\geq 0,\,A_*(E,\gamma)\coloneqq E^2+A_\gamma \,\gamma + A_0,
\end{equation}
where the proportionality constant is positive and
\begin{equation}\label{eq:AcoeffsStability}
    A_0 =  \pi^2 D_* \frac{4\sigma_* D_*''+ 4 D_* \sigma_*'' - \sigma_* D_* \frac{\sigma_*''''}{\sigma_*''}}{3 \sigma_* \left(\sigma_*''\right)^2}\; \; \text{and}\;\; A_\gamma = \frac{6 \sigma_* \sigma_*''- \sigma_*^2 \frac{\sigma_*''''}{\sigma_*''}}{3 \sigma_* \left(\sigma_*''\right)^2}.
\end{equation}
As emphasized in Fig.~\ref{fig:QperturbativeSketch}, this is only a local necessary condition for a second order transition: even when $A_*(E,\gamma)>0$, a distant minimum of $Q_{\gamma,J_c^{\rm loc}}(K)$ may pre-empt the continuous bifurcation. The second necessary condition identified in Sec.~\ref{subsec:classificationFPT} is local convexity of $U_\gamma(J)$ at $J_c^{\rm loc}$. Indeed, if $U_\gamma$ is locally non-convex, its convex envelope replaces that region by a linear coexistence segment, thereby producing a first-order rather than a second-order transition. To evaluate this condition, define $K_{\rm min}(J)$ in a neighborhood of $J_c^{\rm loc}$ by $G_{\gamma,J}(K_{\rm min}(J))=0$; thus, $K_{\rm min}(J)$ follows the nearby local minimum of $Q_{\gamma,J}(K)$. Using Eq.~\eqref{eq:rfrhov1}, the convexity condition becomes
\begin{equation}
    \left.\partial_J^2 Q_{\gamma,J}(K_{\rm min}(J))\right|_{J = J_c^{\rm loc}} \propto  B_*(E,\gamma)/A_*(E,\gamma) \geq 0, \quad B_*(E,\gamma) \coloneqq \left(E^2 + \gamma B_{\gamma} + B_0\right),
\end{equation}
where the proportionality constant is positive and
\begin{equation}\label{eq:BcoeffsStability}
    B_0 = \pi^2 D_* \frac{4\sigma_*'' D_*'' - D_* \sigma_*''''}{\left(\sigma_*''\right)^3}, 
    \quad
    B_\gamma = \frac{2 (\sigma_*'')^2 - \sigma_* \sigma_*''''}{\left(\sigma_*''\right)^3}.
\end{equation}
We conclude that a second-order transition requires both $A_*(E,\gamma)>0$ and $B_*(E,\gamma)>0$. In essence, the $A_*$ condition determines whether the minimizing saddle-point profile can evolve continuously away from the homogeneous profile at $J_c^{\rm loc}$. Provided that such a continuous evolution is possible, the $B_*$ condition, which recovers the perturbative condition of Ref.~\cite{Lecomte_2017}, determines whether $U_{\gamma}(J)$ remains locally convex. We now use these conditions to ask whether monitoring can change the order of a pre-existing DPT.
When $\sigma_*''<0$, this classification must also be restricted to the domain in which the continuous bifurcation exists. We thus rewrite the threshold condition of Eq.~\eqref{eq:GeqCond1} as
\begin{equation}\label{eq:RcoeffsStability}
    R_*(E,\gamma) \coloneqq E^2 + R_0 + \gamma R_\gamma > 0,\;R_0\coloneqq\frac{2\pi^2D_*^2}{\sigma_*\sigma_*''},
    \;
    R_\gamma\coloneqq\frac{2}{\sigma_*''}.
\end{equation}
In fact, the three diagnostics are not independent. Direct substitution of their coefficients gives
\begin{equation}\label{eq:ABRidentity}
    3A_0-B_0=2R_0,\qquad 3A_\gamma-B_\gamma=2R_\gamma,
    \qquad\Longrightarrow\qquad
    3A_*(E,\gamma)-B_*(E,\gamma)=2R_*(E,\gamma).
\end{equation}
Hence, in the accessible domain $R_*(E,\gamma)>0$, $A_*(E,\gamma)<0$ necessarily implies $B_*(E,\gamma)<0$; a region with $A_*(E,\gamma)<0<B_*(E,\gamma)$ can occur only beyond the accessibility threshold. To make this structure more transparent, we introduce the $E$-independent affine function
\begin{equation}\label{eq:CcoeffsStability}
    \begin{aligned}
    C_*(\gamma)
    &\coloneqq A_*(E,\gamma)-R_*(E,\gamma)
    =\frac{B_*(E,\gamma)-R_*(E,\gamma)}{3}=\\
    &=C_0+\gamma C_\gamma
    \quad\Longrightarrow\quad
    \begin{cases}
    A_*(E,\gamma)=R_*(E,\gamma)+C_*(\gamma),\\
    B_*(E,\gamma)=R_*(E,\gamma)+3C_*(\gamma),
    \end{cases}
    \end{aligned}
\end{equation}
where $C_0=A_0-R_0$ and $C_\gamma=A_\gamma-R_\gamma$.
At fixed $\gamma$, if $C_*(\gamma)<0$, increasing $E$ gives the ordered sequence $(A_*(E,\gamma)<0,B_*(E,\gamma)<0)$, $(A_*(E,\gamma)>0,B_*(E,\gamma)<0)$, and $(A_*(E,\gamma)>0,B_*(E,\gamma)>0)$; if $C_*(\gamma)\geq0$, both conditions are positive throughout the accessible domain. As a consequence, the contours $A_*(E,\gamma)=0$ and $B_*(E,\gamma)=0$ can only meet at $R_*(E,\gamma)=0$.
The possible behaviors are shown schematically in Fig.~\ref{fig:SketchStability}.
Because $C_*(\gamma)$ is affine, the four disjoint cases in Fig.~\ref{fig:SketchStability} cover the entire $(C_0,C_\gamma)$ plane. For $C_0<0$ and $C_\gamma>0$ [case (c)], the two first-order sectors shrink and terminate on the accessibility threshold. For $C_0\geq0$ and $C_\gamma<0$ [case (b)], they instead emerge from that threshold as monitoring is increased. They persist for $C_0<0$ and $C_\gamma\leq0$ [case (d)] and are absent for $C_0\geq0$ and $C_\gamma\geq0$ [case (a)]. We emphasize that these curves remain perturbative diagnostic boundaries rather than guaranteed global transition boundaries.

\begin{table}[t]
\centering
\small
\setlength{\tabcolsep}{4pt}
\renewcommand{\arraystretch}{1.1}
\begin{tabular}{cccccc}
\toprule
Models & $R_0$ & $R_\gamma$ & $C_0$ & $C_\gamma$ & Case \\
\midrule
SSEP
& $-\pi^2$ & $-\tfrac{1}{2}$
& $\tfrac{1}{3}\pi^2$ & $0$ & (a) \\
KLS ($\epsilon=0.6$)
& $-4\pi^2$ & $-\tfrac{15}{32}$
& $-\tfrac{11}{3}\pi^2$ & $-\tfrac{15}{128}$ & (d) \\
KLS ($-1<\epsilon<0$)
& $<0$ & $<0$ & $>0$ & $<0$ & (b) \\
KLS ($0\leq\epsilon<\epsilon_1$)
& $<0$ & $<0$ & $>0$ & $\geq0$ & (a) \\
KLS ($\epsilon_1<\epsilon<\epsilon_2$)
& $<0$ & $<0$ & $<0$ & $>0$ & (c) \\
KLS ($\epsilon_2<\epsilon<0.8$)
& $<0$ & $<0$ & $<0$ & $<0$ & (d) \\
KLS ($0.8<\epsilon<1$)
& $>0$ & $>0$ & $>0$ & $>0$ & (a) \\
\bottomrule
\end{tabular}
\caption{Coefficients $R_0$, $R_\gamma$, $C_0$, and $C_\gamma$ entering the local classification of the transition order, and the corresponding case of Fig.~\ref{fig:SketchStability}. All KLS entries are at $\delta=0$, with $\epsilon_1\simeq0.396$ and $\epsilon_2\simeq0.540$ the zeros of $C_0$ and $C_\gamma$, respectively.}
\label{tab:coefficients}
\end{table}

\subsection{Conditions for SSEP and KLS models}\label{app:condssep}

We now apply the local criteria derived above to the monitored SSEP and Katz--Lebowitz--Spohn (KLS) models. We compare their predictions with those extracted from an exact numerical evaluation of $U_\gamma(J)$ using the construction introduced in Sec.~\ref{subsec:weakadditivity1d} and detailed in Appendix~\ref{sec:MethodForRF}. This comparison shows both cases where the local criteria correctly capture the order of the transition and where features beyond the expansion around the homogeneous profile become essential. The transport coefficients of the SSEP and KLS models are given in Eqs.~\eqref{eq:SSEPtranspcoeff} and \eqref{eq:KLStranspcoeff}, respectively, while Table~\ref{tab:coefficients} summarizes the resulting coefficients entering the local criteria.

\subsubsection{SSEP: transition order in the full stationary problem}\label{sec:SSEPextensiveanalysis}
For the SSEP, which is particle-hole symmetric around $n_*=1/2$, substituting Eq.~\eqref{eq:SSEPtranspcoeff} into Eqs.~\eqref{eq:RcoeffsStability} and \eqref{eq:CcoeffsStability} places the model in case (a) of Fig.~\ref{fig:SketchStability}, as summarized in Table~\ref{tab:coefficients}. As shown in Sec.~\ref{app:gencond}, both local conditions are therefore satisfied throughout the region in which a continuous DPT is locally accessible. Equation~\eqref{eq:Ec?} locates this region at $E^2>\pi^2+\gamma/2$. Thus, the local analysis predicts that monitoring shifts the threshold for the DPT, as discussed in Sec.~\ref{subsec:classificationFPT}, but does not change its second-order character.

We first show that, for the SSEP, the region excluded by the local onset condition is also globally free of stationary DPTs. Setting $u=2\rho-1$ and introducing the effective-field combination $E_{\rm eff}^2\coloneqq E^2-\gamma/2$, the difference between the stationary functional evaluated at an arbitrary profile and at the homogeneous profile is
\begin{equation}\label{eq:SSEPglobalBound}
\mathcal U_\gamma(\rho,J)-\mathcal U_\gamma(n_*,J)
=\int_0^1dx\left[
\frac{J^2u^2}{1-u^2}
+\frac{(\partial_xu)^2}{4(1-u^2)}
-\frac{E_{\rm eff}^2}{4}u^2\right].
\end{equation}
Since $u(0)=u(1)=0$, the Poincaré inequality ($\pi^2 \int_0^1 dx\, |u|^2 \leq \int_0^1 dx\, |\partial_x u|^2$) gives
\begin{equation}
\mathcal U_\gamma(\rho,J)-\mathcal U_\gamma(n_*,J)
\geq\frac{\pi^2-E_{\rm eff}^2}{4}\int_0^1dx\,u^2
=-\frac{R_*(E,\gamma)}{4}\int_0^1dx\,u^2.
\end{equation}
Hence, when $R_*(E,\gamma)\leq0$, the homogeneous profile is the global stationary minimizer for every $J$. Thus, for the SSEP, the locally determined accessibility threshold for a DPT is also globally exact within the stationary problem.

It remains to determine the nature of the transition when $R_*(E,\gamma)>0$, which corresponds to $E_{\rm eff}^2>\pi^2$. For the SSEP, the transition can then be studied analytically because $Q_{\gamma,J}(K)$ admits a closed form. Using particle-hole symmetry, we restrict to $\rho\geq n_*=1/2$ and introduce $z=u^2$. The quantity appearing under the square root in Eq.~\eqref{eq:timeconstraint} then factorizes as
\begin{equation}\label{eq:SSEPellipticPolynomial}
2\sigma(\rho)\left[K-V_{\gamma,J}(\rho)\right]
=K(1-z)+J^2+\frac{E^2}{4}(1-z)^2
  +\frac{\gamma}{8}z(1-z)
=\frac{E_{\rm eff}^2}{4}(z_- -z)(z_+ -z).
\end{equation}
We label the roots so that $0\leq z_-\leq z_+\leq1$, with $z_-$ the turning point reached by the physical non-uniform trajectory: the profile leaves $u=0$, reaches $u^*=\sqrt{z_-}$, and returns to $u=0$, while $z_+$ controls the shape of the excursion. Introducing the elliptic parameter $m=z_-/z_+$, we define the following three functions:
\begin{align}\label{eq:SSEPellipticDefinitions}
 \operatorname{Ell}_1(m)
 &\coloneqq\int_0^{\pi/2}\frac{d\theta}
 {\sqrt{1-m\sin^2\theta}}, \;\operatorname{Ell}_2(m)
 \coloneqq\int_0^{\pi/2}d\theta\,
 \sqrt{1-m\sin^2\theta},\\
 &\operatorname{Ell}_3(\nu\mid m)
 \coloneqq\int_0^{\pi/2}\frac{d\theta}
 {(1-\nu\sin^2\theta)\sqrt{1-m\sin^2\theta}}.\nonumber
\end{align}
These are, respectively, the complete elliptic integrals of the first, second, and third kinds. In this notation, Eq.~\eqref{eq:rfrhov0} becomes
\begin{equation}\label{eq:SSEPellipticQ}
 Q_{\gamma,J}(K)=-K+E_{\rm eff}\sqrt{z_+}\bigg[
 \operatorname{Ell}_2(m)
 +\frac{z_- -1}{z_+}\operatorname{Ell}_1(m)
 +\frac{(z_- -1)(z_+ -1)}{z_+}\operatorname{Ell}_3(z_-\mid m)\bigg].
\end{equation}
Matching coefficients in Eq.~\eqref{eq:SSEPellipticPolynomial} gives the first two relations below, while evaluating the unit-length constraint~\eqref{eq:timeconstraint} using the functions defined in Eq.~\eqref{eq:SSEPellipticDefinitions} gives the third:
\begin{equation}\label{eq:SSEPellipticConstraint}
 J^2=\frac{E_{\rm eff}^2}{4}(1-z_-)(1-z_+),\;
 K=\frac{E_{\rm eff}^2}{4}(z_-+z_+-1)-\frac{E^2}{4},\;
 G_{\gamma,J}(K)=\frac{2\operatorname{Ell}_1(z_-/z_+)}
 {E_{\rm eff}\sqrt{z_+}}-1.
\end{equation}
The first two relations can be inverted explicitly:
\begin{equation}\label{eq:zpmKJE}
 z_\pm =1+\frac{2}{E_{\rm eff}^2}\left[
 K+\frac{\gamma}{8}
 \pm\sqrt{\left(K+\frac{\gamma}{8}\right)^2-E_{\rm eff}^2J^2}
 \right].
\end{equation}
Before proceeding, we briefly note an exact connection with the unmonitored problem. Defining $K_{\rm eff}\coloneqq K+\gamma/8$, Eqs.~\eqref{eq:SSEPellipticQ}--\eqref{eq:zpmKJE} reduce to those of the unmonitored open SSEP with external field $E_{\rm eff}$, whose current rate function has been studied in Ref.~\cite{PerezEspigares2018}. More precisely, the corresponding $G_{\gamma,J}(K)$ functions coincide after replacing $(E,K)$ by $(E_{\rm eff},K_{\rm eff})$, while $Q_{\gamma,J}(K;E)=Q_{0,J}(K_{\rm eff};E_{\rm eff})+\gamma/8$. Consequently,
\begin{equation}
 U_\gamma(J;E)=U_0(J;E_{\rm eff})
 +\frac{\gamma}{8}-(E-E_{\rm eff})J,
\end{equation}
so, up to a trivial affine contribution, the effect of monitoring on the rate function amounts to the redefinition $E\to E_{\rm eff}$ of the external field.

Returning to our non-perturbative analysis of the DPT, we first use Eqs.~\eqref{eq:SSEPellipticQ}--\eqref{eq:zpmKJE} to exclude the possibility that a distant minimum of $Q_{\gamma,J}(K)$ gives rise to a first-order transition. Since $0\leq z_-\leq z_+\leq1$, Eq.~\eqref{eq:SSEPellipticConstraint} gives $K+\gamma/8=E_{\rm eff}^2(z_-+z_+-2)/4\leq0$. Hence, we conclude that $z_-$ increases while $z_+$ decreases with $K$. Since $\operatorname{Ell}_1(m)$ is strictly increasing for $0\leq m<1$, both $\operatorname{Ell}_1(m)$ and $z_+^{-1/2}$ increase with $K$ and thus Eq~\eqref{eq:SSEPellipticConstraint} shows that $G_{\gamma,J}(K)=\partial_K Q_{\gamma,J}(K)$ is strictly increasing for all $K$. Consequently, $Q_{\gamma,J}(K)$ has at most one non-uniform stationary point. If $G_{\gamma,J}(V_*)\geq0$, it increases away from the homogeneous endpoint and no distant minimum exists. If $G_{\gamma,J}(V_*)<0$, there is instead a unique non-uniform minimum; as $G_{\gamma,J}(V_*)$ changes sign, this minimum emerges continuously from $z_-=0$. A first-order transition caused by an exchange with a distant time-independent minimum is therefore impossible.

It remains to verify that $U_\gamma(J)$ remains a convex function. Because $Q_{\gamma,J}(K)$ is even in $J$, it is sufficient to consider $J\geq0$. For the values of the current for which the uniform profile $\rho=1/2$ minimizes the action, $U_{\gamma}(J)$ is quadratic and thus strictly convex. Furthermore, at the critical point, local convexity is ensured by $A_*(E,\gamma)>0$ and $B_*(E,\gamma)>0$. We now establish convexity throughout the non-uniform branch. There, $U_{\gamma}(J)=Q_{\gamma,J}(K_{\min}(J))-J E$, with $K_{\min}(J)$ the unique minimum of $Q_{\gamma,J}(K)$, which satisfies $G_{\gamma,J}(K_{\min}(J))=0$. Eq.~\eqref{eq:SSEPellipticConstraint} then provides the one-parameter representation
\begin{equation}\label{eq:SSEPconvexParametrization}
 J(m)=\frac{E_{\rm eff}}{2}\sqrt{[1-z_-(m)][1-z_+(m)]},\qquad
 z_+(m)=\frac{4\operatorname{Ell}_1(m)^2}{E_{\rm eff}^2},\qquad
 z_-(m)=m\,z_+(m).
\end{equation}
Here $0\leq m\leq m_{\max}$, with the endpoint $m_{\max}$ fixed by $4\operatorname{Ell}_1(m_{\max})^2=E_{\rm eff}^2$, at which $z_+=1$ and $J=0$. Direct differentiation of Eq.~\eqref{eq:SSEPconvexParametrization} shows that $J(m)$ is strictly decreasing, so it is one-to-one and admits an inverse $m(J)$. Noting that, at $K_{\min}(J)$, $\partial_KQ_{\gamma,J}=G_{\gamma,J}=0$, the chain rule gives 
\begin{equation}
 \frac{d}{dJ}Q_{\gamma,J}(K_{\min}(J))
 =\int_0^1dx\,\frac{J}{\sigma(\rho(x))}
 =2J\,\frac{\operatorname{Ell}_3(z_-(m(J))\mid m(J))}
 {\operatorname{Ell}_1(m(J))}.
\end{equation}
Finally, direct differentiation of the integral definitions in Eq.~\eqref{eq:SSEPellipticDefinitions}, together with Eq.~\eqref{eq:SSEPconvexParametrization}, gives
\begin{equation}
 \frac{dJ(m)}{dm}<0,\qquad
 \frac{d}{dm}\left[
 J(m)\frac{\operatorname{Ell}_3(z_-(m)\mid m)}
 {\operatorname{Ell}_1(m)}\right]<0
 \qquad (0<m<m_{\max}),
\end{equation}
and therefore
\begin{equation}
 \frac{d^2}{dJ^2}Q_{\gamma,J}\!\left(K_{\min}(J)\right)
 =2\frac{dm}{dJ}\,
 \frac{d}{dm}\left[
 J(m)\frac{\operatorname{Ell}_3(z_-(m)\mid m)}
 {\operatorname{Ell}_1(m)}\right]>0.
\end{equation}
We conclude that the complete time-independent function $U_\gamma(J)=\min_KQ_{\gamma,J}(K)-JE$ is convex and thus the SSEP transition is globally second order within the weak-additivity description.

\subsubsection{KLS: numerical phase diagram and boundary tests}
\label{app:KLSBoundaryTests}
For the KLS model at $\delta=0$ (see Eq.~\eqref{eq:KLStranspcoeff}), which is particle-hole-symmetric around $n_*=1/2$, the local conditions derived above realize all four cases of Fig.~\ref{fig:SketchStability} as $\epsilon$ is varied, as summarized in Table~\ref{tab:coefficients}.
At $\epsilon=0.6$, which we consider in the following, substituting Eq.~\eqref{eq:KLStranspcoeff} into Eqs.~\eqref{eq:RcoeffsStability} and \eqref{eq:CcoeffsStability} places the model in case (d). The local construction then predicts two distinct first-order sectors and a second-order sector. To test this prediction and construct the phase diagram shown in Fig.~\ref{fig:KLSEps060PhaseDiagram}, we globally minimize the stationary functional. We perform this analysis numerically because the complexity of the KLS transport coefficients in Eq.~\eqref{eq:KLStranspcoeff} makes a treatment analogous to the analytical one carried out for the SSEP in Appendix~\ref{sec:SSEPextensiveanalysis} impractical.
For every sampled $(E,\gamma)$ and current $J$, we minimize
$Q_{\gamma,J}(K)$ over the homogeneous endpoint $\rho=1/2$ and the complete numerical
set of non-uniform turning-point branches. At each current, this procedure
yields the stationary function $U_\gamma(J)$ and the turning point
$\rho^*_{\rm min}(J)$ of the corresponding globally minimizing profile.
Particle--hole symmetry gives two equivalent branches,
and we choose the one with $\rho^*_{\rm min}(J)\geq n_*$.

To distinguish first- from second-order DPTs numerically, we construct the
convex envelope $U_\gamma^{**}(J)$. As discussed in
Appendix~\ref{app:gencond}, if $U_\gamma^{**}(J)$ differs from $U_\gamma(J)$,
the latter must be convexified to obtain a valid current rate function, which
is necessarily convex. The resulting affine coexistence segment signals a
first-order DPT. At a second-order DPT, by contrast,
$U_\gamma(J)$ remains convex despite its non-analyticity and therefore
coincides with its convex envelope. It thus makes sense to use
$\max_J[U_\gamma(J)-U_\gamma^{**}(J)]$ as a numerical test: once a DPT is
detected, a non-zero value identifies a first-order transition, whereas a
vanishing value identifies a second-order one. Within the first-order region,
the two mechanisms described in
Appendix~\ref{app:gencond} are
distinguished by the saddle-point amplitude. At the stationary transition
current $J_c$, we define
$\rho^*_{\rm critical}\coloneqq\rho^*_{\rm min}(J_c)$. A finite
$|\rho^*_{\rm critical}-n_*|$ signals a direct exchange between the
homogeneous minimum and a remote non-uniform minimum ($A_*$ mechanism), whereas a vanishing
amplitude shows that the stationary profile emerges continuously and that
the first-order character arises solely from convexification ($B_*$ mechanism).

\begin{figure}[!htbp]
    \centering
    \includegraphics[width=1\linewidth]{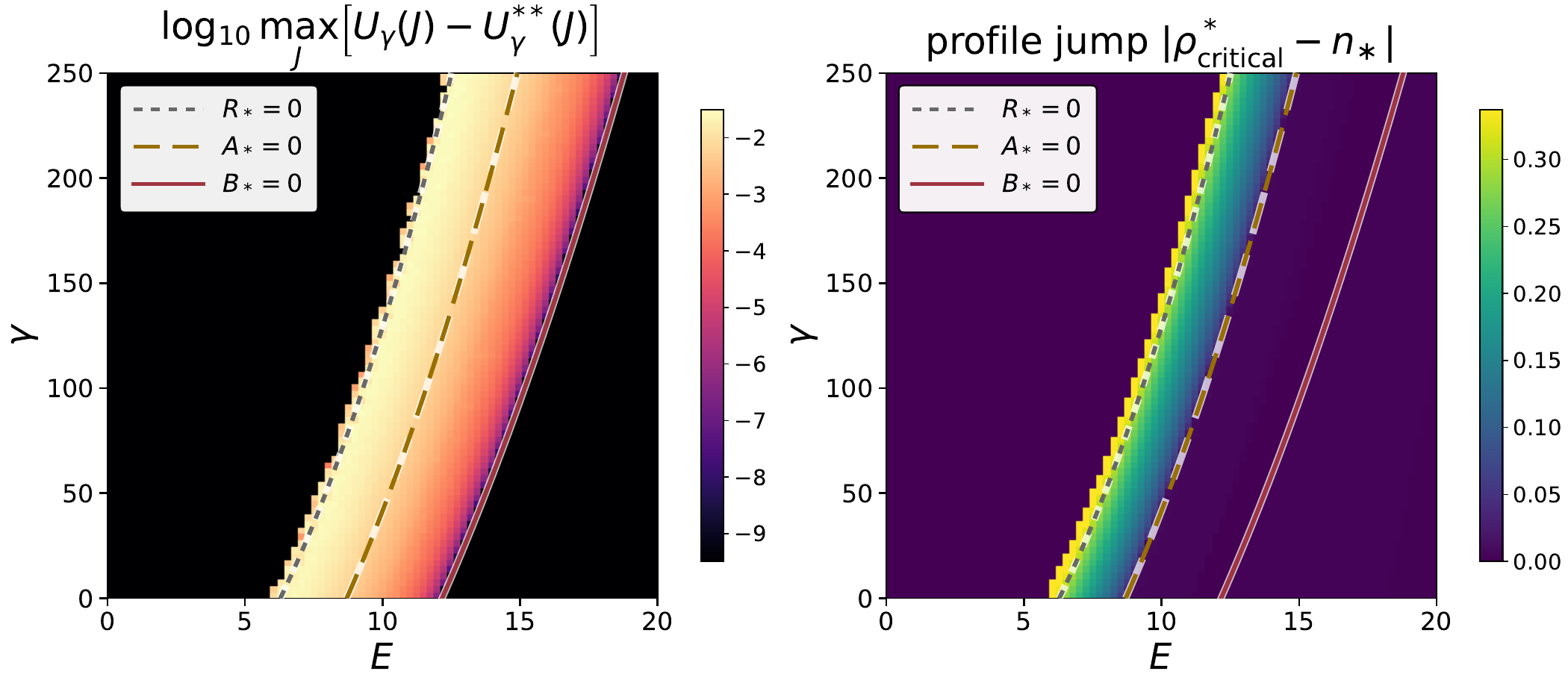}
    \caption{Raw numerical diagnostics underlying
    Fig.~\ref{fig:KLSEps060PhaseDiagram}.  Left: logarithm of the maximum
    convexification gap,
    $\max_J[U_\gamma(J)-U_\gamma^{**}(J)]$.  Right: amplitude
    $|\rho^*_{\rm critical}-n_*|$ of the discontinuous change of the global
    stationary minimizer.  The scan uses an $81\times81$ grid in
    $0\leq E\leq20$ and $0\leq\gamma\leq250$, with global minimization over
    $Q_{\gamma,J}(K)$ at every sampled current.  The gray dotted,
    yellow dashed, and red solid curves are, respectively, the local
    $R_*(E,\gamma)=0$ (or $J_c^{\rm loc}=0$), $A_*(E,\gamma)=0$, and $B_*(E,\gamma)=0$ predictions.  Convexity loss
    persists between $A_*(E,\gamma)=0$ and $B_*(E,\gamma)=0$ even though the stationary-profile
    onset is continuous there, while the finite density jump lies on the $A_*(E,\gamma)<0$ side.}
    \label{fig:KLSEps060DensityDiagnostics}
\end{figure}

We evaluate these two diagnostics over a broad numerical sweep of the
$(E,\gamma)$ plane. The results in
Fig.~\ref{fig:KLSEps060DensityDiagnostics} reproduce the qualitative
structure of case (d) in Fig.~\ref{fig:SketchStability}: the line on which the
profile amplitude $|\rho^*_{\rm critical}-n_*|$ vanishes closely follows $A_*(E,\gamma)=0$, while the disappearance
of the convexification gap closely follows $B_*(E,\gamma)=0$. Thus, at the resolution
of this sweep, both local boundaries appear to remain valid after global
minimization. The small unresolved cells next to these lines reflect the
finite resolution of the broad scan. We therefore test the $B_*(E,\gamma)=0$, $A_*(E,\gamma)=0$,
and $R_*(E,\gamma)=0$ predictions separately at higher resolution below. For convenience, in what follows, we denote the positive-field solutions of
$A_*(E,\gamma)=0$, $B_*(E,\gamma)=0$, and $R_*(E,\gamma)=0$ by
$E_{A_*=0}(\gamma)$, $E_{B_*=0}(\gamma)$, and $E_{R_*=0}(\gamma)$, respectively.

\begin{figure}[!htbp]
    \centering
    \includegraphics[width=0.49\linewidth]{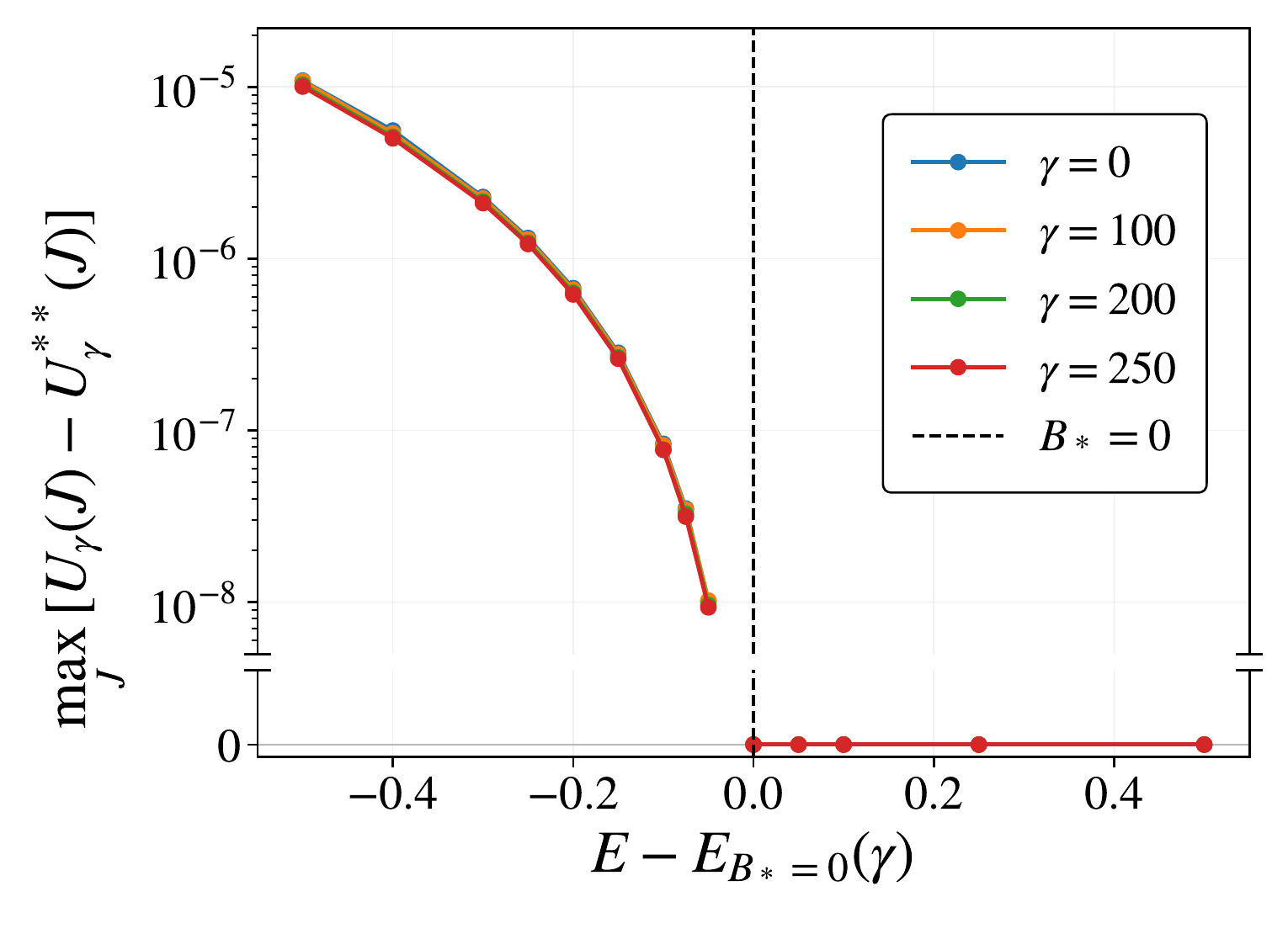}
    \hfill
    \includegraphics[width=0.49\linewidth]{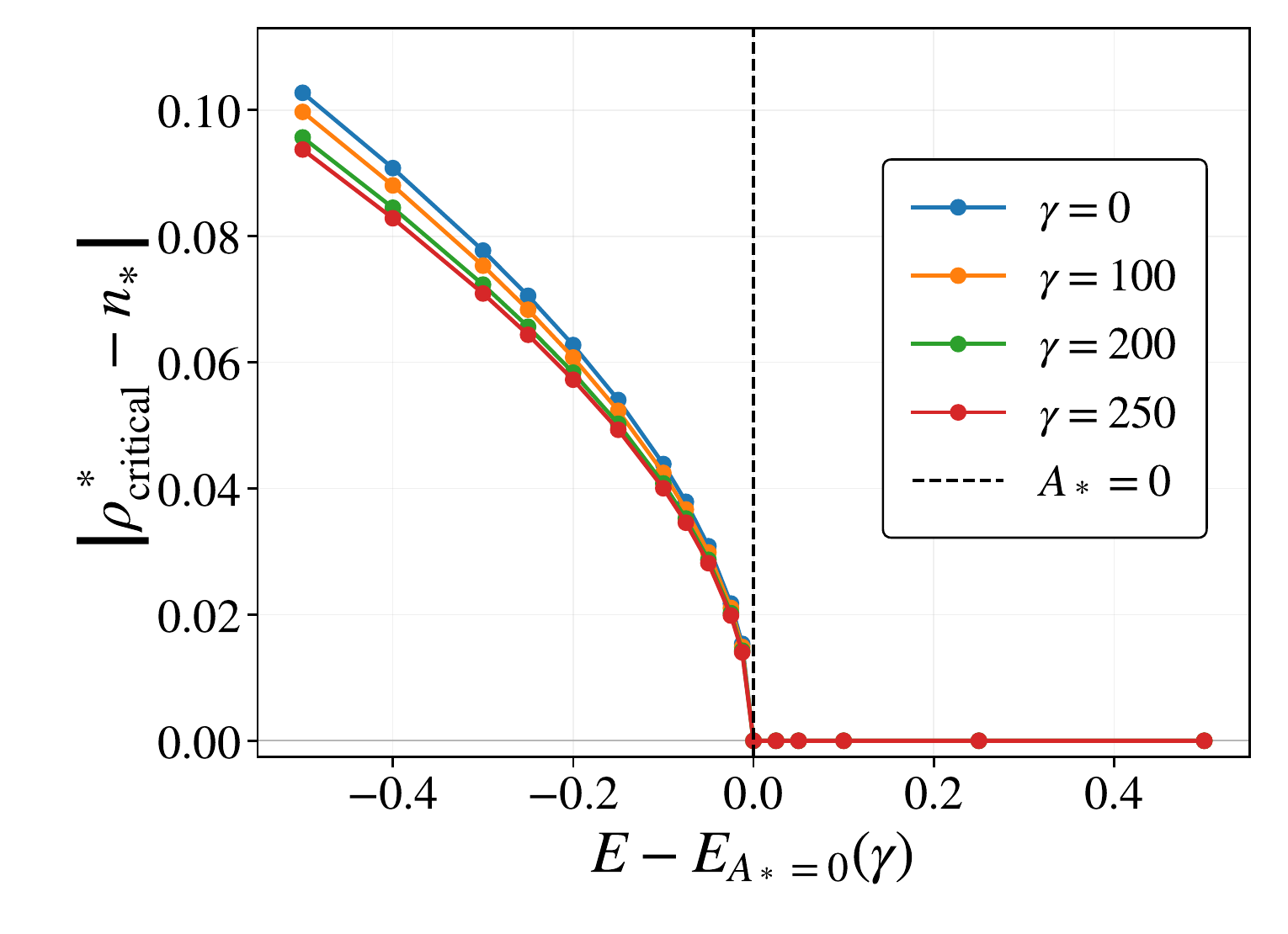}
    \caption{High-resolution tests of the $B_*(E,\gamma)=0$ and $A_*(E,\gamma)=0$ predictions for
    $\gamma\in\{0,100,200,250\}$. Left: the maximum convexification gap is
    positive for $E-E_{B_*=0}(\gamma)<0$ and vanishes at and above $B_*(E,\gamma)=0$;
    the lower strip denotes an exactly vanishing numerical gap. Right: the amplitude
    $|\rho^*_{\rm critical}-n_*|$ approaches zero from the $A_*(E,\gamma)<0$ side and
    remains zero at all sampled points with $A_*(E,\gamma)\geq0$.}
    \label{fig:KLSABBoundaryTests}
\end{figure}

\paragraph{First- versus second-order boundary at $B_*(E,\gamma)=0$.}
The first-/second-order boundary predicted by the local analysis is
$B_*(E,\gamma)=0$. To test it, for each measurement
strength, we globally minimize $Q_{\gamma,J}(K)$ and construct
$U_\gamma^{**}(J)$ on both sides of this line. A non-zero convexification gap
is resolved throughout the $B_*(E,\gamma)<0$ side and decreases to zero at the
predicted boundary, while every sampled point with $B_*(E,\gamma)\geq0$ has an exactly
vanishing numerical gap [left panel of
Fig.~\ref{fig:KLSABBoundaryTests}]. Within the tested $J$-window and the weak-additivity
description, $B_*(E,\gamma)=0$ therefore coincides with the actual boundary between first-order and second-order DPTs.

\paragraph{Change of first-order mechanism at $A_*(E,\gamma)=0$.}
Having identified the first-order region, we next distinguish the two mechanisms that give rise to it through the amplitude $|\rho^*_{\rm critical}-n_*|$. The $A_*$ condition tests
whether the non-uniform stationary profile can meet the homogeneous profile continuously at the local critical current.
For $A_*(E,\gamma)<0$, we vary $J$ until the
homogeneous and a finite-amplitude non-uniform profile have the same globally minimal value of $Q_{\gamma,J}(K)$, always observing a finite value of $|\rho^*_{\rm critical}-n_*|$.
For $A_*(E,\gamma)>0$, we instead track the global minimizer of $Q_{\gamma,J}(K)$ close to
$J_c^{\rm loc}$ and extrapolate its amplitude to $J_c^{\rm loc}$, where we observe that it
vanishes. As shown in the right panel of
Fig.~\ref{fig:KLSABBoundaryTests}, the finite jump approaches zero precisely at $A_*(E,\gamma)=0$ and remains zero on the $A_*(E,\gamma)>0$ side. Thus, $A_*(E,\gamma)=0$ accurately marks the change from a direct discontinuous exchange of stationary minimizers to a continuous stationary onset whose
first-order character is due to convexification.

\paragraph{Global onset and the $R_*(E,\gamma)=0$ prediction.}
Finally, $R_*(E,\gamma)=0$ marks where the locally predicted critical current in Eq.~\eqref{eq:GeqCond1} vanishes and therefore bounds the region in which a continuous DPT is locally admissible. However, it does not necessarily identify the boundary at which a first-order transition disappears.

\begin{figure}[!htbp]
    \centering
    \includegraphics[width=0.32\linewidth]{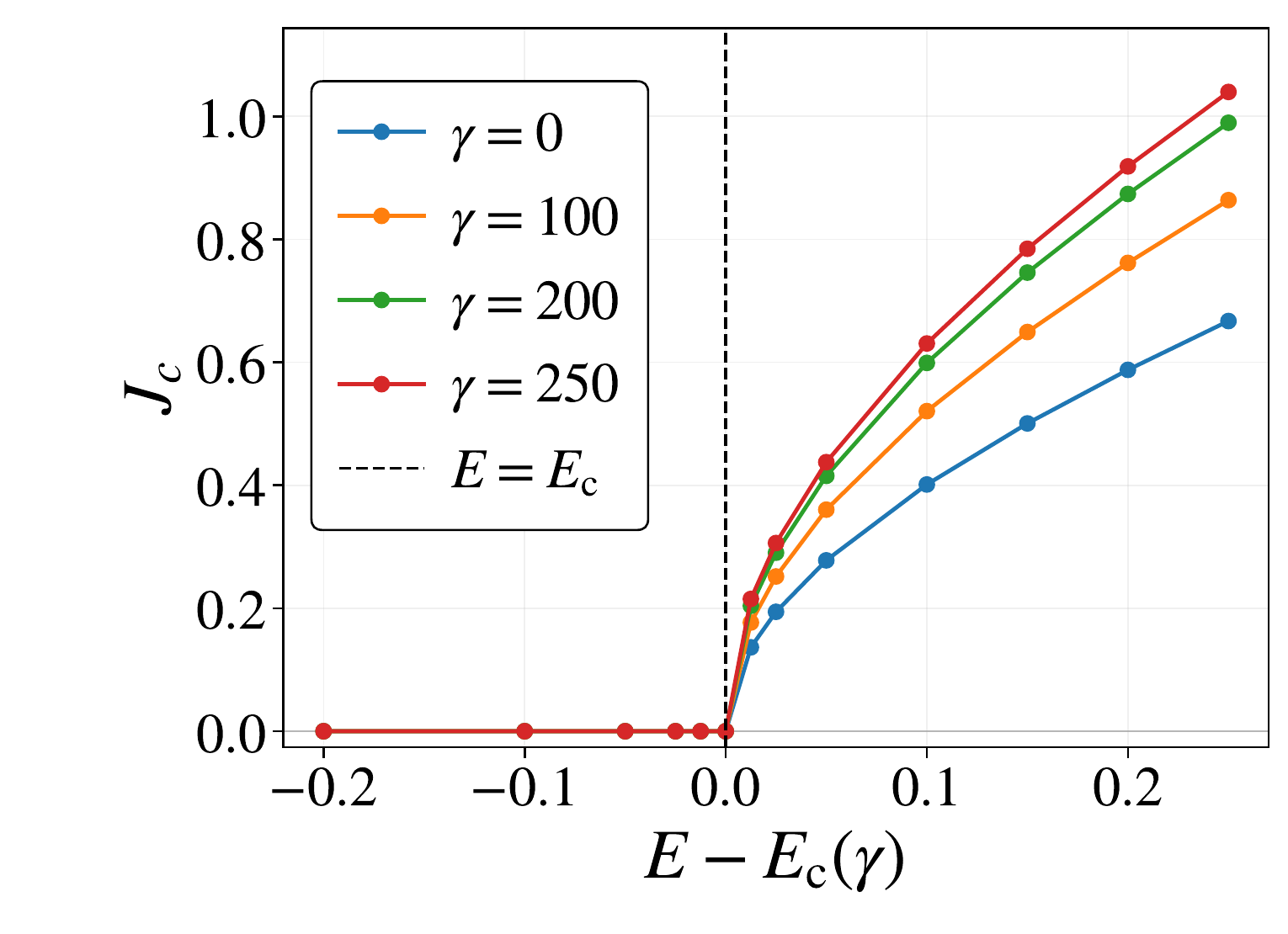}
    \hfill
    \includegraphics[width=0.32\linewidth]{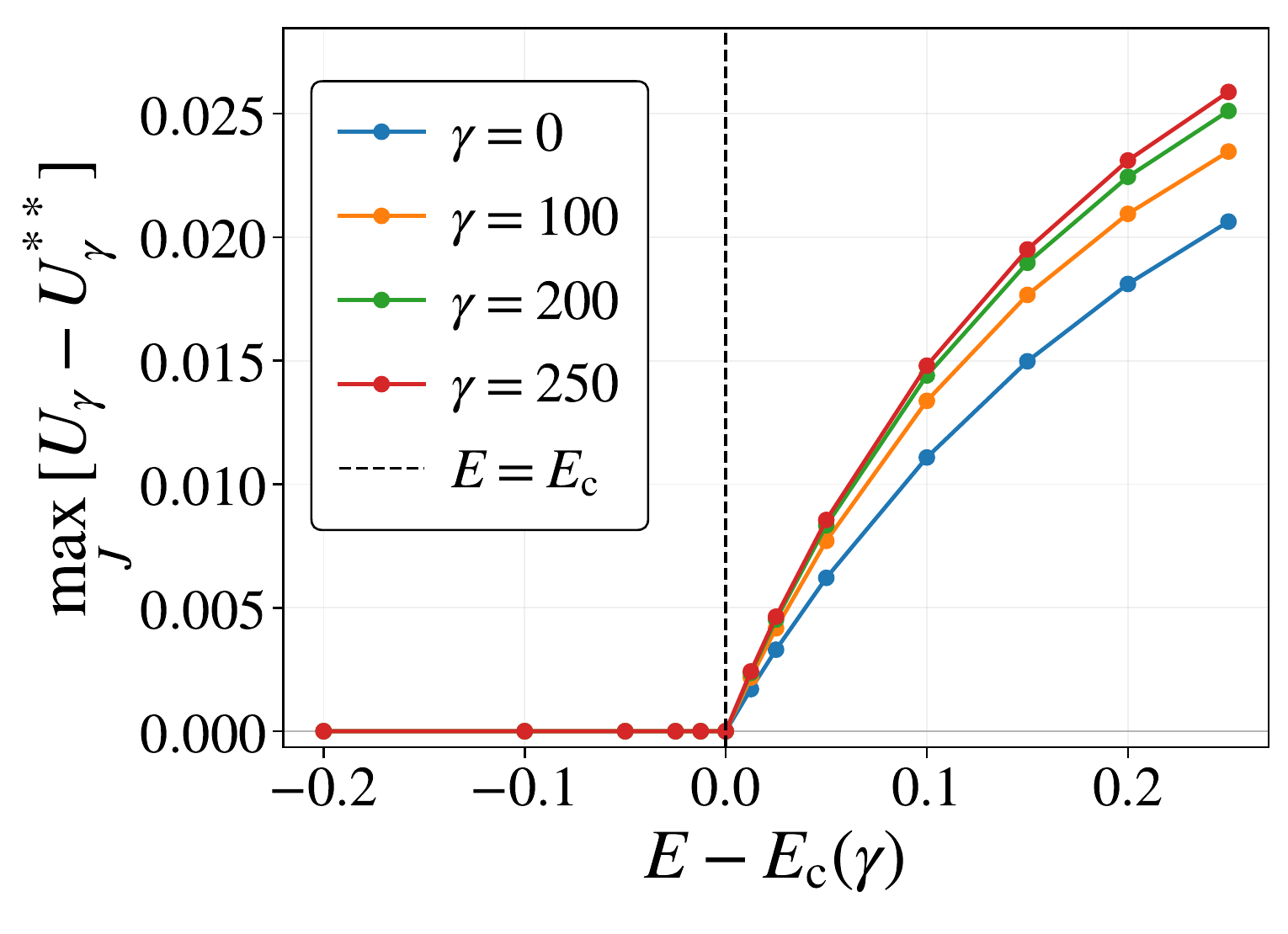}
    \hfill
    \includegraphics[width=0.32\linewidth]{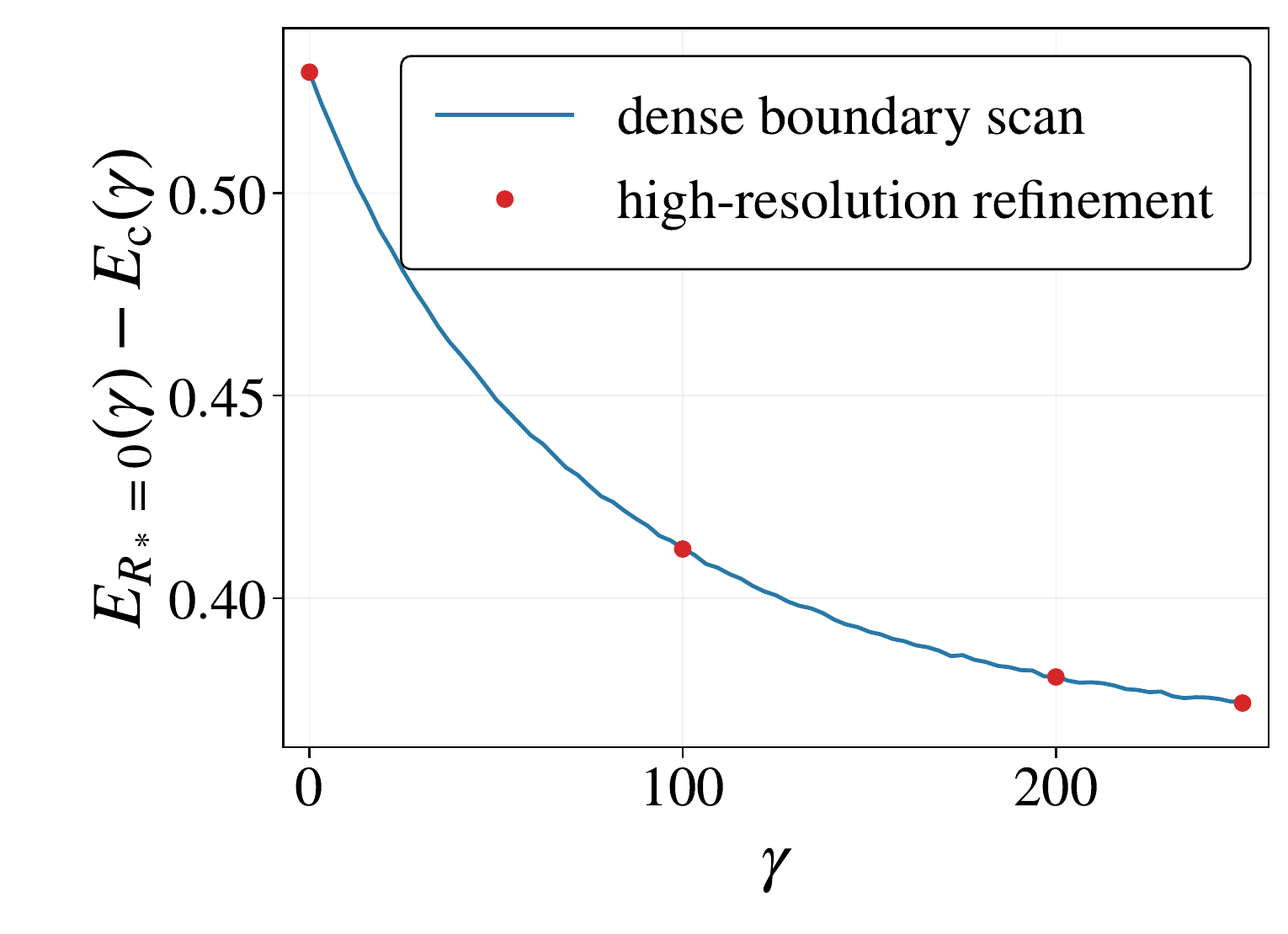}
    \caption{Tests of the $R_*(E,\gamma)=0$ prediction for
    $\gamma\in\{0,100,200,250\}$. Left: the globally determined critical
    current vanishes for $E\leq E_c(\gamma)$ and becomes positive for
    $E>E_c(\gamma)$. Center: the maximum convexification gap across the same
    boundary is zero on the no-DPT side and positive on the DPT side. Right:
    the difference between the local continuous-onset prediction
    $E_{R_*=0}(\gamma)$ and the numerically determined global onset
    $E_c(\gamma)$, obtained from both the brute-force sweep and independent
    high-resolution refinements. Their agreement confirms that the finite
    offset is not a coarse-grid artifact.}
    \label{fig:KLSEps060RBoundaryTests}
\end{figure}

To determine the actual boundary, for each
$\gamma\in\{0,100,200,250\}$ we follow the globally determined transition current as the external field is decreased. As the DPT disappears, $J_c$ is expected to approach zero. We therefore define $E_c(\gamma)$ as the field at which $J_c$ vanishes, corresponding to the threshold above which the DPT emerges. The left panel of
Fig.~\ref{fig:KLSEps060RBoundaryTests} shows this approach to zero. As an independent check, we note that the numerical boundary satisfies $E_c(\gamma)<E_{R_*=0}(\gamma)$ and hence lies in the $R_*(E,\gamma)\leq0$ region, where the local condition rules out a second-order DPT. Any transition in this region must therefore be first order, in which case $U_\gamma(J)$ differs from its convex envelope. We thus examine the convexification gap. As shown in the center panel, the gap vanishes for $E\leq E_c(\gamma)$ and becomes non-zero for $E>E_c(\gamma)$, thereby locating the same global DPT boundary as the critical-current analysis.

The right panel of Fig.~\ref{fig:KLSEps060RBoundaryTests} compares $E_c(\gamma)$, extracted both from the brute-force sweep and from independent high-resolution refinements, with the local prediction $E_{R_*=0}(\gamma)$. Their agreement confirms the numerical boundary, while the positive difference $E_{R_*=0}(\gamma)-E_c(\gamma)$ shows that $R_*(E,\gamma)=0$ is not exact. This is unsurprising: the actual boundary separates a no-DPT region from a finite-amplitude first-order transition, rather than from the continuous transition probed by the local condition. Nevertheless, $R_*(E,\gamma)=0$ remains a reasonably good approximation to the global boundary.

Taken together, these calculations give the phase diagram represented in Fig.~\ref{fig:KLSEps060PhaseDiagram}. The $B_*(E,\gamma)=0$ line separates the convexification-induced first-order sector from the second-order region.
Within the first-order region, $A_*(E,\gamma)=0$ marks the change from a discontinuous stationary minimizer to a continuously emerging branch. The global line $E_c(\gamma)$, rather than $R_*(E,\gamma)=0$, then separates the finite-amplitude first-order DPT from the no-DPT region. Thus, although the local analysis does not determine the existence boundary, the contours $A_*(E,\gamma)=0$ and $B_*(E,\gamma)=0$ correctly locate the changes in transition mechanism and order in the stationary phase diagram at $\epsilon=0.6$, Fig.~\ref{fig:KLSEps060PhaseDiagram}.

\section{Computing \texorpdfstring{$\Phi_0$}{Phi0} in the singular case}\label{app:singularlog}

To determine the finite function $\Phi_0(J)$ appearing in Eq.~\eqref{eq:RFleadinglog}, we take the large-$\gamma$ limit at fixed $J$. In the singular case, Eq.~\eqref{eq:phiJgamma1} is not convenient because $\kappa$ diverges, so we return to Eq.~\eqref{eq:phiJgamma}, which for $E=0$ gives
\begin{equation}\label{eq:phigammaJsingular}
\phi_{\gamma}(J) = \int_0^1 dx \; \left( \frac{\left( J+\partial_x \rho_{\gamma,J} \right)^2}{2 \sigma(\rho_{\gamma,J})} + \frac{\gamma}{2} \left( n_0 + (n_1 - n_0) x - \rho_{\gamma,J} \right)^2 \right).
\end{equation}
In the strict $\gamma=\infty$ limit, the saddle profile is simply the straight line $\rho_\infty(x)=n_0+(n_1-n_0)x$. For large but finite $\gamma$, this remains the correct starting point in the bulk: away from the boundaries $x=0,1$, where $\sigma(\rho)$ is non-zero, the perturbative expansion of Sec.~\ref{sec:bulkexp} still applies. Near the boundaries, however, the denominator in Eq.~\eqref{eq:phigammaJsingular} vanishes, and terms that are subleading in the bulk can become relevant inside a thin boundary region.
This requires a separate scaling analysis near the endpoints.

\begin{figure}[ht]
    \centering
    \includegraphics[width=1\linewidth]{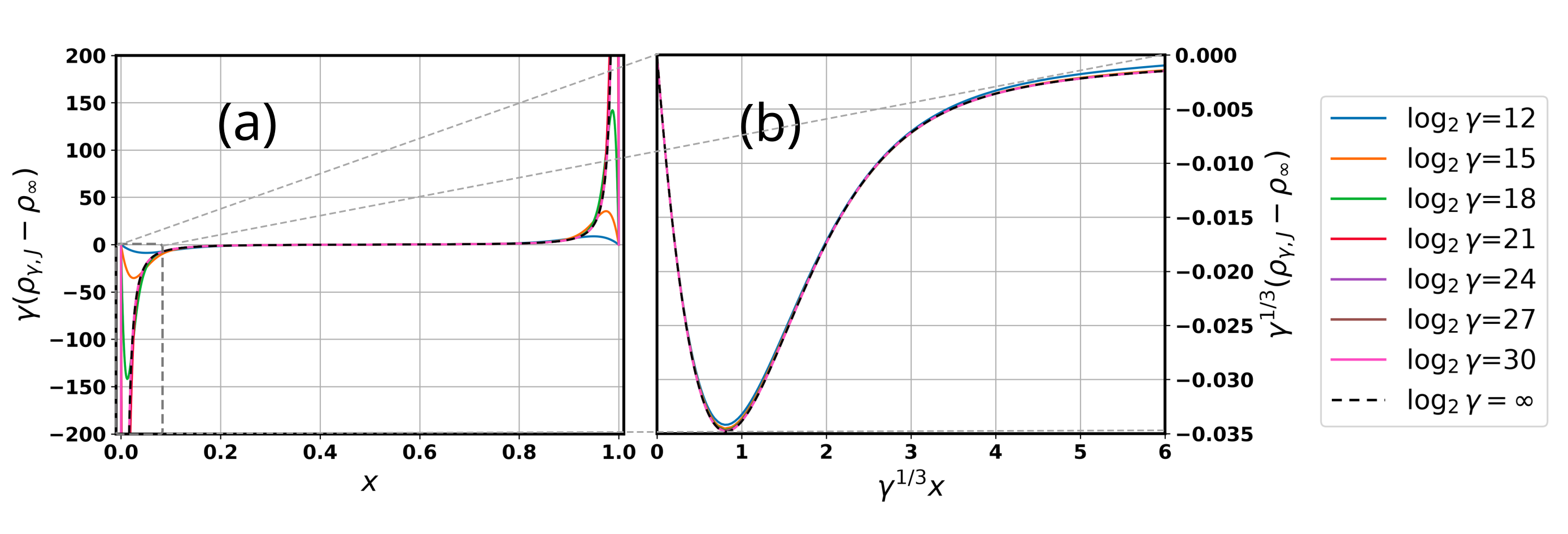}
    \caption{Monitored SSEP with $J=1.1$ and singular boundary conditions $n_0=1-n_1=1$. In (a), we plot the difference between the typical saddle-point density and $\rho_{\infty}$ as $\gamma$ increases, showing convergence to the bulk expansion away from the endpoints. In (b), after the boundary rescaling $x \to y=\gamma^{1/3}x$, we show convergence of the left boundary profile to $\rho^{(\text{L})}_{1/3,J=1.1}(y)$, obtained by solving Eq.~(\ref{eq:rhosing}) numerically.}
    \label{fig:DensityS}
\end{figure}

For convenience, let us address both boundary layers at once and perform the scaling
\begin{equation}\label{eq:singularbdrscaling}
\rho_{\gamma,J}^{(\theta)}(y)=\rho_{\gamma,J}\left(x=\beta_{\gamma}^{(\theta)}(y)\right), \qquad
\beta_{\gamma}^{(\theta)}(y)=\iota_\theta+s_\theta\frac{y}{\sqrt[3]{\gamma}}.
\end{equation}
We then proceed with the ansatz
\begin{equation}\label{eq:rhoJgammasingularB}
    \rho_{\gamma,J}^{(\theta)}(y) = n_{\iota_\theta} + \left(n_{1-\iota_\theta}-n_{\iota_\theta} \right) \gamma^{-1/3} y + \gamma^{-1/3} \rho_{1/3,J}^{(\theta)}(y) + o\left(\gamma^{-1/3}\right).
\end{equation}
Assuming that $\sigma'(n_{\iota_\theta}) \neq 0$, the denominator of the first term in Eq.~(\ref{eq:phigammaJsingular}) vanishes as $\sigma(\rho) \sim \gamma^{-1/3}\sigma'(n_{\iota_\theta})  \left(\left(n_{1-\iota_\theta}-n_{\iota_\theta}\right) y  +  \rho_{1/3,J}^{(\theta)}(y)\right)$. Note that since $\rho_{1/3,J}^{(\theta)}(0)=0$, $\sigma(\rho)$ approaches $0$ linearly in $y$ as $y \to 0$. This implies that the numerator must vanish in the limit $y \to 0$ for any $J$, as otherwise the integral in  Eq.~(\ref{eq:phigammaJsingular}) would diverge, thus imposing
\begin{equation}
    \partial_y \rho_{1/3,J}^{(\theta)}(y) \rvert_{y=0}= -\left(n_{1-\iota_\theta}-n_{\iota_\theta}\right)-s_\theta J.
\end{equation}
Applying the boundary scaling in Eq.~(\ref{eq:singularbdrscaling}) and the ansatz Eq.~(\ref{eq:rhoJgammasingularB}) to Eq.~(\ref{eq:rhof}), one obtains
\begin{equation}\label{eq:rhosing}
\resizebox{0.98\textwidth}{!}{$\displaystyle
  \frac{J^2-\left(n_{1-\iota_\theta}-n_{\iota_\theta}
      + \partial_y\rho_{1/3,J}^{(\theta)}(y)\right)^2}
      {2 \sigma'(n_{\iota_\theta}) \left(
      \left(n_{1-\iota_\theta}-n_{\iota_\theta}\right)y
      + \rho_{1/3,J}^{(\theta)}(y)\right)^2}
   + \frac{\partial_y^2 \rho_{1/3,J}^{(\theta)}(y)}
      {\sigma'(n_{\iota_\theta}) \left(
      \left(n_{1-\iota_\theta}-n_{\iota_\theta}\right)y
      + \rho_{1/3,J}^{(\theta)}(y)\right)}
  =  \rho_{1/3,J}^{(\theta)}(y).$}
\end{equation}
For monitored SSEP with $J=1.1$ and $n_0=1-n_1=1$, Fig.~\ref{fig:DensityS}(b) shows that, after the appropriate boundary rescaling, the saddle-point density profile converges to the solution of Eq.~(\ref{eq:rhosing}). In Fig.~\ref{fig:DensityS}(a), we show the corresponding bulk convergence in the open interval $(0,1)$ to the expansion in Eq.~(\ref{eq:rhopert1}).

Having identified the boundary-layer profile, we split the rate-function integral into the three regions used in Sec.~\ref{sec:asynptmethod}: $[0,\epsilon(\gamma)]$, $[\epsilon(\gamma),1-\epsilon(\gamma)]$, and $[1-\epsilon(\gamma),1]$, with $\epsilon(\gamma)=\epsilon \gamma^{-\nu}$ and $0<\nu<1/3$. The bulk expansion determines the middle interval, but diverges at its endpoints. We therefore subtract the common endpoint/large-$y$ asymptotics from the bulk and boundary contributions before taking their respective limits. The factor $y/(1+y^2)$ used below implements this subtraction while remaining regular at $y=0$.
Applying this procedure and retaining all contributions to the rate function that do not decay to $0$ as $\gamma\to\infty$, we obtain
\begin{multline}\label{eq:RFexpsing0}
\resizebox{0.95\textwidth}{!}{$\displaystyle
    \phi_{\gamma}(J) = \frac{K_{\rm log}}{2}\left(J+n_1-n_0\right)^2\log(\gamma)+\Phi_0(J)+o(1),
    \text{ where } \Phi_0(J)=\frac{\left(J+n_1-n_0\right)^2}{2}K_0+
$}\\
\resizebox{0.96\textwidth}{!}{$\displaystyle
    +\sum_{\theta\in\{\text{L},\text{R}\}} \int_0^\infty dy\;\left(\frac{\left(\rho_{1/3,J}^{(\theta)}(y)\right)^2}{2}+\frac{\left(J+n_1-n_0 + s_\theta\partial_y \rho_{1/3,J}^{(\theta)}(y)\right)^2}{2 \sigma'(n_{\iota_{\theta}}) \left((n_{1-\iota_{\theta}}-n_{\iota_{\theta}}) y + \rho_{1/3,J}^{(\theta)}(y)\right)} -\frac{\left(J+n_1-n_0\right)^2}{2 \sigma'(n_{\iota_{\theta}}) \left(n_{1-\iota_{\theta}}-n_{\iota_{\theta}}\right)} \frac{y}{1+y^2}\right),
$}
\end{multline}
Here, we remind that $\iota_{\mathrm{L}}=0$, $\iota_{\mathrm{R}}=1$, $s_\theta=(-1)^{\iota_\theta}$, and $K_{\rm log}$ is precisely the constant anticipated in Eq.~\eqref{eq:F0singlog}. The function $\Phi_0(J)$ collects all $\gamma$-independent contributions in Eq.~\eqref{eq:RFexpsing0}, with
\begin{equation*}
    K_0=\int_0^1 dx\; \left(\frac{1}{ \sigma(n_0+(n_1-n_0)x)}-\frac{1}{\sigma'(n_0) \left(n_1-n_0\right)x}-\frac{1}{\sigma'(n_1) \left(n_0-n_1\right)(1-x)}\right).
\end{equation*}
The choice of subtraction scheme only changes how finite pieces are split between $K_0$ and the boundary integrals in Eq.~\eqref{eq:RFexpsing0}; these changes cancel in $\Phi_0$, and the same cancellation occurs order by order.

\end{appendix}
\clearpage

\bibliography{biblio.bib}

\begin{thebibliography}{10}
\providecommand{\url}[1]{\texttt{#1}}
\providecommand{\urlprefix}{URL }
\expandafter\ifx\csname urlstyle\endcsname\relax
  \providecommand{\doi}[1]{doi:\discretionary{}{}{}#1}\else
  \providecommand{\doi}{doi:\discretionary{}{}{}\begingroup \urlstyle{rm}\Url}\fi
\providecommand{\eprint}[2][]{\url{#2}}

\bibitem{PhysRevB.100.134203}
R.~Vasseur, A.~C. Potter, Y.-Z. You and A.~W.~W. Ludwig,
\newblock \emph{Entanglement transitions from holographic random tensor networks},
\newblock Phys. Rev. B \textbf{100}, 134203 (2019),
\newblock \doi{10.1103/PhysRevB.100.134203}.

\bibitem{PhysRevX.12.041002}
U.~Agrawal, A.~Zabalo, K.~Chen, J.~H. Wilson, A.~C. Potter, J.~H. Pixley, S.~Gopalakrishnan and R.~Vasseur,
\newblock \emph{Entanglement and charge-sharpening transitions in u(1) symmetric monitored quantum circuits},
\newblock Phys. Rev. X \textbf{12}, 041002 (2022),
\newblock \doi{10.1103/PhysRevX.12.041002}.

\bibitem{PhysRevB.101.104302}
C.-M. Jian, Y.-Z. You, R.~Vasseur and A.~W.~W. Ludwig,
\newblock \emph{Measurement-induced criticality in random quantum circuits},
\newblock Phys. Rev. B \textbf{101}, 104302 (2020),
\newblock \doi{10.1103/PhysRevB.101.104302}.

\bibitem{PhysRevB.109.174307}
Y.~Li, R.~Vasseur, M.~P.~A. Fisher and A.~W.~W. Ludwig,
\newblock \emph{Statistical mechanics model for clifford random tensor networks and monitored quantum circuits},
\newblock Phys. Rev. B \textbf{109}, 174307 (2024),
\newblock \doi{10.1103/PhysRevB.109.174307}.

\bibitem{nahum2025_ClassicalMeasurements}
A.~Nahum and J.~L. Jacobsen,
\newblock \emph{Bayesian critical points in classical lattice models} (2025), \eprint{2504.01264}.

\bibitem{gopalakrishnan2025_ClassicalMeasurements}
S.~Gopalakrishnan, E.~McCulloch and R.~Vasseur,
\newblock \emph{Monitored fluctuating hydrodynamics} (2025), \eprint{2504.02734}.

\bibitem{gerbino2025measurement}
F.~Gerbino, G.~Giachetti, P.~L. Doussal and A.~D. Luca,
\newblock \emph{Measurement-induced phase transition in state estimation of chaotic systems and the directed polymer},
\newblock Phys. Rev. Res. pp.~-- (2025),
\newblock \doi{10.1103/6375-8ncz}.

\bibitem{PhysRevE.111.024135}
S.~W. P.~Kim and A.~Lamacraft,
\newblock \emph{Planted directed polymer: Inferring a random walk from noisy images},
\newblock Phys. Rev. E \textbf{111}, 024135 (2025),
\newblock \doi{10.1103/PhysRevE.111.024135}.

\bibitem{10.5555/897831}
G.~Welch and G.~Bishop,
\newblock \emph{An introduction to the kalman filter},
\newblock Tech. rep., USA (1995).

\bibitem{DBLP:journals/pami/YuilleC00}
A.~L. Yuille and J.~M. Coughlan,
\newblock \emph{Fundamental limits of bayesian inference: Order parameters and phase transitions for road tracking},
\newblock {IEEE} Trans. Pattern Anal. Mach. Intell. \textbf{22}(2), 160 (2000),
\newblock \doi{10.1109/34.825754}.

\bibitem{NishimoriBook2001}
H.~Nishimori,
\newblock \emph{Statistical Physics of Spin Glasses and Information Processing: An Introduction},
\newblock Oxford University Press,
\newblock ISBN 9780198509417,
\newblock \doi{10.1093/acprof:oso/9780198509417.001.0001} (2001), \eprint{https://academic.oup.com/book/5185/book-pdf/54038185/acprof-9780198509400.pdf}.

\bibitem{Ricci_Tersenghi_2019}
F.~Ricci-Tersenghi, G.~Semerjian and L.~Zdeborová,
\newblock \emph{Typology of phase transitions in bayesian inference problems},
\newblock Physical Review E \textbf{99}(4) (2019),
\newblock \doi{10.1103/physreve.99.042109}.

\bibitem{H.Nishimori_1980}
H.~Nishimori,
\newblock \emph{Exact results and critical properties of the ising model with competing interactions},
\newblock Journal of Physics C: Solid State Physics \textbf{13}(21), 4071 (1980),
\newblock \doi{10.1088/0022-3719/13/21/012}.

\bibitem{2022PhRvB.106b4305W}
J.~{Willsher}, S.-W. {Liu}, R.~{Moessner} and J.~{Knolle},
\newblock \emph{{Measurement-induced phase transition in a chaotic classical many-body system}},
\newblock Phys. Rev. B \textbf{106}(2), 024305 (2022),
\newblock \doi{10.1103/PhysRevB.106.024305},
\newblock \eprint{2203.11303}.

\bibitem{ha2024measurement}
H.~Ha, A.~Pandey, S.~Gopalakrishnan and D.~A. Huse,
\newblock \emph{Measurement-induced phase transitions in systems with diffusive dynamics},
\newblock Phys. Rev. B \textbf{110}, L140301 (2024),
\newblock \doi{10.1103/PhysRevB.110.L140301}.

\bibitem{2025arXiv250408888K}
S.~W.~P. {Kim}, C.~{von Keyserlingk} and A.~{Lamacraft},
\newblock \emph{{Measurement-induced phase transitions in quantum inference problems and quantum hidden Markov models}},
\newblock arXiv e-prints arXiv:2504.08888 (2025),
\newblock \doi{10.48550/arXiv.2504.08888},
\newblock \eprint{2504.08888}.

\bibitem{2025arXiv250412385P}
M.~{P{\"u}tz}, S.~J. {Garratt}, H.~{Nishimori}, S.~{Trebst} and G.-Y. {Zhu},
\newblock \emph{{Learning transitions in classical Ising models and deformed toric codes}},
\newblock arXiv e-prints arXiv:2504.12385 (2025),
\newblock \doi{10.48550/arXiv.2504.12385},
\newblock \eprint{2504.12385}.

\bibitem{wiese2026bayesian}
K.~J. Wiese, A.~Das and A.~Nahum,
\newblock \emph{Bayesian phase transition for the critical ising model: Enlarged replica symmetry in the epsilon expansion and in 2d} (2026), \eprint{2604.23346}.

\bibitem{zdeborova2016statistical}
L.~Zdeborov\'a and F.~Krzakala,
\newblock \emph{Statistical physics of inference: thresholds and algorithms},
\newblock Adv. Phys. \textbf{65}, 453 (2016),
\newblock \doi{10.1080/00018732.2016.1211393}.

\bibitem{iba1999nishimori}
Y.~Iba,
\newblock \emph{The nishimori line and bayesian statistics},
\newblock J. Phys. A: Math. Gen. \textbf{32}, 3875 (1999),
\newblock \doi{10.1088/0305-4470/32/21/302}.

\bibitem{RevModPhys.87.593}
L.~Bertini, A.~De~Sole, D.~Gabrielli, G.~Jona-Lasinio and C.~Landim,
\newblock \emph{Macroscopic fluctuation theory},
\newblock Rev. Mod. Phys. \textbf{87}, 593 (2015),
\newblock \doi{10.1103/RevModPhys.87.593}.

\bibitem{PhysRevLett.94.030601}
L.~Bertini, A.~De~Sole, D.~Gabrielli, G.~Jona-Lasinio and C.~Landim,
\newblock \emph{Current fluctuations in stochastic lattice gases},
\newblock Phys. Rev. Lett. \textbf{94}, 030601 (2005),
\newblock \doi{10.1103/PhysRevLett.94.030601}.

\bibitem{Landim_Bertini_2007}
L.~Bertini, A.~D. Sole, D.~Gabrielli, G.~Jona-Lasinio and C.~Landim,
\newblock \emph{Stochastic interacting particle systems out of equilibrium},
\newblock Journal of Statistical Mechanics: Theory and Experiment \textbf{2007}(07), P07014–P07014 (2007),
\newblock \doi{10.1088/1742-5468/2007/07/p07014}.

\bibitem{Eyink1990}
G.~Eyink, J.~L. Lebowitz and H.~Spohn,
\newblock \emph{Hydrodynamics of stationary non-equilibrium states for some stochastic lattice gas models},
\newblock Communications in Mathematical Physics \textbf{132}(1), 253 (1990),
\newblock \doi{10.1007/BF02278011}.

\bibitem{spohn2012large}
H.~Spohn,
\newblock \emph{Large scale dynamics of interacting particles},
\newblock Springer Science \& Business Media (2012).

\bibitem{kipnis2013scaling}
C.~Kipnis and C.~Landim,
\newblock \emph{Scaling limits of interacting particle systems}, vol. 320,
\newblock Springer Science \& Business Media (2013).

\bibitem{kipnis1989hydrodynamics}
C.~Kipnis, S.~Olla and S.~S. Varadhan,
\newblock \emph{Hydrodynamics and large deviation for simple exclusion processes},
\newblock Communications on Pure and Applied Mathematics \textbf{42}(2), 115 (1989).

\bibitem{Marcuzzi_2014}
M.~Marcuzzi, J.~Schick, B.~Olmos and I.~Lesanovsky,
\newblock \emph{Effective dynamics of strongly dissipative rydberg gases},
\newblock Journal of Physics A: Mathematical and Theoretical \textbf{47}(48), 482001 (2014),
\newblock \doi{10.1088/1751-8113/47/48/482001}.

\bibitem{Bauer_2015}
M.~Bauer, D.~Bernard and A.~Tilloy,
\newblock \emph{Computing the rates of measurement-induced quantum jumps},
\newblock Journal of Physics A: Mathematical and Theoretical \textbf{48}(25), 25FT02 (2015),
\newblock \doi{10.1088/1751-8113/48/25/25ft02}.

\bibitem{10.21468/SciPostPhys.3.5.033}
M.~Bauer, D.~Bernard and T.~Jin,
\newblock \emph{{Stochastic dissipative quantum spin chains (I) : Quantum fluctuating discrete hydrodynamics}},
\newblock SciPost Phys. \textbf{3}, 033 (2017),
\newblock \doi{10.21468/SciPostPhys.3.5.033}.

\bibitem{Touchette_2009}
H.~Touchette,
\newblock \emph{The large deviation approach to statistical mechanics},
\newblock Physics Reports \textbf{478}(1–3), 1–69 (2009),
\newblock \doi{10.1016/j.physrep.2009.05.002}.

\bibitem{Touchette_2012}
H.~Touchette,
\newblock \emph{A basic introduction to large deviations: Theory, applications, simulations} (2012), \eprint{1106.4146}.

\bibitem{Derrida2004}
B.~{Derrida}, B.~{Dou{\c{c}}ot} and P.~E. {Roche},
\newblock \emph{{Current Fluctuations in the One-Dimensional Symmetric Exclusion Process with Open Boundaries}},
\newblock Journal of Statistical Physics \textbf{115}, 717 (2004),
\newblock \doi{10.1023/B:JOSS.0000022379.95508.b2},
\newblock \eprint{cond-mat/0310453}.

\bibitem{Derrida_2007}
B.~Derrida,
\newblock \emph{Non-equilibrium steady states: fluctuations and large deviations of the density and of the current},
\newblock Journal of Statistical Mechanics: Theory and Experiment \textbf{2007}(07), P07023 (2007),
\newblock \doi{10.1088/1742-5468/2007/07/P07023}.

\bibitem{Akkermans_2013}
E.~Akkermans, T.~Bodineau, B.~Derrida and O.~Shpielberg,
\newblock \emph{Universal current fluctuations in the symmetric exclusion process and other diffusive systems},
\newblock Europhysics Letters \textbf{103}(2), 20001 (2013),
\newblock \doi{10.1209/0295-5075/103/20001}.

\bibitem{DerridaAdditivityPrinciple}
T.~Bodineau and B.~Derrida,
\newblock \emph{Current fluctuations in nonequilibrium diffusive systems: An additivity principle},
\newblock Phys. Rev. Lett. \textbf{92}, 180601 (2004),
\newblock \doi{10.1103/PhysRevLett.92.180601}.

\bibitem{LecompteActiveLatticeGas}
T.~Agranov, S.~Ro, Y.~Kafri and V.~Lecomte,
\newblock \emph{{Macroscopic fluctuation theory and current fluctuations in active lattice gases}},
\newblock SciPost Phys. \textbf{14}, 045 (2023),
\newblock \doi{10.21468/SciPostPhys.14.3.045}.

\bibitem{10.21468/SciPostPhys.15.4.136}
B.~Doyon, G.~Perfetto, T.~Sasamoto and T.~Yoshimura,
\newblock \emph{{Ballistic macroscopic fluctuation theory}},
\newblock SciPost Phys. \textbf{15}, 136 (2023),
\newblock \doi{10.21468/SciPostPhys.15.4.136}.

\bibitem{PhysRevB.110.L060202}
T.~Jin and D.~G. Martin,
\newblock \emph{Measurement-induced phase transition in a single-body tight-binding model},
\newblock Phys. Rev. B \textbf{110}, L060202 (2024),
\newblock \doi{10.1103/PhysRevB.110.L060202}.

\bibitem{Jacobs01092006}
K.~Jacobs and D.~A. Steck,
\newblock \emph{A straightforward introduction to continuous quantum measurement},
\newblock Contemporary Physics \textbf{47}(5), 279 (2006),
\newblock \doi{10.1080/00107510601101934},
\newblock \eprint{https://doi.org/10.1080/00107510601101934}.

\bibitem{wiseman2009quantum}
H.~M. Wiseman and G.~J. Milburn,
\newblock \emph{Quantum measurement and control},
\newblock Cambridge university press (2009).

\bibitem{barchielli2009quantum}
A.~Barchielli and M.~Gregoratti,
\newblock \emph{Quantum trajectories and measurements in continuous time: the diffusive case}, vol. 782,
\newblock Springer Science \& Business Media (2009).

\bibitem{wiseman}
H.~M. Wiseman and G.~J. Milburn,
\newblock \emph{Quantum measurement and control},
\newblock Cambridge university press (2009).

\bibitem{Bertini_2006}
L.~Bertini, A.~D. Sole, D.~Gabrielli, G.~Jona-Lasinio and C.~Landim,
\newblock \emph{Non equilibrium current fluctuations in stochastic lattice gases},
\newblock Journal of Statistical Physics \textbf{123}(2), 237–276 (2006),
\newblock \doi{10.1007/s10955-006-9056-4}.

\bibitem{Lecomte_2017}
Y.~Baek, Y.~Kafri and V.~Lecomte,
\newblock \emph{Dynamical symmetry breaking and phase transitions in driven diffusive systems},
\newblock Physical Review Letters \textbf{118}(3) (2017),
\newblock \doi{10.1103/physrevlett.118.030604}.

\bibitem{SPITZER1970246}
F.~Spitzer,
\newblock \emph{Interaction of markov processes},
\newblock Advances in Mathematics \textbf{5}(2), 246 (1970),
\newblock \doi{https://doi.org/10.1016/0001-8708(70)90034-4}.

\bibitem{EvansHanney2005}
M.~R. Evans and T.~Hanney,
\newblock \emph{Nonequilibrium statistical mechanics of the zero-range process and related models},
\newblock Journal of Physics A: Mathematical and General \textbf{38}(19), R195 (2005),
\newblock \doi{10.1088/0305-4470/38/19/R01}.

\bibitem{DerridaKMP}
T.~Bodineau and B.~Derrida,
\newblock \emph{Distribution of current in nonequilibrium diffusive systems and phase transitions},
\newblock Physical Review E \textbf{72}(6) (2005),
\newblock \doi{10.1103/physreve.72.066110}.

\bibitem{ZarfatyKMP}
L.~Zarfaty and B.~Meerson,
\newblock \emph{Statistics of large currents in the kipnis–marchioro–presutti model in a ring geometry},
\newblock Journal of Statistical Mechanics: Theory and Experiment \textbf{2016}(3), 033304 (2016),
\newblock \doi{10.1088/1742-5468/2016/03/033304}.

\bibitem{OhadShpielbergNumerical}
O.~Shpielberg, Y.~Don and E.~Akkermans,
\newblock \emph{Numerical study of continuous and discontinuous dynamical phase transitions for boundary-driven systems},
\newblock Phys. Rev. E \textbf{95}, 032137 (2017),
\newblock \doi{10.1103/PhysRevE.95.032137}.

\bibitem{gallavotti1995dynamical}
G.~Gallavotti and E.~G.~D. Cohen,
\newblock \emph{Dynamical ensembles in nonequilibrium statistical mechanics},
\newblock Physical review letters \textbf{74}(14), 2694 (1995).

\bibitem{LebowitzSpohn1999}
J.~L. Lebowitz and H.~Spohn,
\newblock \emph{A {Gallavotti--Cohen}-type symmetry in the large deviation functional for stochastic dynamics},
\newblock Journal of Statistical Physics \textbf{95}, 333 (1999),
\newblock \doi{10.1023/A:1004589714161}.

\bibitem{Kurchan1998}
J.~Kurchan,
\newblock \emph{Fluctuation theorem for stochastic dynamics},
\newblock Journal of Physics A: Mathematical and General \textbf{31}(16), 3719 (1998),
\newblock \doi{10.1088/0305-4470/31/16/003}.

\bibitem{KLS1984}
S.~Katz, J.~L. Lebowitz and H.~Spohn,
\newblock \emph{Nonequilibrium steady states of stochastic lattice gas models of fast ionic conductors},
\newblock Journal of Statistical Physics \textbf{34}(3-4), 497 (1984),
\newblock \doi{10.1007/BF01018556}.

\bibitem{Hager2001MinimalCP}
J.~S. Hager, J.~H.~A. Krug, V.~Popkov and G.~M. Sch{\"u}tz,
\newblock \emph{Minimal current phase and universal boundary layers in driven diffusive systems.},
\newblock Physical review. E, Statistical, nonlinear, and soft matter physics \textbf{63 5 Pt 2}, 056110 (2001).

\bibitem{Krapivsky_2013}
P.~L. Krapivsky,
\newblock \emph{Dynamics of repulsion processes},
\newblock Journal of Statistical Mechanics: Theory and Experiment \textbf{2013}(06), P06012 (2013),
\newblock \doi{10.1088/1742-5468/2013/06/p06012}.

\bibitem{PerezEspigares2018}
C.~P{\'e}rez-Espigares, F.~Carollo, J.~P. Garrahan and P.~I. Hurtado,
\newblock \emph{Dynamical criticality in open systems: Non-perturbative physics, microscopic origin and direct observation},
\newblock Physical Review E \textbf{98}, 060102 (2018),
\newblock \doi{10.1103/PhysRevE.98.060102}.

\end{thebibliography}

\nolinenumbers

\end{document}